\documentclass[lettersize,journal]{IEEEtran}
\usepackage{amsmath,amsfonts}
\usepackage{algorithmic}
\usepackage{array}
\usepackage[caption=false,font=normalsize,labelfont=sf,textfont=sf]{subfig}
\usepackage{textcomp}
\usepackage{stfloats}
\usepackage{url}
\usepackage{verbatim}
\usepackage{graphicx}
\def\BibTeX{{\rm B\kern-.05em{\sc i\kern-.025em b}\kern-.08em
    T\kern-.1667em\lower.7ex\hbox{E}\kern-.125emX}}
\usepackage{balance}
\usepackage{boondox-cal}
\usepackage{cite}
\usepackage{amssymb}
\begin{document}
\title{Last-Pair Swapping Polar Codes: A Structure to Improve Polarization under
Finite-State Modulation}
\author{Yinuo Mei, Yangyong Zhang, and Daiming Qu
\thanks{This work was supported in part by the National Scientific Research Project
of China under Grant JCKY2022206C004. (Corresponding author: Daiming
Qu.)

The authors are with the School of Electronic Information and
Communications, Huazhong University of Science and Technology,
Wuhan 430074, China. (email: D202280923@hust.edu.cn; medy99@126.com; qudaiming@hust.edu.cn).

YangYong Zhang is also with the Wuhan Maritime Communication Research Institute, Wuhan 430205, China}}

\maketitle

\begin{abstract}
A novel structure of polar codes is proposed for
finite-state modulation (FSM), in order to improve polarization under it, and approach the
polarization efficiency that conventional polar codes achieve under memoryless channels. We choose a particular class of FSM for research, termed bijective FSM, and observe an explicit polarization loss under bijective FSM. To eliminate the loss, we propose a novel polar coding structure by substituting the last
kernel of each layer in polar coding structure with a swapping matrix, thereby termed last-pair swapping structure. We prove that under bijective FSM the proposed structure achieves identical polarization efficiency with that of conventional one on memoryless channels, and exceeds that of conventional one under bijective FSM. Furthermore, we give a plausible generalization of last-pair swapping polar code: on a broader class termed sub-injective FSM. Simulation corroborates that under sub-injective FSM polarization efficiency of the proposed polar code exceeds that of conventional one. And simulation
results of error rate are given on continuous phase modulation (CPM)
with additional white Gaussian noise (AWGN) channels, showing a
considerable signal-to-noise power ratio (snr) gain of last-pair swapping polar
code over conventional one, and identical performances between the proposed
polar code under bijective FSM and conventional one on
memoryless channels.
\end{abstract}


\section{Introduction}

\IEEEPARstart{P}{olar} codes \cite{ref1}, as the first kind of deterministic capacity-achieving codes \cite{ref2} that can be constructed, encoded, and decoded with acceptable complexity, have been applied to 5G mobile communications \cite{ref3}. However, constructed for
discrete memoryless channels (DMC), it remains as a problem whether
polar codes could be applied to channels with memory, initially. Whilst
as widely acknowledged, memory channels include many significant
types of channels, which are far from negligible. Therefore, during later research,
efforts are made successively for application of polar codes to channels
with memory. In 2016, E. Şaşoğlu and I. Tal proved polarization holds on
finite-state channels analytically \cite{ref6}. Their theoretical work
revealed the feasibility of polar codes on finite-state channels. Soon
after, Runxin Wang \emph{et al.} accomplished with a decoding algorithm
for polar codes on finite-state channels, termed successive cancellation
trellis decoding (SCTD) \cite{ref7}, which enables decoding of polar codes
on memory channels like that on memoryless ones.

Despite that polarization of polar codes on finite-state channels is proved, its efficiency  or convergence rate remains as a problem. Notably, proof of polarization on memory channels is on an
asymptotic condition, as is on memoryless channels. It
does not guarantee that on finite code length, polarization on memory
channel is identical with that on memoryless channels. In other words,
even though both of them converge to identical limit, their convergency
rate may differ. Unfortunately, just as is noticed in our work, on a
certain category of finite-state channels, polarization is rather slower
in comparison with its counterpart on memoryless channels. Namely, on each finite
code length, the extent for sub-channel capacities to approach poles and
be off average is less than that on memoryless channels. Such
polarization loss weakens performance of polar codes on memory channels.
Therefore, there arises a requirement to modify coding structure of
polar codes for better application on finite-state channels.

Among the various finite-state channels, a salient subclass stands out due to its practical significance and prevalence in the field, which is termed as finite-state modulated channels (FSMC). Each channel of that subclass is composed of a finite-state modulation (FSM) and a memoryless channel. For instance,
continuous phase modulation (CPM) \cite{ref4} along with an memoryless
channel forms an FSMC, as implied in \cite{ref5}. And CPM draws wide concern for its constant envelope and continuous phase, which lead to low peak-to-average power ratio (PAPR) and high
spectral efficiency. Therefore, it is of particular interest to address the challenge of applying polar codes on FSMC.


In this paper, we focus on FSMC and study the application of polar codes on it. Compared to memoryless channels, the distinctiveness of FSMC stems from its modulation with memory: FSM. Therefore, to put the issue in another way, the research focuses on FSM and the application of polar code under it. Furthermore, the behavior of an FSM is determined by its transition function. Then it is necessary to examine the transition function, in order to explore the scope of FSM under which polar code suffers polarization loss. Firstly we find a particular class of FSM termed bijective FSM, and observe an evident polarization loss under it. To find out the cause for such loss, we investigate the coding
structure of conventional polar code on bijective FSM.
By observing capacity polarization of each sub-channel pair on every
layer, we discover that certain pairs fail to polarize.
Focusing on each single pair of them, we notice a structural similarity
between the transition function and Arikan's kernel, and a counteraction occurring
in their concatenation, in certain locations, which eliminates
polarization effect of Arikan's kernel. As a modification, we
substitute the Arikan kernel at these pairs by a swapping matrix, which
brings about a new coding structure we call last-pair swapping structure.
And we prove the polarization loss can be eliminated by our proposed structure.
Furthermore, inspired by the crucial property of bijective FSM which results in a polarization loss and determines the applicability of the proposed structure, termed intrinsic polarization potency, we broaden the scope of FSM that requires last-pair swapping structure: from bijective FSM to a broader class termed sub-injective FSM. Namely, a plausible generalization is made for last-pair swapping polar code. And simulation of polarization efficiency corroborates the superiority of our proposed structure under sub-injective FSM. Finally simulation results of error rate in communication further verify the advantage of the proposed
structure.

The rest of this paper is organized as follows. In
Section II, basic concepts over polar codes and finite-state channels
are presented, along with decoding algorithm. And in Section III the system model is presented. Section
IV focuses on a particular class of FSM termed bijective FSM, observes polarization loss under it, then investigates the coding structure for explanation of polarization
loss and offers a new coding structure as remedy, and proves the improvement. Furthermore in Section V a plausible generalization of last-pair swapping polar code is made. And performance
of frame error rate is compared between the proposed structure and
conventional one in Section VI. Finally, Section VII serves as a summary
of above contents.

\section{Preliminaries}

\subsection{Polar Codes}

Polar codes are a class of linear block codes. Encoding of polar codes
with code length \(N\) and information bit number \(K\), namely code
rate \(R = \frac{K}{N}\), includes assigning information bits
\(c_{0}^{K - 1}\) to an \emph{N}-length source bit sequence \(u_{0}^{N - 1}\),
and mapping source bits \(u_{0}^{N - 1}\) to final codeword
\(x_{0}^{N - 1}\) with a linear operator, which can be represented as a
matrix \(G_{N}\) termed generator matrix. In formula, codeword
\(x_{0}^{N - 1} = u_{0}^{N - 1}G_{N}\), where \(u_{0}^{N - 1}\) is a
combination of information bits \(c_{0}^{K - 1}\) and frozen bits zeros,
namely
\(u_{\mathcal{A}} = c_{0}^{K - 1},u_{\mathcal{A}^{\complement}} = \lbrack 0,\ldots,0\rbrack\).
Here \(\mathcal{A}\) is the index sequence in \(u_{0}^{N - 1}\) to
install information bits, contrarily bits on indices
\(\mathcal{A}^{\complement} = \left\{ 0,1,\ldots,N - 1 \right\}\mathcal{\backslash A}\)
in \(u_{0}^{N - 1}\) are zeros. In other words,
\(u_{\mathcal{A}_{j}} = c_{j},\forall j \in \left\{ 0,1,\ldots,K - 1 \right\};u_{i} = 0,\forall i \in \mathcal{A}^{\complement}\).
Generator matrix of polar codes with code length
\(N = 2^{n},n \in \mathbb{N}^{*}\) is \(G_{N} = B_{N}F^{\otimes n}\),
where \(B_{N}\) is a bit-reversal permutation matrix, and
\(F = \begin{bmatrix}
1 & 0 \\
1 & 1
\end{bmatrix}\) is termed polar kernel, besides, operator \(\otimes n\)
means \emph{n}-th Kronecker power. \cite{ref1}


When decoding, polar code utilizes received signal and decoded source bits
to estimate each source bit. In other words, it constructs \(N\) channels
for every source bit, termed sub-channels, the \emph{i}-th of which is
\(U_{i} \mapsto Y_{0}^{N - 1}U_{0}^{i - 1},i \in \left\{ 0,1,\ldots,N - 1 \right\}\).
So the transition probability of the \emph{i}-th sub-channel is
\(W_{N}^{(i)}\left( y_{0}^{N - 1},u_{0}^{i - 1} \middle| u_{i} \right)\). \(y_i\) denotes the \(i\)-th output of the channel. In this paper, while in non-bold, \(y_i\) refers to a general form output of the channel, which could be a scalar, a vector, a function, a set, etc; and while in bold, \(\boldsymbol{y}_i\) refers specifically to a vector. So does \(Y_i\), the random variable of \(y_i\). The capacity 
of the \(i\)-th sub-channel is
\(I\left( W_{N}^{(i)} \right) = I\left( Y_{0}^{N - 1}U_{0}^{i - 1};U_{i} \right)\).
In this paper, we denote sub-channel capacity \(I\left( W_{N}^{(i)} \right)\) as
\(I_{N}^{(i)}\), and their sequence as \(\mathcal{I}_{N}\), i.e.,
\(\mathcal{I}_{N} := \left\lbrack I_{N}^{(i)} \right\rbrack_{i \in \left\{ 0,\ldots,N - 1 \right\}}\).
Moreover, in a context where a specification of coding structure and
code channel is necessary, we mark the \(i\)-th sub-channel of coding
structure \(\mathcal{P}\) with code length \(N\) on code channel
\(\mathcal{C}\) as \(W_{\left( \mathcal{P,C} \right),N}^{(i)}\), and
similarly, its capacity as \(I_{\left( \mathcal{P,C} \right),N}^{(i)}\),
the sequence of which as
\(\mathcal{I}_{\left( \mathcal{P,C} \right),N}\). Besides, we denote
conventional polar coding structure as \(\mathcal{P}_{c}\).

Due to the recursive property of generator matrix \(G_{N}\), on a
memoryless channel \(W^N\), which means the \(N\)-th extension of single symbol channel \(W\), it is not difficult to derive recursive expression
between transition probability on adjacent layers \cite{ref1}
\begin{align*}
	& W_{2N}^{(2i)}\left( y_{0}^{2N - 1},u_{0}^{2i - 1} \middle| u_{2i} \right) \\
	& =\sum_{u_{2i + 1}\mathbb{\in B}}^{} W_{N}^{(i)}\left( y_{0}^{N - 1},v_{0,0}^{i - 1} \middle| u_{2i} + u_{2i + 1} \right)  \\
	& \cdot W_{N}^{(i)}\left( y_{N}^{2N - 1},v_{1,0}^{i - 1} \middle| u_{2i + 1} \right)
\end{align*}
\begin{align*}
	&W_{2N}^{(2i + 1)}\left( y_{0}^{2N - 1},u_{0}^{2i} \middle| u_{2i + 1} \right) \\
	&= W_{N}^{(i)}\left( y_{0}^{N - 1},v_{0,0}^{i - 1} \middle| u_{2i} + u_{2i + 1} \right)W_{N}^{(i)}\left( y_{N}^{2N - 1},v_{1,0}^{i - 1} \middle| u_{2i + 1} \right)
\end{align*}
where
\(v_{0,i} = u_{2i} + u_{2i + 1},v_{1,i} = u_{2i + 1},i \in \left\{ 0,1,\ldots,N - 1 \right\}\), \(\mathbb{B}:=\{0,1\}\), and \(W_1^{(0)}\) means \(W\).

The fundamental decoding algorithm of polar code is successive
cancellation (SC) \cite{ref1}. SC estimates each source bit \(u_{i}\) by
received signal \(y_{0}^{N - 1}\) and former estimated source
bits \({\widehat{u}}_{0}^{i - 1}\), as
the following expression indicates.
\begin{align*}
	{\widehat{u}}_{i} = \left\{
	\begin{aligned}
 		& \arg{\max_{u_{i} \in \mathbb{B}}{W\left( y_{0}^{N - 1},{\widehat{u}}_{0}^{i - 1} \middle| u_{i} \right)}} & ,i \in \mathcal{A} \\
 		& 0 & ,i \in \mathcal{A}^{\complement}
	\end{aligned} \right.
\end{align*}

However, SC decoding suffers performance loss due to probable decoding
error of former source bits \(u_{0}^{i - 1}\), i.e.,
\({\widehat{u}}_{0}^{i - 1} \neq u_{0}^{i - 1}\), which fails to ensure
the accuracy of former source bits adopted at decoding phase \(i\).
Therefore, I. Tal and A. Vardy proposed a list decoding algorithm, which
saves multiple decoding paths during decoding, each of which records an
estimated bit sequence, but the number of paths is no larger than a specified number. Such algorithm is
termed successive cancellation list (SCL) \cite{ref8}. Considerable
performance improvement and acceptable time and space complexity made SCL a practical
decoding algorithm for polar codes.

\subsection{Finite-State Channels}

Finite state channels refer to memory channels which satisfy
\begin{align*}
	&p_{Y_{n}S_{n}|X_{- \infty}^{n}S_{- \infty}^{n - 1}Y_{- \infty}^{n - 1}}\left( y_{n},s_{n} \middle| x_{- \infty}^{n},s_{- \infty}^{n - 1},y_{- \infty}^{n - 1} \right) \\
	&= p_{Y_{n}S_{n}|X_{n}S_{n - 1}}\left( y_{n},s_{n} \middle| x_{n},s_{n - 1} \right)
\end{align*}
where \(n\) denotes present time, \(S_{n}\) is present state,
\(S_{n}\mathcal{\in S,\forall}n\mathbb{\in Z}\), \(\mathcal{S}\) is a finite
set. \(X_{n}\) is the present input of channels, and \(Y_{n}\)
is the output. Namely, a finite-state channel has finite states, and at
any time, the current output and state depend only on the current input and the
last state \cite{ref9}.

\subsection{Successive Cancellation Trellis
Decoding under Finite-State Channel}

As presented in \cite{ref7}, for polar codes on finite-state channels,
the transition probability of sub-channels in each layer can be derived from their
lower layer, following formulas
\begin{align*}
	&W_{2N}^{(2i + 1)}\left( y_{0}^{2N - 1},u_{0}^{2i},s_{2N - 1} \middle| u_{2i + 1},\ s_{0} \right) \\
	&= \frac{1}{2}\sum_{s_{N - 1}\mathcal{\in S}}^{} W_{N}^{(i)}\left( y_{0}^{N - 1},v_{0,0}^{i - 1},s_{N - 1} \middle| u_{2i} + u_{2i + 1},s_{0} \right) \\
	& \cdot W_{N}^{(i)}\left( y_{N}^{2N - 1},v_{1,0}^{i - 1},s_{2N - 1} \middle| u_{2i + 1},s_{N - 1} \right)
\end{align*}
\begin{align*}
	&W_{2N}^{(2i)}\left( y_{0}^{2N - 1},u_{0}^{2i - 1},s_{2N - 1} \middle| u_{2i},s_{0} \right) \\
	&= \frac{1}{2}\sum_{u_{2i + 1}\in\mathbb{B}}^{}\sum_{s_{N - 1}\mathcal{\in S}}^{}W_{N}^{(i)}\left( y_{0}^{N - 1},v_{0,0}^{i - 1},s_{N - 1} \middle| u_{2i} + u_{2i + 1},s_{0} \right) \\
	&\cdot W_{N}^{(i)}\left( y_{N}^{2N - 1},v_{1,0}^{i - 1},s_{2N - 1} \middle| u_{2i + 1},s_{N - 1} \right)
\end{align*}
where
\(v_{0,j} := u_{2j} + u_{2j + 1},v_{1,j} := u_{2j + 1},\forall j \in \left\{ 0,1,\ldots,N - 1 \right\}\).

And as an initial case, namely \(N = 1\), the transition probability
with state on the channel layer is
\(W_{1}^{(0)}\left( y_{n},s_{n} \middle| x_{n},s_{n - 1} \right) = W\left( y_{n},s_{n} \middle| x_{n},s_{n - 1} \right)\),
which is determined by the channel.

One can calculate transition probability with state in the channel
layer, and utilize formulas above to recursively derive transition probability with state in the
decision layer. The recursive calculation has a form like that in
conventional polar codes, only with state.

Finally in the decision layer, there is
\begin{align*}
	&W_{N}^{(i)}\left( y_{0}^{N - 1},u_{0}^{i - 1} \middle| u_{i} \right) \\
	&= \sum_{s_{0}\mathcal{\in S}}^{}{\sum_{s_{N - 1}\mathcal{\in S}}^{}{W_{N}^{(i)}\left( y_{0}^{N - 1},u_{0}^{i - 1},s_{N - 1} \middle| u_{i},s_{0} \right)}}p_{S_{0}}\left( s_{0} \right)
\end{align*}
which finally eliminates the state and gets transition probability in the
decision layer. And according to the transition probability, the decision of source bits could be made by the same way as SC and SCL. Such decoding is termed successive cancellation trellis decoding (SCTD)\cite{ref7}.

\section{The System Model}

\begin{figure}[!t]
\centering
\includegraphics[width=3.5in]{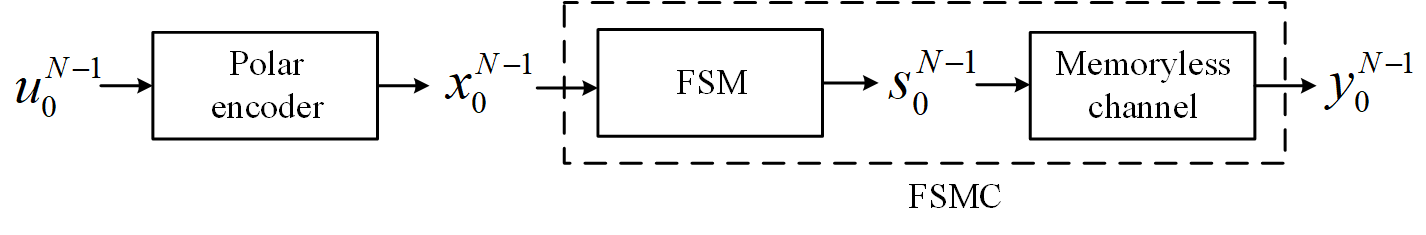}
\caption{System model.}
\label{fig_ChModel}
\end{figure}

\subsection{Definitions about the Channel}

In this section we give the studied system model, which is illustrated in Fig.~\ref{fig_ChModel}, composed of an encoder and a channel. The encoder is a polar encoder, and the channel is a particular class of finite-state channel we name finite-state modulated channels, composed of a finite-state modulation and a
memoryless channel. The input of encoder \(u_{0}^{N - 1}\)
is a source bit sequence, the output of it, also the input of modulator
\(x_{0}^{N - 1}\) is a codeword, and the output of modulator
\(s_{0}^{N - 1}\) is a state sequence, each component of which is termed
state. Each component
\(u_{n},x_{n} \in \mathbb{B}, \ s_{n}\mathcal{\in S}\),
\(\mathbb{B}=\{0,1\}\), and \(\mathcal{S}\) is a finite set. Definition of finite-state modulation is:

\emph{Definition 1:} Finite-State Modulation (FSM). A modulation
\(\mathcal{M:}\mathbb{B}^{N} \mapsto \mathcal{S}^{N}\) is a finite-state
modulation iff there exists a function
\(F:\mathcal{S} \times \mathbb{B} \mapsto \mathcal{S}\) such that
\begin{align*}
	s_{n} = \left\{
	\begin{aligned}
 		& F\left( 0,x_{n} \right) & ,n = 0 \\
 		& F\left( s_{n - 1},x_{n} \right) & ,\forall n \in \left\{ 1,\ldots,N - 1 \right\}
	\end{aligned} \right.
\end{align*}
where \(s_{n}\) denotes the \emph{n}-th component of modulator output
\(s_{0}^{N - 1}\mathcal{= M}\left( x_{0}^{N - 1} \right)\).

\(F\) is termed the transition function. For convenience we also denote \(F\left( s_{n - 1},x_{n} \right)\) as
\(F_{s_{n - 1}}x_{n}\). From above definition we could denote a FSM as a triad \(\left( \mathcal{S},\mathbb{B},F \right)\). The modulator is
deterministic, so as a channel from stochastic sequence
\(X_{0}^{N - 1} \in \mathbb{B}^{N}\) to stochastic sequence
\(S_{0}^{N - 1} \in \mathcal{S}^{N}\) it satisfies
\(p_{S_{n}|S_{n - 1}X_{n}}\left( s_{n} \middle| s_{n - 1},x_{n} \right) = \delta\left( s_{n} - F\left( s_{n - 1},x_{n} \right) \right)\),
\(\delta\) defined as \(\delta(a) = \left\{ \begin{array}{r}
1,a = 0 \\
0,a \neq 0
\end{array} \right.\ \). Therefore, with the definition of FSM and memoryless channel, it is easy to give the definition of
FSMC:

\emph{Definition 2:} Finite-State Modulated Channel (FSMC). A channel from stochastic sequence
\(X_{0}^{N - 1} \in \mathbb{B}^{N}\) to stochastic sequence
\(Y_{0}^{N - 1} \in \mathcal{Y}^{N}\) is a finite-state modulated channel iff
there exists a stochastic sequence \(S_{0}^{N - 1}\mathcal{\in S}\) such
that

\begin{enumerate}
\def\labelenumi{\arabic{enumi})}
\item
  \(\exists F:\mathcal{S} \times \mathbb{B} \mapsto \mathcal{S},\ p_{S_{n}|S_{n - 1}X_{n}}\left( s_{n} \middle| s_{n - 1},x_{n} \right) = \delta\left( s_{n} - F\left( s_{n - 1},x_{n} \right) \right),\forall n \in \left\{ 0,\ldots,N - 1 \right\}\);
\item
  \(\exists W:\mathcal{S \mapsto Y,\ }p_{Y_{0}^{N - 1}|S_{0}^{N - 1}}\left( y_{0}^{N - 1} \middle| s_{0}^{N - 1} \right) = \prod_{n = 0}^{N - 1}{W\left( y_{n} \middle| s_{n} \right)}\).
\end{enumerate}

In Definition 2, 1) represents property of FSM and 2)
represents property of memoryless channel.

A FSMC \(\mathcal{F}\) could be represented as a quad
\(\left( \mathcal{S},\mathbb{B},F,W \right)\). Denoting the modulator as
\(\mathcal{M}\), namely \(\mathcal{M =}\left( \mathcal{S},\mathbb{B},F \right)\),
the FSMC could be abbreviated to \(\left( \mathcal{M,}W \right)\).
Namely \(\mathcal{F}=(\mathcal{M},W)=(\mathcal{S},\mathbb{B},F,W)\).
Considering the deterministic modulator and memoryless state channel,
there are
\(p_{Y_{n}S_{n}|X_{n}S_{n - 1}}\left( y_{n},s_{n} \middle| x_{n},s_{n - 1} \right) = p_{Y_{n}|X_{n}S_{n - 1}^{n}}\left( y_{n} \middle| x_{n},s_{n - 1}^{n} \right)p_{S_{n}|X_{n}S_{n - 1}}\left( s_{n} \middle| x_{n},s_{n - 1} \right) = p_{Y_{n}|S_{n}}\left( y_{n} \middle| s_{n} \right)p_{S_{n}|X_{n}S_{n - 1}}\left( s_{n} \middle| x_{n},s_{n - 1} \right)\),
namely
\(W\left( y_{n},s_{n} \middle| x_{n},s_{n - 1} \right) = W\left( y_{n} \middle| s_{n} \right)\delta\left( s_{n} - F\left( s_{n - 1},x_{n} \right) \right)\),
which is the transition probability with state on channel layer,
i.e., the input of SCTD, mentioned in Section II-C.

\begin{figure}[!t]
\centering
\includegraphics[width=3.5in]{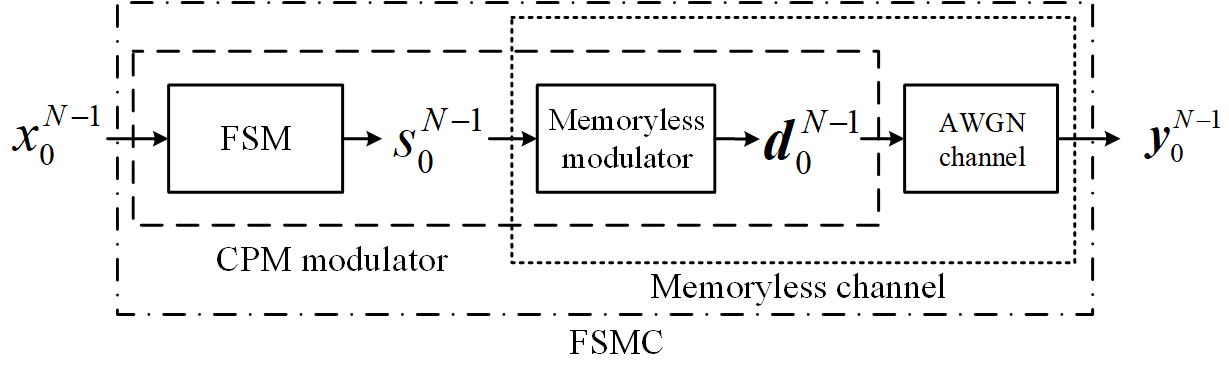}
\caption{CPM with an AWGN channel, as a special case of FSMC.}
\label{fig_CPMblc}
\end{figure}

For instance, CPM with an AWGN channel is a typical case of FSMC. According to \cite{ref5}, CPM modulator with parameters
\((M,h,L)\), namely \((M,h,L)\)-CPM, denoted as \(\mathcal{m}_{(M,h,L)}\), has form
\begin{align} \label{eq1}
	&\boldsymbol{d}_{0}^{N - 1} = \mathcal{m}_{(M,h,L)}\left( x_{0}^{N - 1} \right) \nonumber \\
	&= \left\lbrack A\exp\left( j2\pi h\left\lbrack R_{P}\left( \sum_{i = 0}^{n - L}x_{i} \right) + 2\sum_{i = 0}^{L - 1}{x_{n - i}\boldsymbol{g}_{i}} \right\rbrack \right) \right\rbrack_{n \in \left\{ 0,\ldots,N - 1 \right\}}
\end{align}
\(x_{0}^{N - 1} \in \mathcal{X}^{N}\) is
the codeword input to modulator, where \(\left| \mathcal{X} \right| = M\). In this paper we only discuss binary CPM, namely \(\mathcal{X}=\mathbb{B}, M=2\). And the output
\(\boldsymbol{d}_{0}^{N - 1}\) is sampled signal, where
\(\boldsymbol{g}_{i} := \left\lbrack g\left( k\frac{T}{K} + iT \right) \right\rbrack_{\ k \in \left\{ 1,\ldots,K \right\}}\),
where \(g\) is the shaping pulse
\(g(t) = \left\{ \begin{aligned}
 & 0 & ,t < 0 \\
 & \frac{1}{2LT}t & ,0 \leq t \leq LT \\
 & \frac{1}{2} & ,t > LT
\end{aligned} \right.\ \)
and \(K\) is the sampling factor. \(M,h,L\) are
termed radix, modulation index and pulse length, respectively. \(h\) is required to be rational. \(P\)
represents the denominator of the simplest fractional form of \(h\),
\(R_{P}( \cdot )\) denotes the remainder divided by \(P\).
\(A\) denotes the constant amplitude. Let the \emph{n}-th state be
\(s_{n} = R_{P}\left( \sum_{i = 0}^{n - L}x_{i} \right)M^{L} + \sum_{i = 0}^{L - 1}{x_{n - i}M^{i}} \in \left\{ 0,1,\ \ldots,PM^{L} - 1 \right\}\),
then
\(s_{n} = R_{P}\left( \left\lfloor \frac{s_{n - 1}}{M^{L}} \right\rfloor + \left\lfloor \frac{s_{n - 1}}{M^{L - 1}} \right\rfloor \right) M^{L} + R_{M^{L - 1}}\left( s_{n - 1} \right) M + x_{n}\),
\(\left\lfloor \cdot \right\rfloor\) refers to floor function. Let \(F\) be \(F(s_{n-1},x_n)=R_{P}\left( \left\lfloor \frac{s_{n - 1}}{M^{L}} \right\rfloor + \left\lfloor \frac{s_{n - 1}}{M^{L - 1}} \right\rfloor \right) M^{L} + R_{M^{L - 1}}\left( s_{n - 1} \right) M + x_{n}\), there is \(s_n=F(s_{n-1},x_n), s_0=F(0,x_0)\), so according to Definition 1, \(x_0^{N-1} \mapsto s_0^{N-1}\) forms a FSM. Besides,
\(\boldsymbol{d}_{n} = A\exp\left( j2\pi h\left\lbrack \left\lfloor \frac{s_{n}}{M^{L}} \right\rfloor + 2\sum_{i = 0}^{L - 1}{R_{M}\left( \left\lfloor \frac{s_{n}}{M^{i}} \right\rfloor \right)\boldsymbol{g}_{i}} \right\rbrack \right)\)
only depends on \(s_{n}\), so channel
\(s_{0}^{N - 1} \mapsto \boldsymbol{d}_{0}^{N - 1}\) is memoryless, which is
named memoryless modulator (MM) in \cite{ref5}. Denote the output of
memoryless channel as \(\boldsymbol{y}_{0}^{N - 1}\), since AWGN channel
\(\boldsymbol{d}_{0}^{N - 1} \mapsto \boldsymbol{y}_{0}^{N - 1}\) is memoryless, we know channel
\(s_{0}^{N - 1} \mapsto \boldsymbol{y}_{0}^{N - 1}\) is memoryless.
Therefore, according to Definition 2, CPM with a memoryless channel
forms a FSMC, as Fig.~\ref{fig_CPMblc} shows.

\subsection{A Method to Observe Polarization Efficiency}

With a full presentation of the channel, we now proceed to evaluate its polarization efficiency. Still, before a discussion over the polarization efficiency, it is necessary to choose
an indicator for such efficiency. We choose variance of sub-channel
capacities as an indicator of polarization efficiency, which could be
represented as
\[\sigma_{I}^{2} := \sigma^{2}\left( \mathcal{I}_{N} \right) = \frac{1}{N}\sum_{i = 0}^{N - 1}\left( I_{N}^{(i)} - \overline{I} \right)^{2}\]
where \(\sigma^{2}( \cdot )\) denotes variance function, and average
capacity \(\overline{I}\) is defined as
\begin{align*}
	\overline{I} &:= \mu\left( \mathcal{I}_{N} \right) = \frac{1}{N}\sum_{i = 0}^{N - 1}I_{N}^{(i)} \\
	&= \frac{1}{N}\sum_{i = 0}^{N - 1}{I\left( Y_{0}^{N - 1}U_{0}^{i - 1};U_{i} \right)} = \frac{1}{N}I\left( U_{0}^{N - 1};Y_{0}^{N - 1} \right)
\end{align*}
where \(\mu( \cdot )\) refers to average function.

As could be seen in its definition formula, variance \(\sigma_{I}^{2}\)
represents average square distance between each sub-channel capacity and the mean
capacity, thus being competent to measure the extent that all capacities
are off their average. Namely, \(\sigma_{I}^{2}\) is a qualified
parameter to assess polarization efficiency. Specifically, the larger
variance \(\sigma_{I}^{2}\) is, the more efficiency polarization
achieves \cite{ref:Var}.

Furthermore, for a fixed kind of channel, by altering
the channel parameter, such as the erasure probability of an erasure channel,
the snr of an AWGN channel, \emph{etc.}, we could record capacity variance
\(\sigma_{I}^{2}\) under each capacity average \(\overline{I}\), and
draw a curve of capacity variance w. r. t. capacity average \cite{ref:Var},
hereafter abbreviated to V-A curve. V-A curve offers a panorama of
polarization efficiency of a fixed kind of channel.
\section{Bijective FSM and Its Polarization Improvement}

In this section we will demonstrate that on a particular class of FSMC, polar code suffers a polarization loss compared to memoryless channels, and through our proposed structure the loss will be eliminated. The particular channel type is termed bijective FSMC. The modulation is termed bijective FSM, the definition of which is following.

\emph{Definition 3:} Bijective FSM.
\(\left( \mathcal{S},\mathbb{B},F \right)\) is a bijective FSM, iff
function \(F\) is bijective to each variable, i.e., for any fixed
\(x \in \mathbb{B}\), \(F( \cdot ,x)\) is a bijection (i.e.,
\(\forall s^{\prime}\mathcal{\in S,}x \in \mathbb{B},\exists s \in \mathcal{S}, F(s,x)=s^{\prime}, \forall \widehat{s}\neq s,F\left( \widehat{s},x \right) \neq s^{\prime}\ \mathrm)\),
and for any fixed \(s \in \mathcal{S}\), \(F(s, \cdot )\) is a
bijection (i.e.,
\(\forall s^{\prime},s \in \mathcal{S,\exists}x \in \mathbb{B}, F(s,x)=s^{\prime}, \forall \widehat{x}\neq x, F\left( s,\widehat{x} \right) \neq s^{\prime}\)).

And an FSMC with a bijective FSM is defined as a
bijective FSMC.

\begin{figure}[!t]
	\centering
	\includegraphics[width=3.5in]{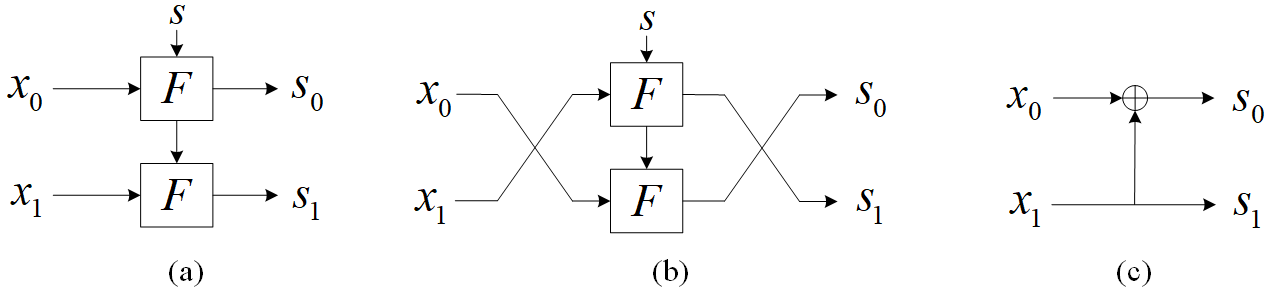}
	\caption{Structure of FSM with length 2 and its swapping, compared with Arikan's kernel. (a) Original structure; (b) Swapped structure; (c) Arikan's kernel.}
	\label{fig_3}
\end{figure}

We observe that a bijective FSM with length 2
has similar structure with Arikan's kernel, through a swapping. General
structure of FSM with length 2 is shown in Fig. 3(a),
indicating that \(s_{0} = F_{s}x_{0},s_{1} = F_{F_{s}x_{0}}x_{1}\). Then
for bijective ones, according to bijectivity of \(F\),
\(s_{0} = F_{s}x_{0}\) is bijective to \(x_{0}\) and
\(s_{1} = F_{F_{s}x_{0}}x_{1}\) is bijective to each
\(x_{i},i \in \left\{ 0,1 \right\}\). If swapping \(x_{0}\) with
\(x_{1}\), and \(s_{0}\) with \(s_{1}\), as Fig. 3(b) shows, we know
\(s_{0} = F_{F_{s}x_{1}}x_{0}\) is bijective to each
\(x_{i},i \in \left\{ 0,1 \right\}\) and \(s_{1} = F_{s}x_{1}\) is
bijective to \(x_{1}\); compared with Arikan's kernel shown in Fig.
3(c), where \({s_{0} = x}_{0} + x_{1}\) is bijective to each
\(x_{i},i \in \left\{ 0,1 \right\}\), and \(s_{1} = x_{1}\) is bijective
to \(x_{1}\), we could say there is a structural similarity in this sense.
As a consequence of such structural similarity, we could say that bijective FSMC have a property of \emph{intrinsic polarization potency}. In other words, through swapping, 
it acts as a structure similar to Arikan's kernel.
Just as Theorem 1 in Section IV-C indicates, through a swapping operation,
polar code on bijective FSMCs can get identical polarization 
efficiency with that of the conventional polar codes on memoryless channels.

\begin{figure}[!t]
	\centering
	\includegraphics[width=2in]{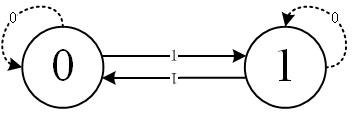}
	\caption{Transition diagram of \(F_1\).}
	\label{fig_M1tran}
\end{figure}

Here we take an instance of bijective FSM from a practical scenario: (2,1/2,1)-CPM, also known as MSK, along with the AWGN channel. Denote MSK as \(\mathcal{m}_1\) and the AWGN channel as \(\mathcal{g}\), in Proposition 1 we prove \((\mathcal{m}_1,\mathcal{g})\) is equivalent to \(\mathcal{F}_1:=(\mathcal{M}_1,\mathcal{G}_1)\). Modulation \(\mathcal{M}_1\) is defined as \(\mathcal{M}_{1}\left( x_{0}^{N - 1} \right) = \left\lbrack R_{2}\left( \sum_{i = 0}^{n}x_{i} \right) \right\rbrack_{n \in \left\{ 0,\ldots,N - 1 \right\}}\), trivially \(\mathcal{M}_1\) is an FSM with state set \(\mathcal{S}_1=\mathbb{B}\) and transition function
\begin{align} \label{2}
	F_{1}(s_{n-1},x_n) := R_{2}(s_{n-1} + x_n)
\end{align}
whose transition diagram is shown in Fig.~\ref{fig_M1tran}. The nodes are states and arrows relate states before and after transition with certain input symbol. Trivially \(F_{1}\) is bijective to each variable, so
according to Definition 3, \(\mathcal{M}_{1}\) is a bijective FSM.

And the definition of \(\mathcal{G}_{1}\) is
\(\mathcal{G}_{1}:\boldsymbol{y}_{n} = \Theta_{1}\left( s_{n} \right) + \boldsymbol{z}_{n}\),
where \(s_{n}\) is the \(n\)-th component of channel input
\(s_{0}^{N - 1}\), \(\boldsymbol{y}_{n}\) is the \(n\)-th component of
channel output \(\boldsymbol{y}_{0}^{N - 1}\), and \(\boldsymbol{z}_{n}\)
denotes AWGN noise. \(\Theta_{1}\) is a mapper from a state to a signal:
\begin{align} \label{3}
	\Theta_{1}\left( s_{n} \right) := A\exp\left( j\pi s_{n} \right) \cdot \boldsymbol{\phi}
\end{align}
where \(\boldsymbol{\phi} := \frac{1}{2}\left\lbrack 1 - \exp\left( j\frac{\pi}{K}m \right) \right\rbrack_{m \in \left\{ 1,\ldots,2K \right\}}\), and \(K \in \mathbb{N}^{*}\) is the sampling factor, namely the number
of sampled points per symbol period, \(A\) is the constant signal
amplitude. From the definition of \(\mathcal{G}_{1}\) we know it is
memoryless. Therefore, \((\mathcal{M}_1,\mathcal{G}_1)\) is a bijective FSMC. By the way, it is easy to verify that \(\mathcal{G}_{1}\) is 
equivalent to a pi/2-BPSK modulator with AWGN channel.

As mentioned above, Proposition 1 affirms the equivalence between
\((\mathcal{m}_{1},\mathcal{g)}\) with
\(\left( \mathcal{M}_{1},\mathcal{G}_{1} \right)\):

\emph{Proposition 1:} The a
posteriori probability of \((\mathcal{m}_{1},\mathcal{g})\) and
\(\left( \mathcal{M}_{1},\mathcal{G}_{1} \right)\) are identical through
an operator on the output, i.e., \( \exists\mathcal{H}, \ A_{\left( \mathcal{m}_{1},\mathcal{g} \right)}\left( x_{0}^{N - 1} \middle| {\widetilde{\boldsymbol{y}}}_{0}^{N - 1} \right) = A_{\left( \mathcal{M}_{1},\mathcal{G}_{1} \right)}\left( x_{0}^{N - 1} \middle| \mathcal{H}\left( {\widetilde{\boldsymbol{y}}}_{0}^{N - 1} \right) \right)\),
where \(A\) denotes the a posteriori probability, and each \(\mathcal{H}\) is an operator.

Namely, we define the equivalence of channels by the equality of their a
posteriori probability through an operator on the output, which is
stated in Proposition 1. The proof of Proposition 1 is in Appendix A.

\subsection{The Polarization Loss of Bijective FSM}
To observe the polarization efficiency on FSMC, we select \(\mathcal{M}_{1}\) as the instance of bijective FSM, and \(\left( \mathcal{M}_{1},\mathcal{G}_{1} \right)\) as the instance of bijective FSMC. The snr is set from -30dB to 20dB at step 0.5dB to plot V-A curves. Here the snr is defined as the ratio of total energy of a code symbol (\(E_{s}\)) to noise power spectral density
(\(N_{0}\)), namely \(\mathrm{snr} := E_{s}/N_{0}\), whose decibel form is
\(\mathrm{snr(dB)} = 10\lg{E_{s}/N_{0}}\)

\begin{figure}[!t]
	\centering
	\includegraphics[width=3.25in]{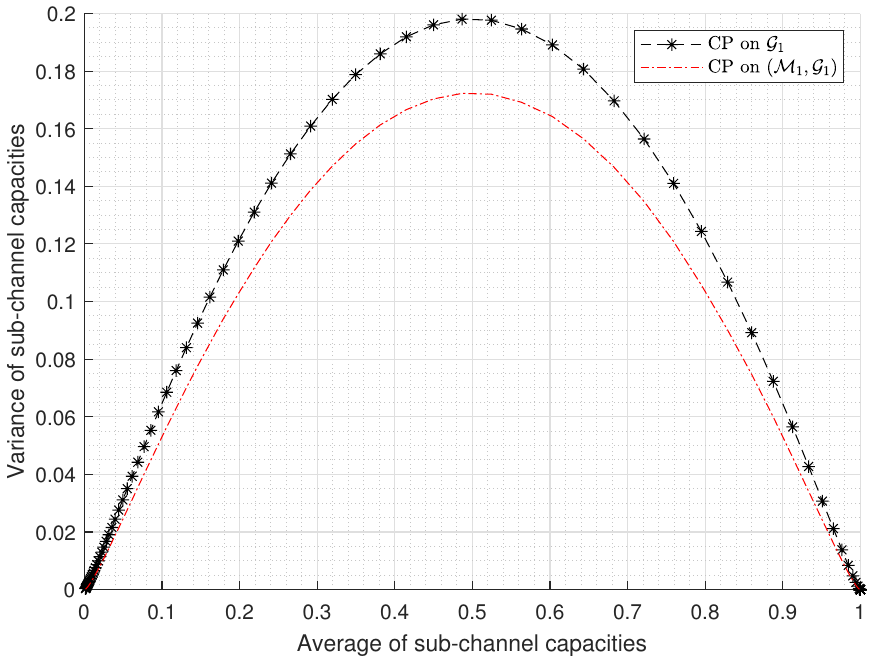}
	\caption{V-A curves of conventional polar code (CP) at code length 256, \(\left( \mathcal{M}_{1},\mathcal{G}_{1} \right)\) vs. \(\mathcal{G}_{1}\).}
	\label{fig_VAcurM1CP}
\end{figure}

Fig.~\ref{fig_VAcurM1CP} shows the V-A curve of
\(\left( \mathcal{M}_{1},\mathcal{G}_{1} \right)\), compared with a memoryless V-A curve of pi/2-BPSK modulation with AWGN, namely \(\mathcal{G}_{1}\). The code length is 256. From Fig.~\ref{fig_VAcurM1CP} we observe an evident gap on polarization efficiency between a bijective FSMC and its relevant memoryless channel. Namely for conventional polar codes,
the instance of bijective FSMC suffer an evident
polarization loss compared to their corresponding memoryless channel.

\subsection{The Reason for the Polarization Inefficiency}

As shown above, conventional polar code fails to polarize so
efficiently as it does on memoryless channels, as a conclusion by
observation of sub-channel capacity variance. However, such variance is
merely an overall assessment of polarization efficiency. To figure out the
reason for this phenomenon, we inspect polarization of each sub-channel
pairs on each layer of conventional polar code with memory channel 
to spot its structural problem with polarization.

\begin{figure}[!t]
\centering
\includegraphics[width=3.8in]{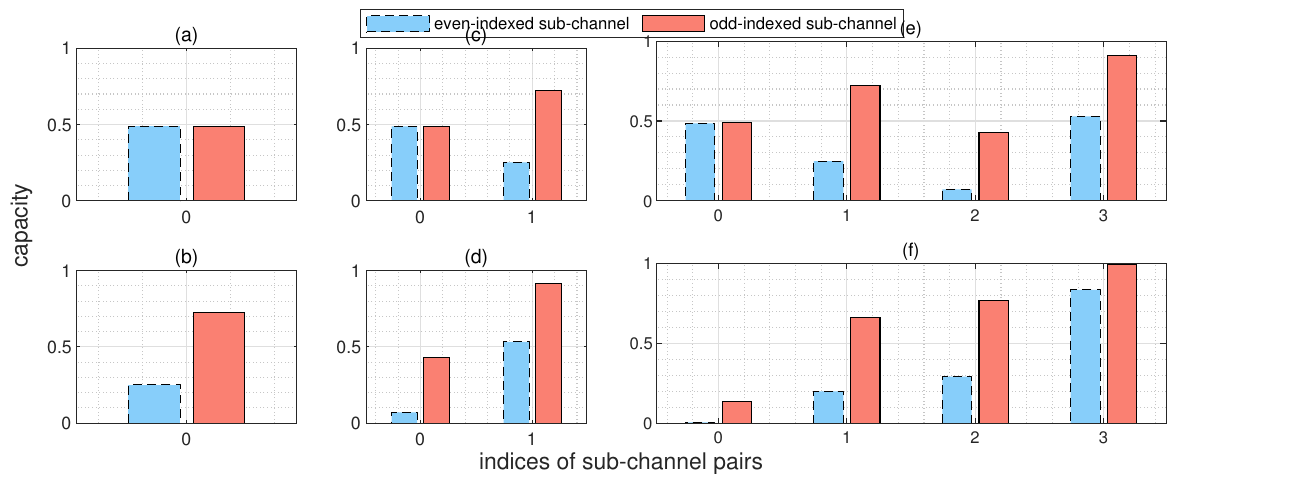}
\caption{Sub-channel capacities of conventional polar
code on memory channel \(\left( \mathcal{M}_{1},\mathcal{G}_{1} \right)\), 
compared to memoryless channel \(\mathcal{G}_{1}\), with snr -3dB, code length N 2, 4 and 8.
(a) memory channel, N=2; (b) memoryless channel, N=2; (c) memory
channel, N=4; (d) memoryless channel, N=4; (e) memory channel, N=8; (f)
memoryless channel, N=8.}
\label{fig_6}
\end{figure}

In Fig. 6 we depict capacity of each sub-channel pair of polar code on
\(\left( \mathcal{M}_{1},\mathcal{G}_{1} \right)\), compared to
memoryless channel \(\mathcal{G}_{1}\), where the \emph{j}-th sub-channel pair of code
length \(N\) is
\(\left\lbrack W_{N}^{(2j)},W_{N}^{(2j + 1)} \right\rbrack\).
Even-indexed sub-channel of the \emph{j}-th sub-channel pair refers to
\(W_{N}^{(2j)}\), and odd-indexed one refers to \(W_{N}^{(2j + 1)}\).
The snr is -3dB. Comparing sub-channel capacities at the same code length, between
memory channel and memoryless channel, viz. Fig. 6(a) to Fig.
6(b), Fig. 6(c) to Fig. 6(d) and Fig. 6(e) to Fig. 6(f), it is evident
that on memory channel, the first sub-channel pair does not polarize at
all. Therefore, we focus on the first sub-channel pair to reveal the
cause for polarization loss.

The simplest case is the initial structure, which contains a single pair of sub-channels only.
Namely the code length is 2. On
\(\mathcal{F}_{1} = \left( \mathcal{M}_{1},\mathcal{G}_{1} \right)\),
the transition function can be described as \(s_{n} = s_{n - 1} + x_{n}\) according
to Section IV-A.  Therefore, the relationship between
state sequence \(s_0^1\) and codeword \(x_0^1\) can be expressed as
\(s_{0}^{1} = x_{0}^{1}T\), where \(T = \begin{bmatrix}
1 & 1 \\
0 & 1
\end{bmatrix}\). And applying conventional polar
coding to source code sequence \(u_0^1\), there is \(x_0^1=u_0^1 F\), where \(F\) is the Arikan's kernel \(F = \begin{bmatrix}
1 & 0 \\
1 & 1
\end{bmatrix}\). As a result,
\(s_{0}^{1} = u_{0}^{1}FT = u_{0}^{1}\begin{bmatrix}
1 & 0 \\
1 & 1
\end{bmatrix}\begin{bmatrix}
1 & 1 \\
0 & 1
\end{bmatrix} = u_{0}^{1}\begin{bmatrix}
1 & 1 \\
1 & 0
\end{bmatrix}\). Notice that the state channel is memoryless, i.e.,
\(p\left( \boldsymbol{y}_{0}^{N - 1} \middle| s_{0}^{N - 1} \right) = \prod_{i = 0}^{N - 1}{p\left( \boldsymbol{y}_{i} \middle| s_{i} \right)}\),
we have
\begin{align*}
	W_{\left( \mathcal{P}_{c},\mathcal{F}_{1} \right),2}^{(1)}\left( \boldsymbol{y}_{0}^{1},u_{0} \middle| u_{1} \right) &= \frac{1}{2}W\left( \boldsymbol{y}_{0}^{1} \middle| u_{0}^{1} \right) \\
	&= \frac{1}{2}W\left( \boldsymbol{y}_{0} \middle| u_{0} + u_{1} \right)W\left( \boldsymbol{y}_{1} \middle| u_{0} \right)
\end{align*}
\begin{align*}
	&W_{\left( \mathcal{P}_{c},\mathcal{F}_{1} \right),\ 2}^{(0)}\left( \boldsymbol{y}_{0}^{1} \middle| u_{0} \right) = \frac{1}{2}\sum_{u_{1}\mathbb{\in B}}^{}{W\left( \boldsymbol{y}_{0}^{1} \middle| u_{0}^{1} \right)} \\
	&= \frac{1}{2}\sum_{u_{1}\mathbb{\in B}}^{}{W\left( \boldsymbol{y}_{0} \middle| u_{0} + u_{1} \right)W\left( \boldsymbol{y}_{1} \middle| u_{0} \right)} \\
	&= \frac{1}{2}W\left( \boldsymbol{y}_{1} \middle| u_{0} \right)\sum_{u_{1}\mathbb{\in B}}^{}{W\left( \boldsymbol{y}_{0} \middle| u_{0} + u_{1} \right)} \\
	&= \frac{1}{2}W\left( \boldsymbol{y}_{1} \middle| u_{0} \right)\sum_{b \in \mathbb{B}}^{}{W\left( \boldsymbol{y}_{0} \middle| b \right)}
\end{align*}

It is exhibited that the expression of
\(W_{\left( \mathcal{P}_{c},\mathcal{F}_{1} \right),2}^{(1)}\) has only
one factor related to its input bit \(u_{1}\), which is determined by
single symbol channel \(W\); so does
\(W_{\left( \mathcal{P}_{c},\mathcal{F}_{1} \right),\ 2}^{(0)}\left( \boldsymbol{y}_{0}^{1} \middle| u_{0} \right)\) to \(u_0\).
Therefore, it is obvious that they are equivalent to single symbol
channel \(W\), thus having identical capacity with \(W\). Specifically,
their capacities are
\begin{align*}
	&I\left( W_{\left( \mathcal{P}_{c},\mathcal{F}_{1} \right),2}^{(1)} \right) = I\left( \boldsymbol{Y}_{0}^{1}U_{0};U_{1} \right)\\
	&=\mathbb{E}\left\lbrack \log_{2}\frac{2W_{\left( \mathcal{P}_{c},\mathcal{F}_{1} \right),2}^{(1)}\left( \boldsymbol{Y}_{0}^{1},U_{0} \middle| U_{1} \right)}{W_{\left( \mathcal{P}_{c},\mathcal{F}_{1} \right),2}^{(1)}\left( \boldsymbol{Y}_{0}^{1},U_{0} \middle| 0 \right) + W_{\left( \mathcal{P}_{c},\mathcal{F}_{1} \right),2}^{(1)}\left( \boldsymbol{Y}_{0}^{1},U_{0} \middle| 1 \right)} \right\rbrack \\
	&=\mathbb{E}\left\lbrack \log_{2}\frac{2W\left( \boldsymbol{Y}_{0} \middle| X_{0} \right)}{W\left( \boldsymbol{Y}_{0} \middle| X_{0} \right) + W\left( \boldsymbol{Y}_{0} \middle| X_{0} \right)} \right\rbrack = I\left( X_{0};\boldsymbol{Y}_{0} \right) = I(W)
\end{align*}
\begin{align*}
	&I\left( W_{\left( \mathcal{P}_{c},\mathcal{F}_{1} \right),\ 2}^{(0)} \right) = I\left( \boldsymbol{Y}_{0}^{1};U_{0} \right)\\
	&=\mathbb{E}\left\lbrack \log_{2}\frac{2W_{\left( \mathcal{P}_{c},\mathcal{F}_{1} \right),\ 2}^{(0)}\left( \boldsymbol{Y}_{0}^{1} \middle| U_{1} \right)}{W_{\left( \mathcal{P}_{c},\mathcal{F}_{1} \right),\ 2}^{(0)}\left( \boldsymbol{Y}_{0}^{1} \middle| 0 \right) + W_{\left( \mathcal{P}_{c},\mathcal{F}_{1} \right),\ 2}^{(0)}\left( \boldsymbol{Y}_{0}^{1} \middle| 1 \right)} \right\rbrack \\
	&=\mathbb{E}\left\lbrack \log_{2}\frac{2W\left( \boldsymbol{Y}_{1} \middle| X_{1} \right)}{W\left( \boldsymbol{Y}_{1} \middle| X_{1} \right) + W\left( \boldsymbol{Y}_{1} \middle| X_{1} \right)} \right\rbrack = I\left( X_{1};\boldsymbol{Y}_{1} \right) = I(W)
\end{align*}

So
\(I\left( W_{\left( \mathcal{P}_{c},\mathcal{F}_{1} \right),\ 2}^{(0)} \right) = I\left( W_{\left( \mathcal{P}_{c},\mathcal{F}_{1} \right),2}^{(1)} \right) = I(W)\).
Namely, sub-channel capacities of conventional polar code on
\(\mathcal{F}_{1}\) cannot polarize while code length is 2.

From above analysis we find the transition matrix \(T = \begin{bmatrix}
1 & 1 \\
0 & 1
\end{bmatrix}\) itself has similar form with polar kernel
\(F = \begin{bmatrix}
1 & 0 \\
1 & 1
\end{bmatrix}\), consequently, concatenation of them suffers from their
counteraction. In other words, the transition function has intrinsic polarization
potency, which counteracts polar codes.

More generally, as with conventional polar code at any code length and on an arbitrary bijective
FSMC, we have Proposition 2. In Proposition 2, as defined in Section II-A, \(I_{\left( \mathcal{P}_c,\mathcal{C} \right),n}^{(i)}\) refers to the \(i\)-th sub-channel capacity of conventional polar code at code length \(n\) on code channel \(\mathcal{C}\); and \(W^{\langle i \rangle}\) means
the \(\langle i \rangle\)-th extension of single symbol channel \(W\).

\emph{Proposition 2:} On any bijective FSMC
\(\mathcal{F}_{b} = \left( \mathcal{S},\mathbb{B},F,W \right)\), for conventional
polar codes at code length
\(n \in \left\{ 2^{m} \middle| m \in \mathbb{N}^{*} \right\}\),
sub-channel capacities have such a relationship with those of
conventional polar codes on \(W^{\left\langle i \right\rangle}\):
\(I_{\left( \mathcal{P}_{c},\mathcal{F}_{b} \right),\ n}^{(i)} = I_{\left( \mathcal{P}_{c},W^{\left\langle i \right\rangle} \right),\ \left\langle i \right\rangle}^{\left( i^{*} \right)}\),
where \(\left\langle i \right\rangle := \left\{ \begin{aligned}
 & 1 & ,i \leq 1 \\
 & \max\left\{ n \middle| n \leq i,\ n = 2^{m},m \in \mathbb{N}^{*} \right\} & ,i \geq 2
\end{aligned} \right.\ ,i^{*} := \left\{ \begin{aligned}
 & 0 & ,i \leq 1 \\
 & i - \left\langle i \right\rangle & ,i \geq 2
\end{aligned} \right.\ \).

Polarization of conventional polar code on any bijective FSMC
revealed by Proposition 2 is illustrated in Fig. 7 where
\({\overline{I}}_{n}^{(i)} := I_{\left( \mathcal{P}_{c},\mathcal{F}_{b} \right),\ n}^{(i)},\ {\widetilde{I}}_{n}^{(i)} := I_{\left( \mathcal{P}_{c},W^{n} \right),\ n}^{(i)}\),
indicating that the first sub-channel pair on each layer does not
polarize. And the proof is in Appendix B. This proposition is consistent
with sub-channel capacities of conventional polar code on
\(\left( \mathcal{M}_{1},\mathcal{G}_{1} \right)\) depicted in Fig. 6.

\begin{figure}[!t]
\centering
\includegraphics[width=3.5in]{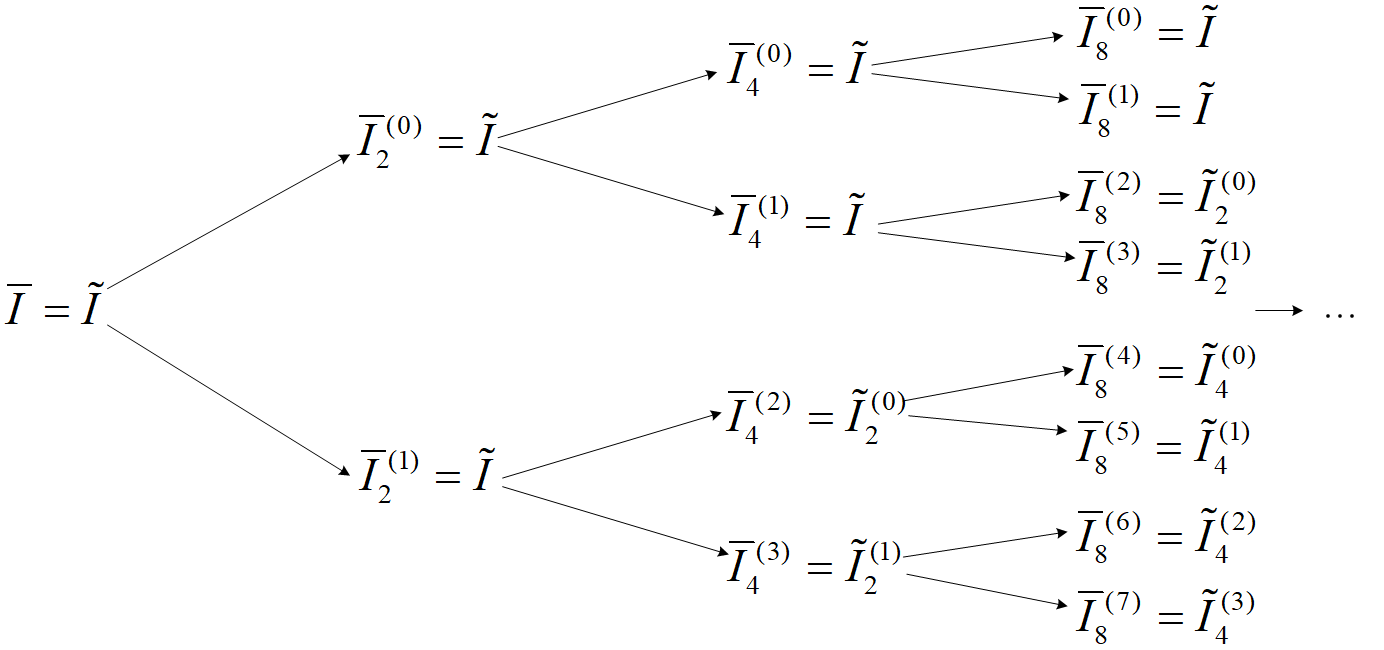}
\caption{Polarization of conventional polar code on any bijective FSMC.}
\label{fig_7}
\end{figure}



The counteraction with polar codes is a common
issue for bijective FSM, resulting in polarization loss. Therefore, to remedy the
polarization loss, we should exploit the intrinsic polarization potency
of bijective FSM and avoid counteraction, which asks for a new coding
structure to fit with bijective FSM.

\subsection{Last-Pair Swapping Structure}

Initially, in the the simplest case which contains a single pair of sub-channels only, namely the code length is 2, inspired by the form of transition matrix and its intrinsic polarization
potency, we attempted to set a new kernel \(F_{r} = \begin{bmatrix}
0 & 1 \\
1 & 0
\end{bmatrix}\) to exploit its intrinsic polarization potency. As a consequence,
\(s_{0}^{1} = u_{0}^{1}F_{r}T = u_{0}^{1}\begin{bmatrix}
0 & 1 \\
1 & 0
\end{bmatrix}\begin{bmatrix}
1 & 1 \\
0 & 1
\end{bmatrix} = u_{0}^{1}\begin{bmatrix}
0 & 1 \\
1 & 1
\end{bmatrix} = u_{0}^{1}\begin{bmatrix}
1 & 0 \\
1 & 1
\end{bmatrix}\begin{bmatrix}
0 & 1 \\
1 & 0
\end{bmatrix} = u_{0}^{1}FF_{r}\), namely coding by \(F_{r}\) on
\(\mathcal{M}_{1}\) equals conventional polar coding with a
swapping. And considering the equivalence of single symbol channels of a
memoryless channel, it is apparent that channel
\(U_{0}^{1} \mapsto \boldsymbol{Y}_{0}^{1}\) 
is equivalent to conventional polar code on memoryless channels.
Furthermore, according to the introduction of Section IV, intrinsic polarization potency
is a nature of general bijective FSMC, so the proposed structure could be
applied to them generally. With such a motivation we are prepared to
give a new coding structure under any code length.

Before the presentation of a remedy for polar code structure, it is
necessary to have a review on conventional polar coding structure, which
can be recursively described as: elements in source bit sequence of
length \(N\) are transformed by Arikan's kernel \(F = \begin{bmatrix}
1 & 0 \\
1 & 1
\end{bmatrix}\) in pairs, and the sequence got are divided into its even
subsequence and odd one by \(R_{N}\), which are separately sent to polar
code encoder of length \(N/2\). \(R_{N}\) is a reverse shuffle matrix to
separate even and odd subsequences of a row vector, i.e.,
\(u_{0}^{N - 1}R_{N} = \left\lbrack u_{0},u_{2},\ldots,u_{N - 2},u_{1},u_{3},\ldots,u_{N - 1} \right\rbrack\).
The recursive formula of generator matrix is
\(G_{N} = \left( I_{N/2} \otimes F \right)R_{N}\left( I_{2} \otimes G_{N/2} \right)\)
and the initial one is \(G_{2} = F\). Such structure could be depicted
as Fig. 8.

\begin{figure}[!t]
\centering
\includegraphics[width=3.5in]{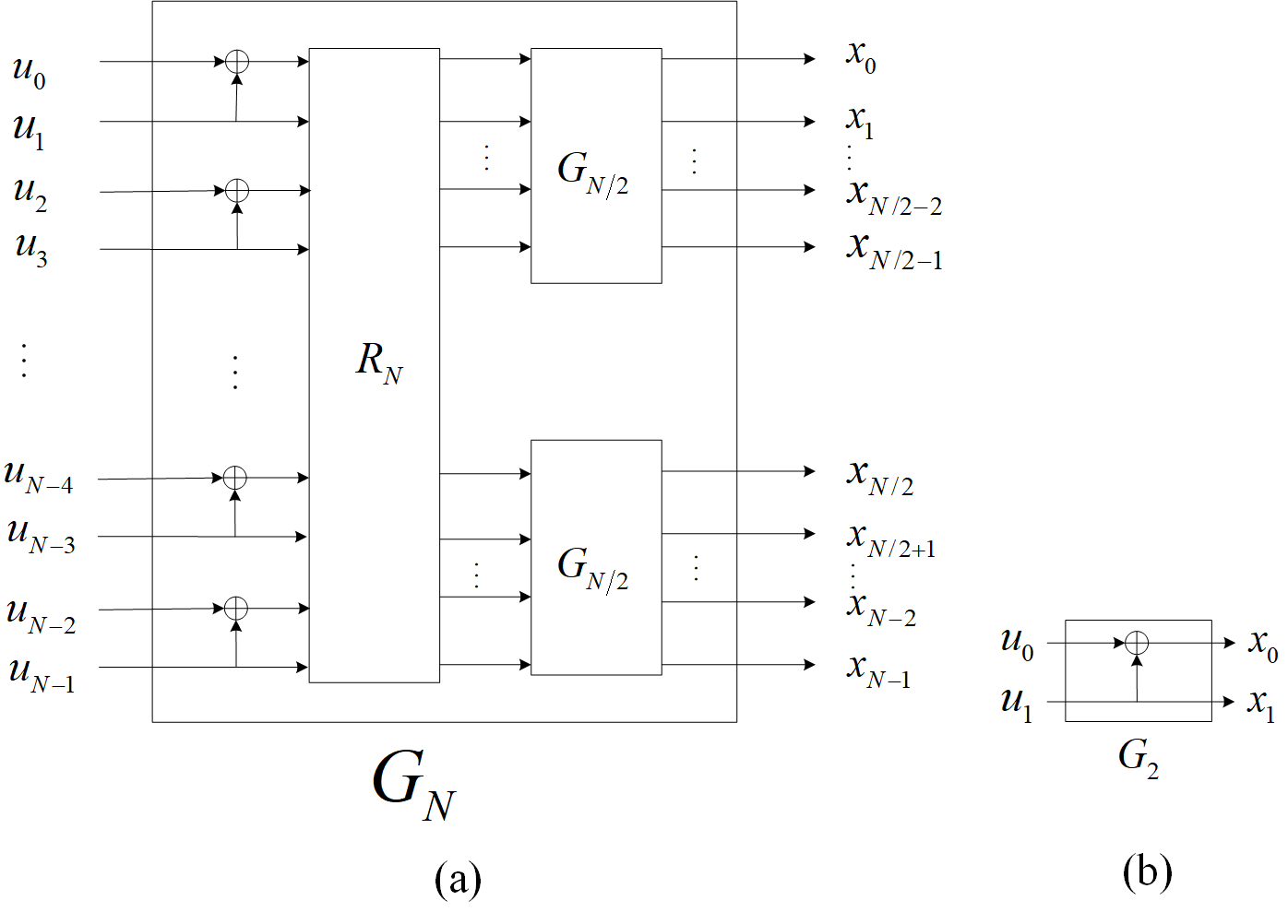}
\caption{Structure of conventional polar codes. (a) Recursive structure; (b) Initial structure.}
\label{fig_8}
\end{figure}

As a modification to conventional polar code structure, we substitute
the last kernel of each layer from Arikan's kernel
\(F = \begin{bmatrix}
1 & 0 \\
1 & 1
\end{bmatrix}\) to a swapping matrix
\[F_{r} = \begin{bmatrix}
0 & 1 \\
1 & 0
\end{bmatrix}\]
as an inspiration from the similarity between bijective transition matrix and
polar kernel, which is discussed in the introduction of Section IV. The proposed
structure sets part of kernels in encoding structure as swapping kernel,
so it is termed \emph{last-pair swapping structure}, denoted as
\(\mathcal{P}_{s}\). And polar code with last-pair swapping structure is termed last-pair swapping polar code. Fig. 9. depicts the recursive and initial
structures of last-pair swapping polar code, indicating that the recursive
formula of generator matrix of last-pair swapping polar code is
\(G_{N} = \begin{bmatrix}
I_{N/2 - 1} \otimes F & O_{(N - 2) \times 2} \\
O_{2 \times (N - 2)} & F_{r}
\end{bmatrix}R_{N}(I_{2} \otimes G_{N/2})\), where \(O_{m \times n}\) refers to an
\(m \times n\) zero matrix.

\begin{figure}[!t]
\centering
\includegraphics[width=3.5in]{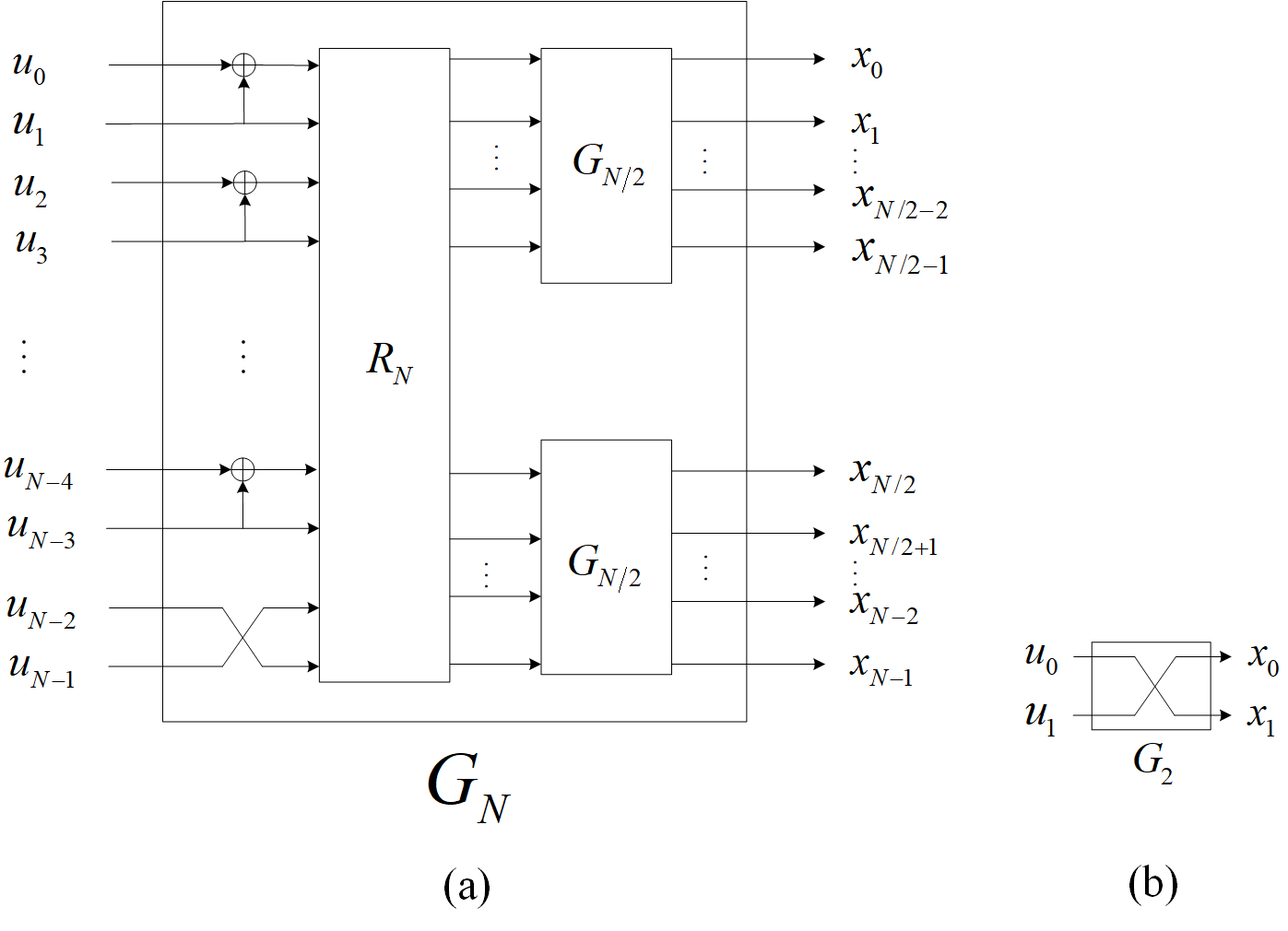}
\caption{Structure of last-pair swapping polar codes. (a) Recursive structure; (b) Initial structure.}
\label{fig_9}
\end{figure}

The consideration for our proposed structure's design derives from a
phenomenon that if modifying coding structure of lower layers, we
observe the last sub-channel pair of each present layer does not
polarize. As a solution, we determine to substitute every layer's last
kernel with \(\begin{bmatrix}
0 & 1 \\
1 & 0
\end{bmatrix}\) to guarantee that polarization on every layer is ideal,
namely polarization is ideal on arbitrary code length. Such substitution
result in last-pair swapping structure proposed above.

With the proposed structure applied to bijective FSMC,
sub-channel capacities are amended as Fig. 10, on
\(\left( \mathcal{M}_{1},\mathcal{G}_{1} \right)\) channel, compared to conventional polar code on memoryless channel \(\mathcal{G}_{1}\). The snr is -3dB. Comparing
sub-channel capacities on the same code length, between the proposed polar
code on memory channel and conventional polar code on memoryless
channel, viz. Fig. 10(a) to Fig. 10(b), Fig. 10(c) to Fig. 10(d)
and Fig. 10(e) to Fig. 10(f), it is evident that on memory channel,
every sub-channel pair polarizes as memoryless channel. Therefore, the
loss in polarization efficiency of a single pair is alleviated by the proposed
coding structure.

\begin{figure}[!t]
	\centering
	\includegraphics[width=3.8in]{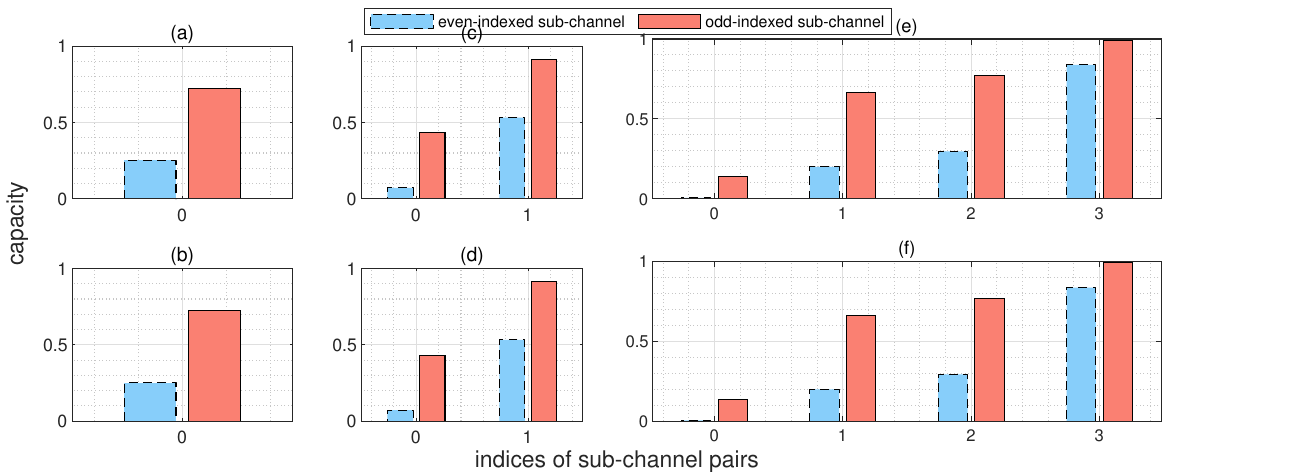}
	\caption{Sub-channel capacities of the proposed polar
		code on memory channel \(\left( \mathcal{M}_{1},\mathcal{G}_{1} \right)\), 
		compared to memoryless channel \(\mathcal{G}_{1}\), with snr -3dB, code length N 2, 4 and 8. (a)
		the proposed polar code, memory channel, N=2; (b) conventional polar code,
		memoryless channel, N=2; (c) the proposed polar code, memory channel, N=4;
		(d) conventional polar code, memoryless channel, N=4; (e) the proposed
		polar code, memory channel, N=8; (f) conventional polar code, memoryless
		channel, N=8.}
	\label{fig_10}
\end{figure}

Rigorously, on bijective FSMCs, last-pair swapping 
structure achieves identical polarization efficiency with that of conventional polar
codes on memoryless channels, which is expressed in Theorem 1.

\emph{Theorem 1:} On any bijective FSMC
\(\mathcal{F}_{b} = \left( \mathcal{S},\mathbb{B},F,W \right)\), at code
length \(n \in \left\{ 2^{m} \middle| m \in \mathbb{N}^{*} \right\}\),
each sub-channel of last-pair swapping polar code has identical capacity
with the same indexed sub-channel of conventional polar code on memoryless channel
\(W^{n}\), i.e.,
\(I_{\left( \mathcal{P}_{s},\mathcal{F}_{b} \right),\ n}^{(i)} = I_{\left( \mathcal{P}_{c},W^{n} \right),\ n}^{(i)}\).

The proof of Theorem 1 is in Appendix C.

Furthermore, on bijective FSMCs, polarization efficiency of
last-pair swapping structure is superior to that of conventional polar
code, which is formulated in Theorem 2.

\emph{Theorem 2}: On any bijective FSMC
\(\mathcal{F}_{b} = \left( \mathcal{S},\mathbb{B},F,W \right)\), at code
length \(n \in \left\{ 2^{m} \middle| m \in \mathbb{N}^{*} \right\}\),
sub-channel capacity variance of last-pair swapping polar code is no less
than that of conventional polar code, i.e.
\begin{align*}
	\forall n \in \left\{ 2^{m} \middle| m \in \mathbb{N}^{*} \right\},\ \sigma^{2}\left( \mathcal{I}_{\left( \mathcal{P}_{s},\mathcal{F}_{b} \right),n} \right) \geq \sigma^{2}\left( \mathcal{I}_{\left( \mathcal{P}_{c},\mathcal{F}_{b} \right),n} \right)
\end{align*}
and the inequality is strict iff capacity of \(W\) is not 0 or 1, namely
\(I(W) \in (0,1)\).

The proof of Theorem 2 is in Appendix D.

\subsection{Polarization Efficiency of the Proposed Structure}

\begin{figure}[!t]
	\centering
	\includegraphics[width=3.25in]{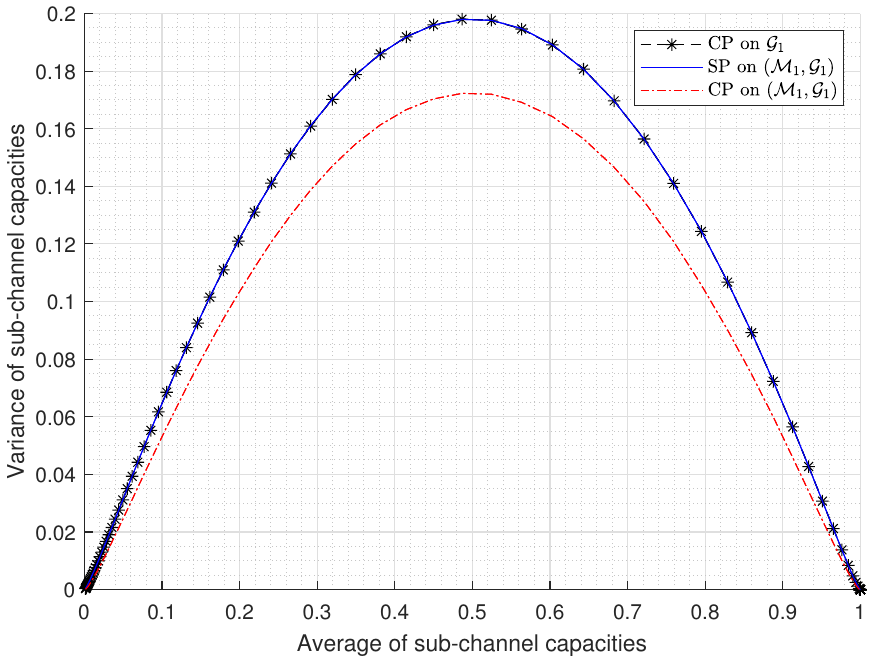}
	\caption{V-A curves of conventional polar code (CP) and last-pair swapping polar code (SP) on different code channels at code length 256, \(\left( \mathcal{M}_{1},\mathcal{G}_{1} \right)\) vs. \(\mathcal{G}_{1}\).}
	\label{fig_VAcurM1SP}
\end{figure}

While the proposed structure is applied to polar coding, polarization efficiency
is significantly promoted on bijective FSMCs, as the following results illustrate.
Fig.~\ref{fig_VAcurM1SP} show V-A curves of conventional and last-pair swapping polar code on
\(\left( \mathcal{M}_{1},\mathcal{G}_{1} \right)\), compared with the V-A curve of conventional polar code on pi/2-BPSK modulation with AWGN, namely \(\mathcal{G}_{1}\). The code length is 256. The advantage of the proposed structure over the conventional one on bijective
FSMC is evident in this figure, consistent with Theorem 2. Furthermore, by applying last-pair swapping polar code to bijective FSM, the polarization efficiency is resumed exactly to memoryless level, which confirms Theorem 1.

\section{A Plausible Generalization of Proposed Structure}

\subsection{Sub-Injective FSM and Its Intrinsic Polarization Potency}

With rigorous proofs in Section IV, we demonstrate that last-pair swapping structure could be applied to bijective FSM to improve polarization under it. And the improvement of our proposed structure relies on the intrinsic polarization potency of the transition function of bijective FSM, which is discussed throughout Section IV. The intrinsic polarization potency acts as a crucial property for the applicability of last-pair swapping structure. Hereby there arises a natural motivation to explore general types of FSM with intrinsic polarization potency, so that last-pair swapping polar code could be generalized to a broader scope. Without loss of intrinsic polarization potency, we could make an
extension of bijective FSM, and give the definition of sub-injective FSM in
Definition 4.

\emph{Definition 4:} Sub-Injective FSM.
\(\left( \mathcal{S},\mathbb{B},F \right)\) is a sub-injective FSM,
iff there exists a finite partition
\(\mathcal{Q :=}\left\{ Q_{0},Q_{1},\ldots,Q_{n - 1} \right\}\) of state
space \(\mathcal{S}\), such that

\begin{enumerate}
	\def\labelenumi{\arabic{enumi})}
	\item
	\(\forall Q \in \mathcal{Q,}x \in \mathbb{B},\exists Q^{\prime}\mathcal{\in Q,}\left\{ F(s,x) \middle| s \in Q \right\} \subseteq Q^{\prime}\);
	\item
	Function \(F^{\mathcal{Q}}\)
	is injective to each variable, i.e., for any fixed
	\(x \in \mathbb{B}\), \(F^{\mathcal{Q}}( \cdot ,x)\) is an
	injection (i.e.,
	\(\forall Q,\widehat{Q}\mathcal{\in Q,}x \in \mathbb{B},\ Q \neq \widehat{Q} \rightarrow F^{\mathcal{Q}}(Q,x) \neq F^{\mathcal{Q}}\left( \widehat{Q},x \right)\)),
	and for any fixed \(Q \in \mathcal{Q}\),
	\(F^{\mathcal{Q}}(Q, \cdot )\) is an injection (i.e.,
	\(\forall Q \in \mathcal{Q,\ }x,\widehat{x} \in \mathbb{B},x \neq \widehat{x} \rightarrow F^{\mathcal{Q}}(Q,x) \neq F^{\mathcal{Q}}\left( Q,\widehat{x} \right)\)), where \(F^{\mathcal{Q}}\mathcal{:Q} \times \mathbb{B} \mapsto \mathcal{Q}\) is defined as
	\(F^{\mathcal{Q}}(Q,x) := \{ s^{\prime}|\exists Q^{\prime}\mathcal{\in Q,}\left\{ F(s,x) \middle| s \in Q \right\} \cup \{ s^{\prime}\} \subseteq Q^{\prime}\}\).
\end{enumerate}

And an FSMC with a sub-injective FSM is defined as a
sub-injective FSMC. For
convenience we also denote \(F^{\mathcal{Q}}(Q,x)\) as
\(F_{Q}^{\mathcal{Q}}x\). And define \(\left\langle s \right\rangle\) as
the sub-space to which \(s\) belongs, i.e.,
\(\left\langle s \right\rangle := \left\{ s^{\prime} \middle| \exists Q \in \mathcal{Q,}\left\{ s,s^{\prime} \right\} \subseteq Q \right\}\),
from 1) in Definition 4 we know
\(\left\langle F_{s}x \right\rangle = F_{\left\langle s \right\rangle}^{\mathcal{Q}}x\).

Sub-injectivity for a channel defined here is a property that
the state space could be reduced to several sub-spaces, transform from
one of which with a symbol to another is injective to both sub-space and
symbol. Such property also ensures intrinsic polarization potency.
Now observe a sub-injective FSM with length 2. In Fig.
3(a), while the FSM is sub-injective, there are
\(\left\langle s_{0} \right\rangle = \left\langle F_{s}x_{0} \right\rangle = F_{\left\langle s \right\rangle}^{\mathcal{Q}}x_{0},\left\langle s_{1} \right\rangle = \left\langle F_{F_{s}x_{0}}x_{1} \right\rangle = F_{\left\langle F_{s}x_{0} \right\rangle}^{\mathcal{Q}}x_{1} = F_{F_{\left\langle s \right\rangle}^{\mathcal{Q}}x_{0}}^{\mathcal{Q}}x_{1}\).
According to injectivity of \(F^{\mathcal{Q}}\),
\(\left\langle s_{0} \right\rangle = F_{\left\langle s \right\rangle}^{\mathcal{Q}}x_{0}\)
is injective to \(x_{0}\) and
\(\left\langle s_{1} \right\rangle = F_{F_{\left\langle s \right\rangle}^{\mathcal{Q}}x_{0}}^{\mathcal{Q}}x_{1}\)
is injective to each \(x_{i},i \in \left\{ 0,1 \right\}\), then
\(s_{0} = F_{s}x_{0}\) and \(s_{1} = F_{F_{s}x_{0}}x_{1}\) are also
injective to them, for the intersection of distinct sets in partition
\(\mathcal{Q}\) is empty. If swapping \(x_{0}\) with \(x_{1}\), and
\(s_{0}\) with \(s_{1}\), as Fig. 3(b) shows, we know
\(s_{0} = F_{F_{s}x_{1}}x_{0}\) is injective to each
\(x_{i},i \in \left\{ 0,1 \right\}\) and \(s_{1} = F_{s}x_{1}\) is
injective to \(x_{1}\); compared with Arikan's kernel shown in Fig.
3(c), where \(s_{0} = x_{0} + x_{1}\) is injective to each
\(x_{i},i \in \left\{ 0,1 \right\}\), and \(s_{1} = x_{1}\) is injective
to \(x_{1}\), such structural similarity leads to intrinsic polarization
potency.

Trivially, for any bijective FSMC, let \(\mathcal{Q = S}\),
then 1) in Definition 4 is satisfied, and \(F^{\mathcal{Q}} = F\), which
is injective to each variable, so 2) is also satisfied. So any
bijective FSM is a sub-injective FSM. Namely, bijective FSM is a special case of sub-injective FSM.

\begin{figure}[!t]
	\centering
	\includegraphics[width=2.75in]{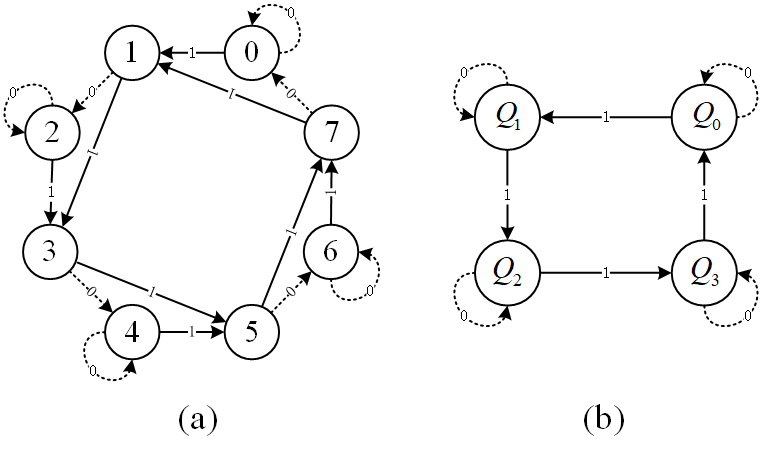}
	\caption{Transition diagrams. (a) \(F_2\); (b) \(F_2^{\mathcal{Q}}\).}
	\label{fig_M2tran}
\end{figure}

For sub-injective FSM, we also take an example from a practical scenario: (2,1/4,1)-CPM with an AWGN channel. Denoting (2,1/4,1)-CPM as \(\mathcal{m}_2\), the channel is \((\mathcal{m}_2,\mathcal{g})\), according to formula (1), trivially
it is equivalent to
\(\mathcal{F}_{2} := \left( \mathcal{M}_{2},\mathcal{G}_{2} \right)\).
\(\mathcal{M}_{2}\) is defined as
\(\mathcal{M}_{2}\left( x_{0}^{N - 1} \right) = \left\lbrack 2R_{4}\left( \sum_{i = 0}^{n - 1}x_{i} \right) + x_{n} \right\rbrack_{n \in \left\{ 0,\ldots,N - 1 \right\}}\), which is evidently an FSM whose state set is
\(\mathcal{S}_{2} = \left\{ 0,1,2,3,4,5,6,7 \right\}\), and transition function is
\begin{align*}
F_{2}(s_{n-1},x_n) = R_{8}\left( s_{n-1} + x_n + R_{2}(s_{n-1}) \right)
\end{align*}
whose transition diagram is shown in Fig.~\ref{fig_M2tran}(a). By dividing the state space to
\(\mathcal{Q} = \left\{ \left\{ 0,7 \right\},\left\{ 1,2 \right\},\left\{ 3,4 \right\},\left\{ 5,6 \right\} \right\}\),
namely sub-space
\(Q_{0} = \left\{ 0,7 \right\},Q_{1} = \left\{ 1,2 \right\},\ Q_{2} = \left\{ 3,4 \right\},\ Q_{3} = \left\{ 5,6 \right\}\),
mapping \(F_{2}^{\mathcal{Q}}\) among them could be formulated as
\(F_{2}^{\mathcal{Q}}\left( Q_{i},x \right) = Q_{R_{4}(i + x)}\), which
is injective to \(Q_{i}\) and \(x\). Fig.~\ref{fig_M2tran}(b) shows its transition
diagram. According to Definition 4, \(\mathcal{M}_{2}\) is a
sub-injective FSM.

And definition of \(\mathcal{G}_{2}\) is
\(\mathcal{G}_{2}:\boldsymbol{y}_{n} = \Theta_{2}\left( s_{n} \right) + \boldsymbol{z}_{n}\),
where \(s_{n}\) is the \(n\)-th component of channel input
\(s_{0}^{N - 1}\), \(\boldsymbol{y}_{n}\) is the \(n\)-th component of
channel output \(\boldsymbol{y}_{0}^{N - 1}\), and \(\boldsymbol{z}_{n}\)
denotes AWGN noise. \(\Theta_{2}\) is a mapper from a state to a signal
\begin{align*}
	\Theta_{2}\left( s_{n} \right) := A\exp\left( j\frac{\pi}{2}\left( 2R_{2}\left( s_{n} \right)\boldsymbol{g} + \left\lfloor \frac{s_{n}}{2} \right\rfloor \right) \right)
\end{align*}
where \(\boldsymbol{g} := \left\lbrack \frac{1}{2K}m \right\rbrack_{m \in \left\{ 1,\ldots,K \right\}}\), and \(K \in \mathbb{N}^{*}\) is the sampling factor, 
\(A\) is the constant signal amplitude. From the definition of
\(\mathcal{G}_{2}\) we know it is memoryless. Therefore,
\(\left( \mathcal{M}_{2},\mathcal{G}_{2} \right)\) is a
sub-injective FSMC.

\subsection{Intuitive Applicability of Proposed Structure on Sub-Injective FSM}

As for general sub-injective FSMC, despite demonstrations of
specific instances, analytical proof is not yet obtained. So we give
the following conjecture for future work to verify.

\emph{Conjecture 1}: On any sub-injective FSMC
\(\mathcal{F}_{si} = \left( \mathcal{S},\mathbb{B},F,W \right)\), at code
length \(n \in \left\{ 2^{m} \middle| m \in \mathbb{N}^{*} \right\}\),
variance of sub-channel capacities of last-pair swapping polar code is no
less than that of conventional polar code,
i.e.
\begin{align*}
	\forall n \in \left\{ 2^{m} \middle| m \in \mathbb{N}^{*} \right\},\ \sigma^{2}\left( \mathcal{I}_{\left( \mathcal{P}_{s},\mathcal{F}_{si} \right),n} \right) \geq \sigma^{2}\left( \mathcal{I}_{\left( \mathcal{P}_{c},\mathcal{F}_{si} \right),n} \right)
\end{align*}

The basic idea to support Conjecture 1 lies in a similarity between
structures of conventional polar code and sub-injective FSM. As discussed in Section V-A, sub-injective FSM
has intrinsic polarization potency due to its structure, so last-pair
swapping structure fully exploit this intrinsic polarization potency. However, conventional
polar code suffers from the counteraction between the encoder and sub-injective FSM, which leads to a polarization loss. Therefore, the advantage of last-pair swapping polar
code over conventional one arises.

\begin{figure}[!t]
	\centering
	\includegraphics[width=3.25in]{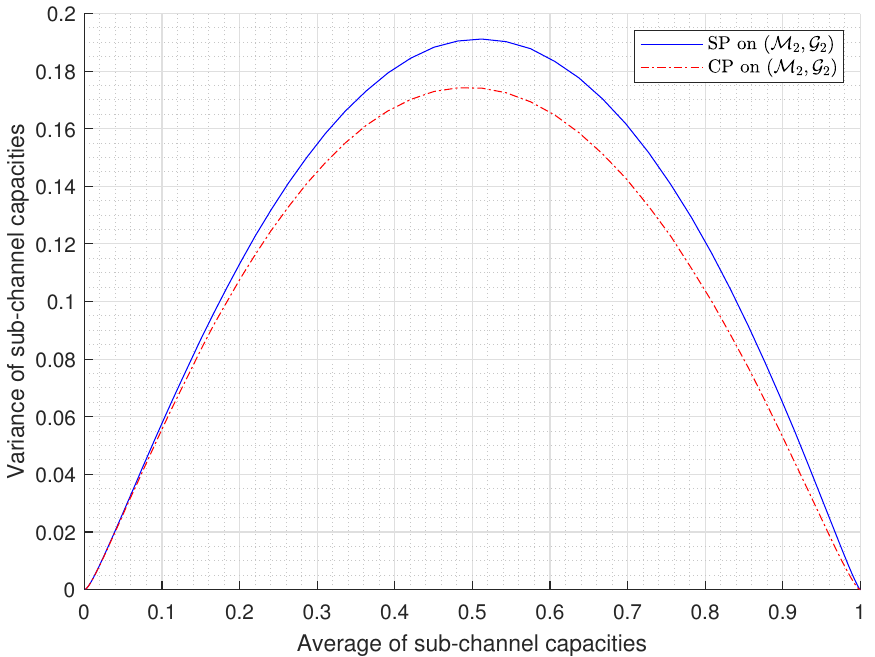}
	\caption{V-A curves of conventional polar code (CP) and last-pair swapping polar code (SP) on \(\left( \mathcal{M}_{2},\mathcal{G}_{2} \right)\) at code length 256.}
	\label{fig_VAcurM2SP}
\end{figure}

Fig.~\ref{fig_VAcurM2SP} shows the polarization improvement on sub-injective FSM, which corroborates Conjecture 1.
On \(\left( \mathcal{M}_{2},\mathcal{G}_{2} \right)\) and at code length 256, through the comparison of V-A curve between the conventional polar code and the last-pair swapping one, an advantage of the proposed structure over the conventional one on sub-injective FSMC is evident.

\section{Simulation Results}

After discussion over polarization efficiency, it is necessary to finally
assess the performance of the proposed polar code compared to the
conventional one by error rate.

\begin{figure}[!t]
\centering
\includegraphics[width=3.5in]{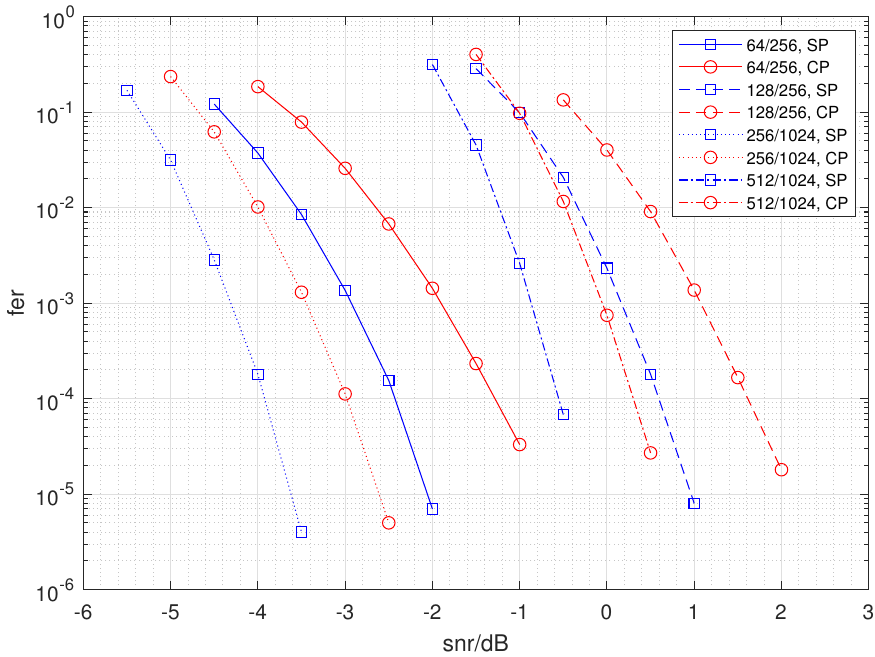}
\caption{Frame error rate on MSK with AWGN channel,
i.e., \(\left( \mathcal{M}_{1},\mathcal{G}_{1} \right)\), under
code length 256, 1024 and code rate 1/4, 1/2, last-pair swapping polar code
(SP) compared with conventional polar code (CP). \(\mathrm{snr} = E_{s}/N_{0}\ \mathrm{(in\ dB)}\).}
\label{fig_ferM1}
\end{figure}



The instance of bijective FSMC is MSK with AWGN channel
\(\left( \mathcal{m}_{1}\mathcal{,g} \right)\), i.e.,
\(\left( \mathcal{M}_{1},\mathcal{G}_{1} \right)\). Fig.~\ref{fig_ferM1} shows frame error rate (FER) of
last-pair swapping polar code and conventional polar code on MSK with AWGN
channel, under two code lengths 256, 1024 and two code rates 1/4, 1/2. The sampling factor
\(K\) is set as 16. And the snr is \(E_{s}/N_{0}\), as
Section IV-A states.
It is seen that the proposed structure exceeds conventional structure in FER
performance. The performance superiority illustrated by Fig.~\ref{fig_ferM1} confirms Theorem 2,
which reveals a polarization efficiency gain accomplished by the proposed
polar code.

\begin{figure}[!t]
\centering
\includegraphics[width=3.5in]{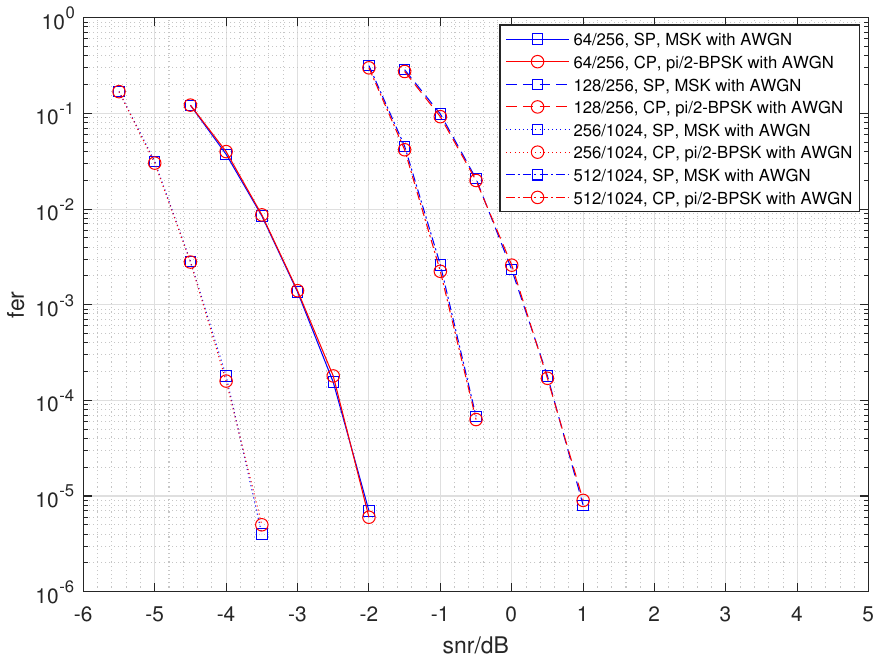}
\caption{Frame error rate of last-pair swapping polar code (SP) on MSK modulation with AWGN,
i.e., \(\left( \mathcal{M}_{1},\mathcal{G}_{1} \right)\),
compared to conventional polar code (CP) on pi/2-BPSK modulation with
AWGN, under code length 256, 1024 and code rate 1/4, 1/2. \(\mathrm{snr} = E_{s}/N_{0}\ \mathrm{(in\ dB)}\).}
\label{fig_ferM1G1}
\end{figure}

Moreover, given that \(\left( \mathcal{M}_{1},\mathcal{G}_{1} \right)\)
is a bijective FSMC, the performance of last-pair swapping polar
code on \(\left( \mathcal{M}_{1},\mathcal{G}_{1} \right)\) is also
compared with that of conventional polar code on its corresponding
memoryless channel \(\mathcal{G}_{1}\), to verify Theorem 1 that
polarization efficiency of last-pair swapping polar code on bijective FSMC is identical with that of conventional polar code on
corresponding memoryless channel. It is well known that MSK is
equivalent to a cascade of differential coding and pi/2-BPSK modulation
\cite{ref4}, the former of which is \(\mathcal{M}_{1}\), so pi/2-BPSK with
AWGN channel is equivalent to \(\mathcal{G}_{1}\), acting as
the corresponding memoryless channel of
\(\left( \mathcal{M}_{1},\mathcal{G}_{1} \right)\). Therefore, Fig.~\ref{fig_ferM1G1} compares the
FER of last-pair swapping polar code on MSK with AWGN channel to
conventional polar code on pi/2-BPSK with AWGN channel, under two code lengths
1024, 256 and two code rates 1/4, 1/2. As indicated in this figure, on a bijective FSMC, last-pair
swapping polar code has accomplished identical performance of conventional polar code on corresponding memoryless channel. This result is consistent with Theorem
1.

\begin{figure}[!t]
\centering
\includegraphics[width=3.5in]{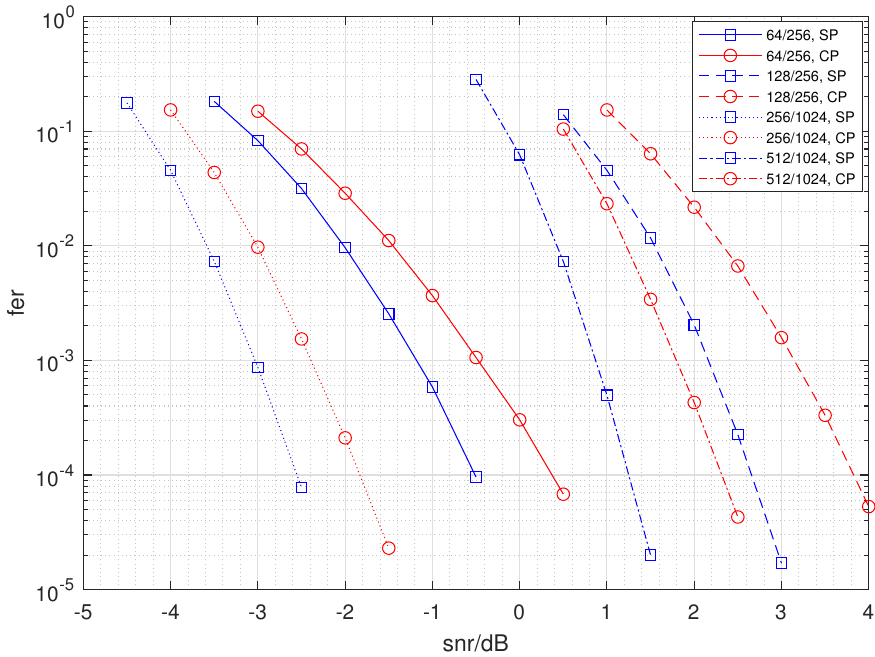}
\caption{Frame error rate on (2, 1/4, 1)-CPM with AWGN channel,
i.e., \(\left( \mathcal{M}_{2},\mathcal{G}_{2} \right)\), under
code length 256, 1024 and code rate 1/4, 1/2, last-pair swapping polar code
(SP) compared with conventional polar code (CP).
\(\mathrm{snr} = E_{s}/N_{0}\ \mathrm{(in\ dB)}\).}
\label{fig_ferM2}
\end{figure}



And the instance of general
sub-injective FSMC is (2,1/4,1)-CPM with AWGN channel
\(\left( \mathcal{m}_{2}\mathcal{,g} \right)\), i.e.,
\(\left( \mathcal{M}_{2},\mathcal{G}_{2} \right)\). Under code length 256,
1024 and code rate 1/4, 1/2, Fig. 14 shows comparation between the proposed
polar code and conventional one on (2, 1/4, 1)-CPM with AWGN channel. As
Fig.~\ref{fig_ferM2} illustrates, the proposed polar code exceeds conventional one in FER
performance. The performance superiority illustrated by Fig.~\ref{fig_ferM2} supports Conjecture 1,
which infers a polarization efficiency gain accomplished by the proposed
polar code.

\section{Conclusion}

In this paper, we research on polar codes under FSMC. A polarization loss is observed 
under bijective FSMC, to solve this we modify the
structure of polar codes, specifically, substitute the last
kernel of each layer in polar coding structure by a swapping matrix, 
to propose a new structure termed last-pair swapping structure. And we
give an analytical proof of the equivalence in polarization efficiency
between last-pair swapping polar codes on bijective FSMC and
conventional ones on corresponding memoryless channels, which exceeds that of
conventional polar codes on bijective FSMC. 
Furthermore, a plausible generalization of last-pair swapping polar 
code on sub-injective FSMC is made. We corroborate the advantage in polarization
efficiency of last-pair swapping polar code over the conventional one on
sub-injective FSMC. The proposed polar code would have promising application prospects 
for FSMC, such as communications with CPM modulation.

\appendix

\subsection{Proof of Proposition 1}

\emph{Proof:}

We will prove the operation mentioned in Proposition 1,
i.e., \(\mathcal{H}\) can be
\(\mathcal{D}_{\boldsymbol{\phi}}\left( \mathcal{V}( \cdot ) \right)\),
namely
\(A_{\left( \mathcal{m}_{1}, \mathcal{g} \right)}\left( x_{0}^{N - 1} \middle| {\widetilde{\boldsymbol{y}}}_{0}^{N - 1} \right) = A_{\left( \mathcal{M}_{1},\mathcal{G}_{1} \right)}\left( x_{0}^{N - 1} \middle| \mathcal{D}_{\boldsymbol{\phi}}\left( \mathcal{V}\left( {\widetilde{\boldsymbol{y}}}_{0}^{N - 1} \right) \right) \right)\),
where operators \(\mathcal{D}_{\boldsymbol{\phi}}\) and \(\mathcal{V}\) are
defined below. \(x_{0}^{N - 1}\) is the input codeword of
\(\mathcal{(m}_{1}\mathcal{,g)}\) and
\({\widetilde{\boldsymbol{y}}}_{0}^{N - 1}\) is its output received signal.
In A.1 we give the relationship between \(\mathcal{m}_{1}\) and
\(\mathcal{M}_{1}\) as formula (4) and prove it, to finish the proof to above formula in
A.3.

\subsubsection*{A.1
	\(\mathcal{m}_{1}\left( x_{0}^{N-1} \right)\)
	and
	\(\mathcal{M}_{1}\left( x_{0}^{N-1} \right)\) are
	related}

In this subsection, we will prove that the relationship between
\(\mathcal{m}_{1}\left( x_{0}^{N - 1} \right)\) and
\(\mathcal{M}_{1}\left( x_{0}^{N - 1} \right)\) are related through the
following equation:
\begin{align} \label{8}
	\mathcal{V}\left( \mathcal{m}_{1}\left( x_{0}^{N - 1} \right) \right) &= \sum_{n = 0}^{N - 1}{A\exp\left( j\pi s_{n} \right)  \boldsymbol{\phi}_{(n)}}, \nonumber \\
	&s_{0}^{N - 1} = \mathcal{M}_{1}\left( x_{0}^{N - 1} \right)
\end{align}

In above formula, \(\mathcal{V}\) is defined as an operator to join all
vectors in a vector sequence together, i.e.,
\(\mathcal{V}\left( \boldsymbol{c}_{0}^{N - 1} \right) := \left\lbrack c_{0}\lbrack 0\rbrack,c_{0}\lbrack 1\rbrack,\ldots,c_{0}\lbrack K - 1\rbrack,c_{1}\lbrack 0\rbrack,\ldots c_{1}\lbrack K - 1\rbrack,\ldots,c_{N - 1}\lbrack K - 1\rbrack \right\rbrack\),
where
\(\boldsymbol{c}_{n} \in \mathbb{C}^{K},\forall n \in \left\{ 0,\ldots,N - 1 \right\}\);
and \(\boldsymbol{\phi}_{(n)}\) is defined as a \(NK\)-length sequence
\(\phi_{(n)}\lbrack k\rbrack = \left\{ \begin{aligned}
	& \phi\lbrack k - nK\rbrack & ,k \in \left\{ nK + 1,\ldots,2(n + 1)K \right\} \\
	& 0 & ,\mathrm{otherwise}
\end{aligned} \right.\ \), namely zero-padding \(\boldsymbol{\phi}\) to
\(NK\)-length and right-shifting it by \(nK\) points. The expression of
\(\boldsymbol{\phi}\) is formula (3). Construct linear space \((\mathbb{C}^{NK},\mathbb{R},+,\cdot)\), and define an inner product in it as \(\left\langle \boldsymbol{\phi}_{(n_1)},\boldsymbol{\phi}_{(n_2)} \right\rangle = \Re\left\{\boldsymbol{\phi}_{(n_1)} \boldsymbol{\phi}_{(n_2)}^H\right\} \), it is easy to verify that distinct
\(\boldsymbol{\phi}_{(n)}\) are orthogonal, i.e.,
\(\left\langle \boldsymbol{\phi}_{\left( n_{1} \right)},\boldsymbol{\phi}_{\left( n_{2} \right)} \right\rangle = \left\| \boldsymbol{\phi} \right\|^{2}\delta\left( n_{1} - n_{2} \right),\forall n_{1},n_{2} \in \left\{ 0,\ldots,N - 1 \right\}\).

Formula (4) is proved as below:

Denote \(\mathcal{V}(\mathcal{m}_{1}\left( x_{0}^{N - 1} \right))\) as
\(\overline{\boldsymbol{d}}\), and
\(\mathcal{m}_{1}\left( x_{0}^{N - 1} \right)\) as
\(\boldsymbol{d}_{0}^{N - 1}\), namely
\(\overline{d}\lbrack nK + k\rbrack = \boldsymbol{d}_{n}\lbrack k\rbrack.\)
And denote \(\mathcal{M}_{1}\left( x_{0}^{N - 1} \right)\) as
\(s_{0}^{N - 1}\). According to formula (1) we have
\begin{align*}
	\boldsymbol{d}_{0}^{N - 1} &= \mathcal{m}_{1}\left( x_{0}^{N - 1} \right) = \mathcal{m}_{(2,1/2,1)}\left( x_{0}^{N - 1} \right) \\
	&= \left\lbrack A\exp\left( j\pi\left\lbrack R_{2}\left( \sum_{i = 0}^{n - 1}x_{i} \right) + 2x_{n}\boldsymbol{g} \right\rbrack \right) \right\rbrack_{n \in \left\{ 0,\ldots,N - 1 \right\}}
\end{align*}
so
\begin{align*}
	\overline{d}\lbrack nK + k\rbrack &= A\exp\left( j\pi\left\lbrack R_{2}\left( \sum_{i = 0}^{n - 1}x_{i} \right) + 2x_{n}g\lbrack k\rbrack \right\rbrack \right) \\
	&= A\exp\left( j\pi\left\lbrack s_{n - 1} + x_{n}\frac{1}{K}k \right\rbrack \right)
\end{align*}

According to formula (2), if \(x_{n} = 0\), \(s_{n} = s_{n - 1}\),
\(\overline{d}\lbrack nK + k\rbrack = A\exp\left( j\pi s_{n - 1} \right) = A\exp\left( j\pi s_{n - 1} \right)  \frac{1}{2}\left( 1 + \exp\left( j\frac{\pi}{K}k \right) + 1 - \exp\left( j\frac{\pi}{K}k \right) \right) = A\exp\left( j\pi s_{n} \right)  \left( \phi\lbrack k + K\rbrack + \phi\lbrack k\rbrack \right) = A\exp\left( j\pi s_{n - 1} \right)  \phi\lbrack k + K\rbrack\boldsymbol{+}A\exp\left( j\pi s_{n} \right)  \phi\lbrack k\rbrack\);
else, \(x_{n} = 1\), \(s_{n} = R_{2}\left( s_{n - 1} + 1 \right)\),
\(\overline{d}\lbrack nK + k\rbrack = A\exp\left( j\pi s_{n - 1} \right)  \exp\left( j\frac{\pi}{K}k \right) = A\exp\left( j\pi s_{n - 1} \right)  \frac{1}{2}\left( 1 + \exp\left( j\frac{\pi}{K}k \right) - 1 + \exp\left( j\frac{\pi}{K}k \right) \right) = A\exp\left( j\pi s_{n - 1} \right)  \left( \phi\lbrack k + K\rbrack - \phi\lbrack k\rbrack \right) = A\exp\left( j\pi s_{n - 1} \right)  \phi\lbrack k + K\rbrack\boldsymbol{+}A\exp\left( j\pi s_{n} \right)  \phi\lbrack k\rbrack\).
In summary,
\(\overline{d}\lbrack nK + k\rbrack = A\exp\left( j\pi s_{n - 1} \right)  \phi\lbrack k + K\rbrack\boldsymbol{+}A\exp\left( j\pi s_{n} \right)  \phi\lbrack k\rbrack\).

And according to the definition of \(\boldsymbol{\phi}_{(n)}\), we know
\(\phi_{\left( n^{\prime} \right)}\lbrack nK + k\rbrack = \left\{ \begin{aligned}
	& \phi\lbrack k\rbrack & ,n^{\prime} = n \\
	& \phi\lbrack k + K\rbrack & ,n^{\prime} = n - 1 \\
	& 0 & ,\mathrm{otherwise}
\end{aligned} \right.\ \), so
\begin{align*}
	\overline{d}\lbrack nK + k\rbrack &= A\exp\left( j\pi s_{n - 1} \right)  \phi_{(n - 1)}\lbrack nK + k\rbrack \\
	&+ A\exp\left( j\pi s_{n} \right)  \phi_{(n)}\lbrack nK + k\rbrack \\
	&= \sum_{n^{\prime} = 0}^{N - 1}{A\exp\left( j\pi s_{n^{\prime}} \right)  \phi_{\left( n^{\prime} \right)}\lbrack nK + k\rbrack}
\end{align*}

In other words,
\(\overline{\boldsymbol{d}} = \sum_{n = 0}^{N - 1}{A\exp\left( j\pi s_{n} \right)  \boldsymbol{\phi}_{(n)}}\).
So
\(\mathcal{V}\left( \mathcal{m}_{1}\left( x_{0}^{N - 1} \right) \right) = \sum_{n = 0}^{N - 1}{A\exp\left( j\pi s_{n} \right)  \boldsymbol{\phi}_{(n)},s_{0}^{N - 1} = \mathcal{M}_{1}\left( x_{0}^{N - 1} \right)}\).

\subsubsection*{A.2 Definition of operator
	\(\mathcal{D}_{\boldsymbol{\phi}}\)}
\(\mathcal{D}_{\boldsymbol{\phi}}\) is an operator acting on an
\(NK\)-length complex number sequence to produce an \(N\)-length vector
sequence, the elements of which are \(2K\)-length complex vectors.
\(\mathcal{D}_{\boldsymbol{\phi}}\) is defined as
\(\mathcal{D}_{\boldsymbol{\phi}}\left( \boldsymbol{c} \right) := \left\lbrack \frac{1}{\left\| \boldsymbol{\phi} \right\|^{2}}\left\langle \boldsymbol{c},\boldsymbol{\phi}_{(n)} \right\rangle\boldsymbol{\phi} \right\rbrack_{n \in \left\{ 0,\ldots,N - 1 \right\}},\forall\boldsymbol{c} \in \mathbb{C}^{NK}\).
\(\mathcal{D}_{\boldsymbol{\phi}}\) means extracting a sequence's orthogonal
component on each \(\boldsymbol{\phi}_{(n)}\), and taking each component
sequence only on its support set to list them as a vector sequence.

Specifically, according to the knowledge of Hilbert space, for any
\(\boldsymbol{c} \in \mathbb{C}^{NK}\), it has a unique orthogonal
projection in the subspace spanned by
\(\boldsymbol{\Phi} := \left\{ \boldsymbol{\phi}_{(n)} \middle| n \in \left\{ 0,\ldots,N - 1 \right\} \right\}\),
i.e.,
\(D := \left\{ \sum_{n = 0}^{N - 1}{k_{n}\boldsymbol{\phi}_{(n)}} \middle| k_{0}^{N - 1} \in \mathbb{R}^{N} \right\}\),
which is
\(\sum_{n = 0}^{N - 1}{\frac{1}{\left\| \boldsymbol{\phi} \right\|^{2}}\left\langle \boldsymbol{c},\boldsymbol{\phi}_{(n)} \right\rangle\boldsymbol{\phi}_{(n)}}\),
and the \(n\)-th term of this summation, i.e.,
\(\frac{1}{\left\| \boldsymbol{\phi} \right\|^{2}}\left\langle \boldsymbol{c},\boldsymbol{\phi}_{(n)} \right\rangle\boldsymbol{\phi}_{(n)}\)
is the orthogonal component of \(\boldsymbol{c}\) on basis
\(\boldsymbol{\phi}_{(n)}\). Furthermore, from the definition of
\(\boldsymbol{\phi}_{(n)}\) we know its support set is
\(\left\{ nK + 1,\ldots,2(n + 1)K \right\}\), namely the non-zero part
of \(\boldsymbol{\phi}_{(n)}\) is a \(2K\)-length vector, so after
extracting each orthogonal component, we only take this non-zero part,
i.e.,
\(\frac{1}{\left\| \boldsymbol{\phi} \right\|^{2}}\left\langle \boldsymbol{c},\boldsymbol{\phi}_{(n)} \right\rangle\boldsymbol{\phi}\),
and list all of them as a vector sequence
\(\left\lbrack \frac{1}{\left\| \boldsymbol{\phi} \right\|^{2}}\left\langle \boldsymbol{c},\boldsymbol{\phi}_{(n)} \right\rangle\boldsymbol{\phi} \right\rbrack_{n \in \left\{ 0,\ldots,N - 1 \right\}}\).
Such operator on \(\boldsymbol{c}\) is \(\mathcal{D}_{\boldsymbol{\phi}}\).

\subsubsection*{A.3 A posteriori probability of
	\(\left( \mathcal{m}_{\boldsymbol{1}}\mathcal{,g} \right)\) and that of
	\(\left( \mathcal{M}_{\boldsymbol{1}}\boldsymbol{,}\mathcal{G}_{\boldsymbol{1}} \right)\)
	are
	equivalent}
In this subsection, we will prove that a posteriori probability of
\(\left( \mathcal{m}_{1}\mathcal{,g} \right)\) and
\(\left( \mathcal{M}_{1},\mathcal{G}_{1} \right)\) are identical through
operator
\(\mathcal{D}_{\boldsymbol{\phi}}\left( \mathcal{V}( \cdot ) \right)\),
namely
\begin{align} \label{9}
	&A_{\left( \mathcal{m}_{1},\mathcal{g} \right)}\left( x_{0}^{N - 1} \middle| {\widetilde{\boldsymbol{y}}}_{0}^{N - 1} \right) \nonumber \\
	&= A_{\left( \mathcal{M}_{1},\mathcal{G}_{1} \right)}\left( x_{0}^{N - 1} \middle| \mathcal{D}_{\boldsymbol{\phi}}\left( \mathcal{V}\left( {\widetilde{\boldsymbol{y}}}_{0}^{N - 1} \right) \right) \right)
\end{align}

Equation (5) is proved as below:

According to Bayes formula, the a posteriori probability can be derived
from the likelihood probability. Therefore, to begin with, we derive the
likelihood probability relationship below. The likelihood probability of
\(\left( \mathcal{m}_{1}\mathcal{,g} \right)\) and
\(\left( \mathcal{M}_{1},\mathcal{G}_{1} \right)\) are

\[W_{\left( \mathcal{m}_{1}, \mathcal{g} \right)}\left( {\widetilde{\boldsymbol{y}}}_{0}^{N - 1} \middle| x_{0}^{N - 1} \right) = \left( \frac{1}{2\pi\sigma^{2}} \right)^{NK}e^{- \frac{\left\| {\widetilde{\boldsymbol{y}}}_{0}^{N - 1} - \mathcal{m}_{1}\left( x_{0}^{N - 1} \right) \right\|^{2}}{2\sigma^{2}}}\]

\[W_{\left( \mathcal{M}_{1},\mathcal{G}_{1} \right)}\left( \boldsymbol{y}_{0}^{N - 1} \middle| x_{0}^{N - 1} \right) = \left( \frac{1}{2\pi\sigma^{2}} \right)^{2NK}e^{- \frac{\left\| \boldsymbol{y}_{0}^{N - 1} - \Theta_{1}^{N}\left( \mathcal{M}_{1}\left( x_{0}^{N - 1} \right) \right) \right\|^{2}}{2\sigma^{2}}}\]

For \(\left( \mathcal{m}_{1}\mathcal{,g} \right)\), with operator
\(\mathcal{V}\) we transform its likelihood probability formula to
\(W_{\left( \mathcal{m}_{1}, \mathcal{g} \right)}\left( {\widetilde{\boldsymbol{y}}}_{0}^{N - 1} \middle| x_{0}^{N - 1} \right) = \left( \frac{1}{2\pi\sigma^{2}} \right)^{NK}e^{- \frac{\left\| \mathcal{V}\left( {\widetilde{\boldsymbol{y}}}_{0}^{N - 1} \right)\mathcal{- V}\left( \mathcal{m}_{1}\left( x_{0}^{N - 1} \right) \right) \right\|^{2}}{2\sigma^{2}}}\),
and denote
\(\overline{\boldsymbol{y}} = \mathcal{V}({\widetilde{\boldsymbol{y}}}_{0}^{N - 1})\),
we have

\[W_{\left( \mathcal{m}_{1}, \mathcal{g} \right)}\left( {\widetilde{\boldsymbol{y}}}_{0}^{N - 1} \middle| x_{0}^{N - 1} \right) = \left( \frac{1}{2\pi\sigma^{2}} \right)^{NK}e^{- \frac{\left\| \overline{\boldsymbol{y}}\mathcal{- V}\left( \mathcal{m}_{1}\left( x_{0}^{N - 1} \right) \right) \right\|^{2}}{2\sigma^{2}}}\]

For any \(\overline{\boldsymbol{y}} \in \mathbb{C}^{NK}\), denote its
orthogonal projection in subspace \(D\) as
\({\overline{\boldsymbol{y}}}_{s}\), namely
\({\overline{\boldsymbol{y}}}_{s} := \sum_{n = 0}^{N - 1}{\frac{1}{\left\| \boldsymbol{\phi} \right\|^{2}}\left\langle \overline{\boldsymbol{y}},\boldsymbol{\phi}_{(n)} \right\rangle\boldsymbol{\phi}_{(n)}}\),
and denote its orthogonal component as \({\overline{\boldsymbol{y}}}_{z}\),
i.e.,
\({\overline{\boldsymbol{y}}}_{z} = \overline{\boldsymbol{y}} - {\overline{\boldsymbol{y}}}_{s}\),
which is orthogonal to any sequence in \(D\). According to formula (4), there is
\(\mathcal{V}\left( \mathcal{m}_{1}\left( x_{0}^{N - 1} \right) \right) = \sum_{n = 0}^{N - 1}{A\exp\left( j\pi s_{n} \right)\boldsymbol{\phi}_{(n)}} \in D\),
so for
\(\overline{\boldsymbol{y}}\mathcal{- V}\left( \mathcal{m}_{1}\left( x_{0}^{N - 1} \right) \right) = {\overline{\boldsymbol{y}}}_{s} + {\overline{\boldsymbol{y}}}_{z} - \mathcal{V}\left( \mathcal{m}_{1}\left( x_{0}^{N - 1} \right) \right) = {\overline{\boldsymbol{y}}}_{s} - \mathcal{V}\left( \mathcal{m}_{1}\left( x_{0}^{N - 1} \right) \right) + {\overline{\boldsymbol{y}}}_{z}\),
we have
\(\left\langle {\overline{\boldsymbol{y}}}_{s} - \mathcal{V}\left( \mathcal{m}_{1}\left( x_{0}^{N - 1} \right) \right),{\overline{\boldsymbol{y}}}_{z} \right\rangle = 0\),
then
\(\left\| \overline{\boldsymbol{y}}\mathcal{- V}\left( \mathcal{m}_{1}\left( x_{0}^{N - 1} \right) \right) \right\|^{2} = \left\| {\overline{\boldsymbol{y}}}_{s} - \mathcal{V}\left( \mathcal{m}_{1}\left( x_{0}^{N - 1} \right) \right) \right\|^{2} + \left\| {\overline{\boldsymbol{y}}}_{z} \right\|^{2}\).
And for
\(\left\| {\overline{\boldsymbol{y}}}_{s} - \mathcal{V}\left( \mathcal{m}_{1}\left( x_{0}^{N - 1} \right) \right) \right\|^{2}\),
according to the orthogonality of each component on
\(\boldsymbol{\phi}_{(n)}\), we have
\(\left\| {\overline{\boldsymbol{y}}}_{s} - \mathcal{V}\left( \mathcal{m}_{1}\left( x_{0}^{N - 1} \right) \right) \right\|^{2} = \sum_{n = 0}^{N - 1}\left\| \frac{1}{\left\| \boldsymbol{\phi} \right\|^{2}}\left\langle \overline{\boldsymbol{y}},\boldsymbol{\phi}_{(n)} \right\rangle\boldsymbol{\phi}_{(n)} - A\exp\left( j\pi s_{n} \right)\boldsymbol{\phi}_{(n)} \right\|^{2} = \sum_{n = 0}^{N - 1}\left\| \frac{1}{\left\| \boldsymbol{\phi} \right\|^{2}}\left\langle \overline{\boldsymbol{y}},\boldsymbol{\phi}_{(n)} \right\rangle\boldsymbol{\phi} - A\exp\left( j\pi s_{n} \right)\boldsymbol{\phi} \right\|^{2} = \left\| \mathcal{D}_{\boldsymbol{\phi}}\left( \overline{\boldsymbol{y}} \right)\  - \Theta_{1}^{N}\left( \mathcal{M}_{1}\left( x_{0}^{N - 1} \right) \right) \right\|^{2}\).
So
\begin{align*}
	&\left\| \overline{\boldsymbol{y}}\mathcal{- V}\left( \mathcal{m}_{1}\left( x_{0}^{N - 1} \right) \right) \right\|^{2} \\
	&= \left\| \mathcal{D}_{\boldsymbol{\phi}}\left( \overline{\boldsymbol{y}} \right) - \Theta_{1}^{N}\left( \mathcal{M}_{1}\left( x_{0}^{N - 1} \right) \right) \right\|^{2} + \left\| {\overline{\boldsymbol{y}}}_{z} \right\|^{2}
\end{align*}

Substitute it into the formula of likelihood probability of
\(\left( \mathcal{m}_{1}\mathcal{,g} \right)\), there
is \(W_{\left( \mathcal{m}_{1}, \mathcal{g} \right)}\left( {\widetilde{\boldsymbol{y}}}_{0}^{N - 1} \middle| x_{0}^{N - 1} \right) = \left( \frac{1}{2\pi\sigma^{2}} \right)^{NK}e^{- \frac{\left\| \overline{\boldsymbol{y}}\mathcal{- V}\left( \mathcal{m}_{1}\left( x_{0}^{N - 1} \right) \right) \right\|^{2}}{2\sigma^{2}}} = \left( \frac{1}{2\pi\sigma^{2}} \right)^{NK}e^{- \frac{\left\| \mathcal{D}_{\boldsymbol{\phi}}\left( \overline{\boldsymbol{y}} \right) - \Theta_{1}^{N}\left( \mathcal{M}_{1}\left( x_{0}^{N - 1} \right) \right) \right\|^{2} + \left\| {\overline{\boldsymbol{y}}}_{z} \right\|^{2}}{2\sigma^{2}}} = \left( \frac{1}{2\pi\sigma^{2}} \right)^{2NK}e^{- \frac{\left\| \mathcal{D}_{\boldsymbol{\phi}}\left( \overline{\boldsymbol{y}} \right) - \Theta_{1}^{N}\left( \mathcal{M}_{1}\left( x_{0}^{N - 1} \right) \right) \right\|^{2}}{2\sigma^{2}}}  \left( \frac{1}{2\pi\sigma^{2}} \right)^{- NK}e^{- \frac{\left\| {\overline{\boldsymbol{y}}}_{z} \right\|^{2}}{2\sigma^{2}}} = W_{\left( \mathcal{M}_{1},\mathcal{G}_{1} \right)}\left( \mathcal{D}_{\boldsymbol{\phi}}\left( \overline{\boldsymbol{y}} \right) \middle| x_{0}^{N - 1} \right)  \left( \frac{1}{2\pi\sigma^{2}} \right)^{- NK}e^{- \frac{\left\| {\overline{\boldsymbol{y}}}_{z} \right\|^{2}}{2\sigma^{2}}}\),
then with Bayes formula, the a posteriori probability is
\begin{align*}
	&A_{\left( \mathcal{m}_{1}, \mathcal{g} \right)}\left( x_{0}^{N - 1} \middle| {\widetilde{\boldsymbol{y}}}_{0}^{N - 1} \right) = \frac{W_{\left( \mathcal{m}_{1}, \mathcal{g}_{1} \right)}\left( \overline{\boldsymbol{y}} \middle| x_{0}^{N - 1} \right)}{\sum_{\chi_{0}^{N - 1} \in \mathbb{B}^{N}}^{}{W_{\left( \mathcal{m}_{1}, \mathcal{g}_{1} \right)}\left( \overline{\boldsymbol{y}} \middle| \chi_{0}^{N - 1} \right)}} \\
	&= \frac{W_{\left( \mathcal{M}_{1},\mathcal{G}_{1} \right)}\left( \mathcal{D}_{\boldsymbol{\phi}}\left( \overline{\boldsymbol{y}} \right) \middle| x_{0}^{N - 1} \right)  \left( \frac{1}{2\pi\sigma^{2}} \right)^{- NK}e^{- \frac{\left\| {\overline{\boldsymbol{y}}}_{z} \right\|^{2}}{2\sigma^{2}}}}{\sum_{\chi_{0}^{N - 1} \in \mathbb{B}^{N}}^{}{W_{\left( \mathcal{M}_{1},\mathcal{G}_{1} \right)}\left( \mathcal{D}_{\boldsymbol{\phi}}\left( \overline{\boldsymbol{y}} \right) \middle| \chi_{0}^{N - 1} \right)  \left( \frac{1}{2\pi\sigma^{2}} \right)^{- NK}e^{- \frac{\left\| {\overline{\boldsymbol{y}}}_{z} \right\|^{2}}{2\sigma^{2}}}}} \\
	&= \frac{W_{\left( \mathcal{M}_{1},\mathcal{G}_{1} \right)}\left( \mathcal{D}_{\boldsymbol{\phi}}\left( \overline{\boldsymbol{y}} \right) \middle| x_{0}^{N - 1} \right)}{\sum_{\chi_{0}^{N - 1} \in \mathbb{B}^{N}}^{}{W_{\left( \mathcal{M}_{1},\mathcal{G}_{1} \right)}\left( \mathcal{D}_{\boldsymbol{\phi}}\left( \overline{\boldsymbol{y}} \right) \middle| \chi_{0}^{N - 1} \right)}} \\
	&= A_{\left( \mathcal{M}_{1},\mathcal{G}_{1} \right)}\left( x_{0}^{N - 1} \middle| \mathcal{D}_{\boldsymbol{\phi}}\left( \overline{\boldsymbol{y}} \right) \right) \\
	&= A_{\left( \mathcal{M}_{1},\mathcal{G}_{1} \right)}\left( x_{0}^{N - 1} \middle| \mathcal{D}_{\boldsymbol{\phi}}\left( \mathcal{V}\left( {\widetilde{\boldsymbol{y}}}_{0}^{N - 1} \right) \right) \right)
\end{align*}
briefly
\begin{align*}
	&A_{\left( \mathcal{m}_{1}, \mathcal{g} \right)}\left( x_{0}^{N - 1} \middle| {\widetilde{\boldsymbol{y}}}_{0}^{N - 1} \right) \\
	&= A_{\left( \mathcal{M}_{1},\mathcal{G}_{1} \right)}\left( x_{0}^{N - 1} \middle| \mathcal{D}_{\boldsymbol{\phi}}\left( \mathcal{V}\left( {\widetilde{\boldsymbol{y}}}_{0}^{N - 1} \right) \right) \right)
\end{align*}

Therefore, let operator \(\mathcal{H}\) be
\(\mathcal{D}_{\boldsymbol{\phi}}\left( \mathcal{V}( \cdot ) \right)\),
namely
\(\mathcal{H}\left( {\widetilde{\boldsymbol{y}}}_{0}^{N - 1} \right) = \mathcal{D}_{\boldsymbol{\phi}}\left( \mathcal{V}\left( {\widetilde{\boldsymbol{y}}}_{0}^{N - 1} \right) \right)\),
it holds that
\(A_{\left( \mathcal{m}_{1}, \mathcal{g} \right)}\left( x_{0}^{N - 1} \middle| {\widetilde{\boldsymbol{y}}}_{0}^{N - 1} \right) = A_{\left( \mathcal{M}_{1},\mathcal{G}_{1} \right)}\left( x_{0}^{N - 1} \middle| \mathcal{H}\left( {\widetilde{\boldsymbol{y}}}_{0}^{N - 1} \right) \right)\),
proof of Proposition 1 is finished.

\subsection{Proof of Proposition 2}

To prove Proposition 2, we firstly put up with a condition about two
channels denoted as \emph{A-condition} in Definition 5, and prove it
acts as a sufficient condition for identical capacity in Lemma 1. Then
we infer two conclusions in Lemma 2, 3 to prove that A-condition is
satisfied between sub-channels of conventional polar code on bijective
FSMCs and those of conventional polar code on memoryless
channels, in Lemma 5. Finally with major premise Lemma 1 and minor
premise Lemma 5, we finish the proof to Proposition 2.

For convenience, we abbreviate Lemma x to Le. x, and Corollary y to Co.
y, where x and y are positive integers. And while applying them to an
equation, we use the form \(\ldots\overset{Le. x}{=}\ldots\) or
\(\ldots\overset{Co. y}{=}\ldots\).

And for convenience we define
\[W_{n}^{(i)} := W_{\left( \mathcal{P}_{c},\mathcal{F}_{b} \right),\ n}^{(i)},{\widetilde{W}}_{n}^{(i)} := W_{\left( \mathcal{P}_{c},W^{n} \right),\ n}^{(i)}\]
\[I_{n}^{(i)} := I\left( W_{n}^{(i)} \right),{\widetilde{I}}_{n}^{(i)} := I\left( {\widetilde{W}}_{n}^{(i)} \right)\]
\[F_{s}^{- 1}t := the\ u\ that\ F_{s}u = t\]
\[F_{s}^{n}u_{0}^{n - 1} := \left\{ \begin{aligned}
	& F_{s}u_{0} & ,n = 1 \\
	& F_{F_{s}^{n - 1}u_{0}^{n - 2}}u_{n - 1} & ,n \geq 2
\end{aligned} \right.\ \]
where \(F_{s}u\) means \(F(s,u)\), just as Section III denotes.

In addition, notice
that in Definition 1 we fix the initial state as 0, namely
\(s_{0} = F\left( 0,x_{0} \right)\); on the other hand, if the initial
state is not fixed as 0, but arbitrary \(s \in \mathcal{S}\), namely
\(s_{0} = F\left( s,x_{0} \right)\), we overload the notation
\(W_{N}^{(i)}\) with state
\(W_{N}^{(i)}\left( y_{0}^{N - 1},u_{0}^{i - 1} \middle| u_{i},s \right)\).

\emph{Definition 5:} A-condition. For any two channels \(w_{1},w_{2}\)
with likelihood probability distribution
\(w_{1}\left( y_{0}^{n - 1},u_{0}^{i - 1} \middle| u_{i} \right),\ y_{0}^{n - 1} \in \mathcal{Y}^{n},u_{0}^{i} \in \mathbb{B}^{i + 1},n,i\mathbb{\in N}\)
and
\(w_{2}\left( z_{0}^{m - 1},v_{0}^{j - 1} \middle| v_{j} \right),\ z_{0}^{m - 1} \in \mathcal{Y}^{m},v_{0}^{j} \in \mathbb{B}^{j + 1},m,j\mathbb{\in N}\),
\(\mathcal{Y}\) is any set, call \(w_{1}\) and \(w_{2}\) satisfy
A-condition, denoted as \(w_{1}\overset{A}{\sim}w_{2}\), if and only if
there exists a sequence
\(\mathcal{J \subseteq}\left\{ 0,\ldots,n - 1 \right\},\left| \mathcal{J} \right| = m\),
a mapping \(\rho:\mathbb{B}^{i} \mapsto \mathbb{B}^{j}\), a mapping
\(d\left( u_{0}^{i} \right):\mathbb{B}^{i + 1}\mathbb{\mapsto B}\)
bijective to \(u_{i}\), and a function
\(w:\mathcal{Y}^{n - m} \times \mathbb{B}^{i} \mapsto \lbrack 0, + \infty)\)
such that
\begin{align*}
	&\forall y_{0}^{n - 1} \in \mathcal{Y}^{n},u_{0}^{i} \in \mathbb{B}^{i + 1}, w_{1}\left( y_{0}^{n - 1},u_{0}^{i - 1} \middle| u_{i} \right) \\
	&= w_{2}\left( y_{\mathcal{J}},\rho\left( u_{0}^{i - 1} \right) \middle| d\left( u_{0}^{i} \right) \right)w\left( y_{\overline{\mathcal{J}}},u_{0}^{i - 1} \right)
\end{align*}
\begin{align*}
	\forall v_{0}^{j - 1} \in \mathbb{B}^{j},&\sum_{y_{\overline{\mathcal{J}}} \in \mathcal{Y}^{m},u_{0}^{i - 1} \in \mathbb{B}^{i}}^{}{\delta\left( \rho\left( u_{0}^{i - 1} \right) - v_{0}^{j - 1} \right)w\left( y_{\overline{\mathcal{J}}},u_{0}^{i - 1} \right)} \\
	&= 1
\end{align*}
where
\(\overline{\mathcal{J}} := \left\{ 0,\ldots,n - 1 \right\}\mathcal{\backslash J}\).

Notice that while the channel output \(y_i\) is continuous, the summation symbol \(\Sigma\) refers to
integration.

\emph{Lemma 1:} Channels satisfying A-condition have identical capacity.
Namely, for any \(w_{1},w_{2}\), if \(w_{1}\overset{A}{\sim}w_{2}\),
\(I\left( w_{1} \right) = I\left( w_{2} \right)\).

\emph{Proof:}

\(w_{1}\overset{A}{\sim}w_{2}\), so according to Definition 5, there
exists a sequence
\(\mathcal{J \subseteq}\left\{ 0,\ldots,n - 1 \right\},\left| \mathcal{J} \right| = m\),
a surjection \(\rho:\mathbb{B}^{i} \mapsto \mathbb{B}^{j}\) and a
mapping
\(d\left( u_{0}^{i} \right):\mathbb{B}^{i + 1}\mathbb{\mapsto B}\)
bijective to \(u_{i}\) such that
\begin{align*}
	&\forall y_{0}^{n - 1} \in \mathcal{Y}^{n},u_{0}^{i} \in \mathbb{B}^{i + 1}, w_{1}\left( y_{0}^{n - 1},u_{0}^{i - 1} \middle| u_{i} \right) \\
	&= w_{2}\left( y_{\mathcal{J}},\rho\left( u_{0}^{i - 1} \right) \middle| d\left( u_{0}^{i} \right) \right)w\left( y_{\overline{\mathcal{J}}},u_{0}^{i - 1} \right)
\end{align*}
\begin{align*}
	\forall v_{0}^{j - 1} \in \mathbb{B}^{j},&\sum_{y_{\overline{\mathcal{J}}} \in \mathcal{Y}^{m},u_{0}^{i - 1} \in \mathbb{B}^{i}}^{}{\delta\left( \rho\left( u_{0}^{i - 1} \right) - v_{0}^{j - 1} \right)w\left( y_{\overline{\mathcal{J}}},u_{0}^{i - 1} \right)} \\
	&= 1
\end{align*}
where
\(\overline{\mathcal{J}} := \left\{ 0,\ldots,n - 1 \right\}\mathcal{\backslash J}\).

Therefore
\begin{align*}
	&w_{1}\left( y_{0}^{n - 1},u_{0}^{i - 1} \middle| 0 \right) + w_{1}\left( y_{0}^{n - 1},u_{0}^{i - 1} \middle| 1 \right) \\
	&= \left\lbrack w_{2}\left( y_{\mathcal{J}},\rho\left( u_{0}^{i - 1} \right) \middle| 0 \right) + w_{2}\left( y_{\mathcal{J}},\rho\left( u_{0}^{i - 1} \right) \middle| 1 \right) \right\rbrack w\left( y_{\overline{\mathcal{J}}},u_{0}^{i - 1} \right)
\end{align*}

So mutual information
\begin{align*}
	&\eta_{1}\left( y_{0}^{n - 1},u_{0}^{i - 1};u_{i} \right) \\
	&= \log_{2}\frac{2w_{1}\left( y_{0}^{n - 1},u_{0}^{i - 1} \middle| u_{i} \right)}{w_{1}\left( y_{0}^{n - 1},u_{0}^{i - 1} \middle| 0 \right) + w_{1}\left( y_{0}^{n - 1},u_{0}^{i - 1} \middle| 1 \right)} \\
	&= \log_{2}\frac{2w_{2}\left( y_{\mathcal{J}},\rho\left( u_{0}^{i - 1} \right) \middle| d\left( u_{0}^{i} \right) \right)}{w_{2}\left( y_{\mathcal{J}},\rho\left( u_{0}^{i - 1} \right) \middle| 0 \right) + w_{2}\left( y_{\mathcal{J}},\rho\left( u_{0}^{i - 1} \right) \middle| 1 \right)} \\
	&= \eta_{2}\left( y_{\mathcal{J}},\rho\left( u_{0}^{i - 1} \right);d\left( u_{0}^{i} \right) \right)
\end{align*}

Define \(p\) as probability distribution of \(w_{1}\), and \(q\) as that
of \(w_{2}\), and define
\(Z_{0}^{m - 1} = Y_{\mathcal{J}},\ V_{0}^{j} = \left\lbrack \rho\left( U_{0}^{i - 1} \right),d\left( U_{0}^{i} \right) \right\rbrack\),
then
\begin{align*}
	I\left( w_{1} \right) &= \mathbb{E}_{p}\left\lbrack \eta_{1}\left( Y_{0}^{n - 1},U_{0}^{i - 1};U_{i} \right) \right\rbrack \\
	&= \mathbb{E}_{p}\left\lbrack \eta_{2}\left( Y_{\mathcal{J}},\rho\left( U_{0}^{i - 1} \right);d\left( U_{i} \right) \right) \right\rbrack \\
	&= \mathbb{E}_{p}\left\lbrack \eta_{2}\left( Z_{0}^{m - 1},V_{0}^{j - 1};V_{j} \right) \right\rbrack
\end{align*}
\begin{align*}
	&p_{Z_{0}^{m - 1}V_{0}^{j}}\left( z_{0}^{m - 1},v_{0}^{j} \right) \\
	&= \sum_{y_{\mathcal{J}} = z_{0}^{m - 1},u_{0}^{i} \in \mathbb{B}^{i + 1}}^{}\delta\left( \left\lbrack \rho\left( u_{0}^{i - 1} \right),d\left( u_{0}^{i} \right) \right\rbrack - v_{0}^{j} \right) \\
	&\cdot p_{Y_{0}^{n - 1}U_{0}^{i}}\left( y_{0}^{n - 1},u_{0}^{i} \right) \\
	&= \sum_{y_{\mathcal{J}} = z_{0}^{m - 1},u_{0}^{i} \in \mathbb{B}^{i + 1}}^{}\frac{1}{2}\delta\left( \left\lbrack \rho\left( u_{0}^{i - 1} \right),d\left( u_{0}^{i} \right) \right\rbrack - v_{0}^{j} \right) \\
	& \cdot w_{1}\left( y_{0}^{n - 1},u_{0}^{i - 1} \middle| u_{i} \right) \\
	&= \sum_{y_{\mathcal{J}} = z_{0}^{m - 1},\ u_{0}^{i} \in \mathbb{B}^{i + 1}}^{}\frac{1}{2}\delta\left( \left\lbrack \rho\left( u_{0}^{i - 1} \right),d\left( u_{0}^{i} \right) \right\rbrack - v_{0}^{j} \right) \\
	& \cdot w_{2}\left( y_{\mathcal{J}},\rho\left( u_{0}^{i - 1} \right) \middle| d\left( u_{0}^{i} \right) \right)w\left( y_{\overline{\mathcal{J}}},u_{0}^{i - 1} \right) \\
	&= \sum_{y_{\mathcal{J}} = z_{0}^{m - 1},\ u_{0}^{i} \in \mathbb{B}^{i + 1}}^{}\frac{1}{2}\delta\left( \left\lbrack \rho\left( u_{0}^{i - 1} \right),d\left( u_{0}^{i} \right) \right\rbrack - v_{0}^{j} \right) \\
	& \cdot w_{2}\left( z_{0}^{m - 1},v_{0}^{j - 1} \middle| v_{j} \right)w\left( y_{\overline{\mathcal{J}}},u_{0}^{i - 1} \right) \\
	&= \frac{1}{2}w_{2}\left( z_{0}^{m - 1},v_{0}^{j - 1} \middle| v_{j} \right) \\
	& \cdot \sum_{y_{\mathcal{J}} = z_{0}^{m - 1},\ u_{0}^{i} \in \mathbb{B}^{i + 1}}^{}{\delta\left( \left\lbrack \rho\left( u_{0}^{i - 1} \right),d\left( u_{0}^{i} \right) \right\rbrack - v_{0}^{j} \right)w\left( y_{\overline{\mathcal{J}}},u_{0}^{i - 1} \right)} \\
	&= \frac{1}{2}w_{2}\left( z_{0}^{m - 1},v_{0}^{j - 1} \middle| v_{j} \right) \\
	& \cdot \sum_{y_{\mathcal{J}} = z_{0}^{m - 1},\ u_{0}^{i} \in \mathbb{B}^{i + 1}}^{}\delta\left( \rho\left( u_{0}^{i - 1} \right) - v_{0}^{j - 1} \right) \\
	& \cdot \delta\left( u_{i} - d^{- 1}\left( v_{j};u_{0}^{i - 1} \right) \right)w\left( y_{\overline{\mathcal{J}}},u_{0}^{i - 1} \right) \\
	&= \frac{1}{2}w_{2}\left( z_{0}^{m - 1},v_{0}^{j - 1} \middle| v_{j} \right) \\
	& \cdot \sum_{y_{\overline{\mathcal{J}}} \in \mathcal{Y}^{n - m},u_{0}^{i - 1} \in \mathbb{B}^{i}}^{}{\delta\left( \rho\left( u_{0}^{i - 1} \right) - v_{0}^{j - 1} \right)w\left( y_{\overline{\mathcal{J}}},u_{0}^{i - 1} \right)} \\
	&= \frac{1}{2}w_{2}\left( z_{0}^{m - 1},v_{0}^{j - 1} \middle| v_{j} \right) = q_{Z_{0}^{l - 1}V_{0}^{j}}\left( z_{0}^{m - 1},v_{0}^{j} \right)
\end{align*}

So
\begin{align*}
	I\left( w_{1} \right) &= \mathbb{E}_{p}\left\lbrack \eta_{2}\left( Z_{0}^{l - 1},V_{0}^{j - 1};V_{j} \right) \right\rbrack \\
	&= \mathbb{E}_{q}\left\lbrack \eta_{2}\left( Z_{0}^{l - 1},V_{0}^{j - 1};V_{j} \right) \right\rbrack = I\left( w_{2} \right)
\end{align*}

\emph{Lemma 2:} Every bijection \(d:\mathbb{B \mapsto B}\) satisfies
\(\forall u \in \mathbb{B,}d(u) = u + d_{0}\), where \(d_{0} := d(0)\).

\emph{Proof:} \(d\) is an injection to \(u\), so
\(d(u + 1) = d(u) + 1\), and \(d(u + 0) = d(u) + 0\), namely
\(\forall v \in \mathbb{B,}d(u + v) = d(u) + v\). Therefore,
\(d(u) = d(0 + u) = d(0) + u = u + d_{0}\).

From Lemma 2 we immediately get Corollary 1 and 2:

\emph{Corollary 1:} For any function \(g(u_{0}^{k - 1})\):
\(\mathbb{B}^{k}\mathbb{\mapsto B}\), \(k \in \mathbb{N}^{*}\) which is
bijective to \(u_{k - 1}\), it holds that
\(\forall u_{0}^{k - 1} \in \mathbb{B}^{k},\ g\left( u_{0}^{k - 1} \right) = u_{k - 1} + g_{0}\left( u_{0}^{k - 2} \right)\),
where
\(g_{0}\left( u_{0}^{k - 2} \right) := g\left( \left\lbrack u_{0}^{k - 2},0 \right\rbrack \right)\).

\emph{Proof:} For any fixed \(u_{0}^{k - 2} \in \mathbb{B}^{k - 1}\),
\(g\left( u_{0}^{k - 1} \right)\) is a bijection to \(u_{k}\), so
according to Lemma 2,
\(g\left( u_{0}^{k - 1} \right) = u_{k - 1} + g\left( \left\lbrack u_{0}^{k - 2},0 \right\rbrack \right) = u_{k - 1} + g_{0}\left( u_{0}^{k - 2} \right)\).

\emph{Corollary 2:} For any function
\(F:\mathbb{B}^{2}\mathbb{\mapsto B}\) that \(F(s,u)\) is bijective to
both \(s\) and \(u\), it holds that
\(\forall s,u \in \mathbb{B,}F(s,u) = s + u + F_{00}\), where
\(F_{00} := F(0,0)\).

\emph{Proof:} According to Lemma 2,
\(F(s,u) = F(s,0) + u = F(0,0) + s + u = s + u + F_{00}\).

Lemma 3 given below gives the relationship of likelihood probability of
each sub-channel with each initial state, i.e.,
\(W_{n}^{(j)}\left( y_{0}^{n - 1},u_{0}^{j - 1} \middle| u_{j},s \right)\)
and that with initial state 0, i.e.,
\(W_{n}^{(j)}\left( y_{0}^{n - 1},{\widetilde{u}}_{0}^{j - 1} \middle| {\widetilde{u}}_{j} \right)\),
namely
\(W_{n}^{(j)}\left( y_{0}^{n - 1},{\widetilde{u}}_{0}^{j - 1} \middle| {\widetilde{u}}_{j} \right) = W_{n}^{(j)}\left( y_{0}^{n - 1},{\widetilde{u}}_{0}^{j - 1} \middle| {\widetilde{u}}_{j},0 \right)\),
which indicating that through a transformation on \(u_{0}^{j}\) to
produce \({\widetilde{u}}_{0}^{j}\), the former equals the latter.

\emph{Lemma 3:} For conventional polar code, on any bijective FSMC \(\left( \mathcal{S},\mathbb{B},F,W \right)\), at every code length
\(n = 2^{m},m \in \mathbb{N}^{*}\), transition probability of each
sub-channel \(W_{n}^{(j)},j \in \left\{ 0,\ldots,n - 1 \right\}\)
satisfies

\begin{align*}
	&W_{n}^{(j)}\left( y_{0}^{n - 1},u_{0}^{j - 1} \middle| u_{j},s \right) \\
	&= \left\{
	\begin{aligned}
		& W_{n}^{(j)}\left( y_{0}^{n - 1} \middle| F_{0}^{- 1}F_{s}u_{j} \right) & ,j = 0 \\
		& W_{n}^{(j)}\left( y_{0}^{n - 1},\left\lbrack F_{0}^{- 1}F_{s}u_{0},u_{1}^{j - 1} \right\rbrack \middle| u_{j} \right) & ,j \in \left\{ 1,\ldots,n - 1 \right\}
	\end{aligned} \right.
\end{align*}

\emph{Proof:}

Let \(p_{m}\) be the statement:

For conventional polar code, on any bijective FSMCs
\(\left( \mathcal{S},\mathbb{B},F,W \right)\), at code length \(n = 2^{m}\),
transition probability of each sub-channel
\(W_{n}^{(j)},j \in \left\{ 0,\ldots,n - 1 \right\}\) satisfies
\begin{align*}
	&W_{n}^{(j)}\left( y_{0}^{n - 1},u_{0}^{j - 1} \middle| u_{j},s \right) \\
	&= \left\{
	\begin{aligned}
		& W_{n}^{(j)}\left( y_{0}^{n - 1} \middle| F_{0}^{- 1}F_{s}u_{j} \right) & ,j = 0 \\
		& W_{n}^{(j)}\left( y_{0}^{n - 1},\left\lbrack F_{0}^{- 1}F_{s}u_{0},u_{1}^{j - 1} \right\rbrack \middle| u_{j} \right) & ,j \in \left\{ 1,\ldots,n - 1 \right\}
	\end{aligned} \right.
\end{align*}

We give an induction on \(m\).

\begin{enumerate}
	\def\labelenumi{\arabic{enumi})}
	\item
	For \(m = 1\), \(n = 2\), the synthesized channel is \(W_{2}\)
\end{enumerate}
\begin{align*}
	&W_{2}^{(1)}\left( y_{0}^{1},u_{0} \middle| u_{1},s \right) \\
	&= \frac{1}{2}W\left( y_{0} \middle| F_{s}\left( u_{0} + u_{1} \right) \right)W\left( y_{1} \middle| F_{s}^{2}\left\lbrack u_{0} + u_{1},u_{1} \right\rbrack \right) \\
	&\overset{Co. 2}{=}\frac{1}{2}W\left( y_{0} \middle| s + u_{0} + u_{1} + F_{00} \right)W\left( y_{1} \middle| s + u_{0} \right) \\
	&= W_{2}^{(1)}\left( y_{0}^{1},F_{0}^{- 1}F_{s}u_{0} \middle| u_{1} \right)
\end{align*}
\begin{align*}
	&W_{2}^{(0)}\left( y_{0}^{1} \middle| u_{0},s \right) = \sum_{u_{1}}^{}{W_{2}^{(1)}\left( y_{0}^{1},u_{0} \middle| u_{1},s \right)} \\
	&= \sum_{u_{1}}^{}{W_{2}^{(1)}\left( y_{0}^{1},F_{0}^{- 1}F_{s}u_{0} \middle| u_{1} \right)} = W_{2}^{(0)}\left( y_{0}^{1} \middle| F_{0}^{- 1}F_{s}u_{0} \right)
\end{align*}

So \(p_{1}\) is true.

\begin{enumerate}
	\def\labelenumi{\arabic{enumi})}
	\setcounter{enumi}{1}
	\item
	For any \(m \geq 1\), assume that \(p_{m}\) is true, then for
	\(p_{m + 1}\):
\end{enumerate}

For each sub-channel pair, i.e.,
\(\left\lbrack W_{2n}^{(2i)},W_{2n}^{(2i + 1)} \right\rbrack,\ i \in \left\{ 0,1,\ldots,n - 1 \right\}\)

\(F(s + u)\overset{Co. 2}{=}s + u + F_{00}\), so
\(\forall a_{0}^{n - 1} \in \mathbb{B}^{n},s \in \mathbb{B,}F_{s}^{n}a_{0}^{n - 1} = s + \sum_{i = 0}^{n - 1}\left( a_{i} + F_{00} \right) = s + \sum_{i = 0}^{n - 1}a_{i} + nF_{00} = s + a_{0}^{n - 1}\boldsymbol{1} + nF_{00}\).
\(F_{s}^{n}\left( \left\lbrack u_{0} + u_{1},v_{0,1}^{n - 1} \right\rbrack G_{n} \right) = s + \left\lbrack u_{0} + u_{1},v_{0,1}^{n - 1} \right\rbrack G_{n}\boldsymbol{1 +}nF_{00}\),
notice \(G_{n}\boldsymbol{1 =}\delta\),
\(F_{s}^{n}\left( \left\lbrack u_{0} + u_{1},v_{0,1}^{n - 1} \right\rbrack G_{n} \right) = s + u_{0} + u_{1}\).
Here \(G_{n}\) refers to the generator matrix of conventional polar code
at code length \(n\).

Namely
\begin{align*}
	&\forall i \in \left\{ 0,\ldots,n - 1 \right\},\ W_{n}^{(i)}\left( y_{0}^{n - 1},v_{0,0}^{i - 1},t \middle| u_{2i} + u_{2i + 1},s \right) \\
	&= \left\{
	\begin{aligned}
		& W_{n}^{(i)}\left( y_{0}^{n - 1},v_{0,0}^{i - 1} \middle| u_{2i} + u_{2i + 1},s \right) & ,t = s + u_{0} + u_{1} \\
		& 0 & ,t \neq s + u_{0} + u_{1}
	\end{aligned} \right.
\end{align*}

So we have:

For the first sub-channel pair
\(\left\lbrack W_{2n}^{(0)},W_{2n}^{(1)} \right\rbrack\)
\begin{align*}
	&W_{2n}^{(1)}\left( y_{0}^{2n - 1},u_{0} \middle| u_{1},s \right) \\
	&= \frac{1}{2}\sum_{t \in \mathcal{S}}^{}{W_{n}^{(0)}\left( y_{0}^{n - 1},t \middle| u_{0} + u_{1},s \right)W_{n}^{(0)}\left( y_{n}^{2n - 1} \middle| u_{1},t \right)} \\
	&= \frac{1}{2}W_{n}^{(0)}\left( y_{0}^{n - 1} \middle| u_{0} + u_{1},s \right)W_{n}^{(0)}\left( y_{n}^{2n - 1} \middle| u_{1},s + u_{0} + u_{1} \right)
\end{align*}

According to the hypothesis that \(p_{m}\) is true, indicating
\(W_{n}^{(0)}\left( y_{0}^{n - 1} \middle| u_{0} + u_{1},s \right) = W_{n}^{(0)}\left( y_{0}^{n - 1} \middle| F_{0}^{- 1}F_{s}\left( u_{0} + u_{1} \right) \right)\),
\(W_{n}^{(0)}\left( y_{n}^{2n - 1} \middle| u_{1},s + u_{0} + u_{1} \right) = W_{n}^{(0)}\left( y_{n}^{2n - 1} \middle| F_{0}^{- 1}F_{s + u_{0} + u_{1}}u_{1} \right)\),
we know
\begin{align*}
	&W_{2n}^{(1)}\left( y_{0}^{2n - 1},u_{0} \middle| u_{1},s \right) \\
	&= \frac{1}{2}W_{n}^{(0)}\left( y_{0}^{n - 1} \middle| F_{0}^{- 1}F_{s}\left( u_{0} + u_{1} \right) \right) \\
	&\cdot W_{n}^{(0)}\left( y_{n}^{2n - 1} \middle| F_{0}^{- 1}F_{s + u_{0} + u_{1}}u_{1} \right) \\
	&\overset{Co. 2}{=}\frac{1}{2}W_{n}^{(0)}\left( y_{0}^{n - 1} \middle| s + u_{0} + u_{1} \right)W_{n}^{(0)}\left( y_{n}^{2n - 1} \middle| s + u_{0} \right) \\
	&= W_{2n}^{(2i + 1)}\left( y_{0}^{2n - 1},F_{0}^{- 1}F_{s}u_{0} \middle| u_{1} \right)
\end{align*}

and
\begin{align*}
	&W_{2n}^{(0)}\left( y_{0}^{2n - 1} \middle| u_{0},s \right) = \sum_{u_{1}}^{}{W_{2n}^{(1)}\left( y_{0}^{2n - 1},u_{0} \middle| u_{1},s \right)} \\
	&= \sum_{u_{1}}^{}{W_{2n}^{(2i + 1)}\left( y_{0}^{2n - 1},F_{0}^{- 1}F_{s}u_{0} \middle| u_{1} \right)} \\
	&= W_{2n}^{(1)}\left( y_{0}^{2n - 1} \middle| F_{0}^{- 1}F_{s}u_{0} \right)
\end{align*}

For non-first sub-channels
\(\left\lbrack W_{2n}^{(2i)},W_{2n}^{(2i + 1)} \right\rbrack,i \in \left\{ 1,\ldots,n - 1 \right\}\),
there is
\begin{align*}
	&W_{2n}^{(2i + 1)}\left( y_{0}^{2n - 1},u_{0}^{2i} \middle| u_{2i + 1},s \right) \\
	&= \frac{1}{2}\sum_{t \in \mathcal{S}}^{}W_{n}^{(i)}\left( y_{0}^{n - 1},v_{0,0}^{i - 1},t \middle| u_{2i} + u_{2i + 1},s \right) \\
	&\cdot W_{n}^{(i)}\left( y_{n}^{2n - 1},v_{1,0}^{i - 1} \middle| u_{2i + 1},t \right) \\
	&= \frac{1}{2}W_{n}^{(i)}\left( y_{0}^{n - 1},v_{0,0}^{i - 1} \middle| u_{2i} + u_{2i + 1},s \right) \\
	&\cdot W_{n}^{(i)}\left( y_{n}^{2n - 1},v_{1,0}^{i - 1} \middle| u_{2i + 1},s + u_{0} + u_{1} \right)
\end{align*}
where
\(v_{0,0}^{i - 1} = \left\lbrack u_{2j} + u_{2j + 1} \right\rbrack_{j \in \left\{ 0,1,\ldots,i - 1 \right\}},v_{1,0}^{i - 1} = \left\lbrack u_{2j + 1} \right\rbrack_{j \in \left\{ 0,1,\ldots,i - 1 \right\}}\).
According to the hypothesis that \(p_{m}\) is true, indicating
\(W_{n}^{(i)}\left( y_{0}^{n - 1},v_{0,0}^{i - 1} \middle| u_{2i} + u_{2i + 1},s \right) = W_{n}^{(i)}\left( y_{0}^{n - 1},\left\lbrack F_{0}^{- 1}F_{s}v_{0,0},\ v_{0,1}^{i - 1} \right\rbrack \middle| u_{2i} + u_{2i + 1} \right)\),
\(W_{n}^{(i)}\left( y_{n}^{2n - 1},v_{1,0}^{i - 1} \middle| u_{2i + 1},s + u_{0} + u_{1} \right) = W_{n}^{(i)}\left( y_{n}^{2n - 1},\left\lbrack F_{0}^{- 1}F_{s + u_{0} + u_{1}}v_{0,1},\ v_{0,1}^{i - 1} \right\rbrack \middle| u_{2i + 1} \right)\),
we know
\begin{align*}
	&W_{2n}^{(2i + 1)}\left( y_{0}^{2n - 1},u_{0}^{2i} \middle| u_{2i + 1},s \right) \\
	&= \frac{1}{2}W_{n}^{(i)}\left( y_{0}^{n - 1},\left\lbrack F_{0}^{- 1}F_{s}v_{0,0},\ v_{0,1}^{i - 1} \right\rbrack \middle| u_{2i} + u_{2i + 1} \right) \\
	&\cdot W_{n}^{(i)}\left( y_{n}^{2n - 1},\left\lbrack F_{0}^{- 1}F_{s + u_{0} + u_{1}}v_{0,1},\ v_{0,1}^{i - 1} \right\rbrack \middle| u_{2i + 1} \right) \\
	&\overset{Co. 2}{=}\frac{1}{2}W_{n}^{(i)}\left( y_{0}^{n - 1},\left\lbrack s + u_{0} + u_{1},\ v_{0,1}^{i - 1} \right\rbrack \middle| u_{2i} + u_{2i + 1} \right) \\
	&\cdot W_{n}^{(i)}\left( y_{n}^{2n - 1},\left\lbrack s + u_{0},\ v_{0,1}^{i - 1} \right\rbrack \middle| u_{2i + 1} \right) \\
	&= W_{2n}^{(2i + 1)}\left( y_{0}^{2n - 1},\left\lbrack s + u_{0},u_{1}^{2i} \right\rbrack \middle| u_{2i + 1} \right) \\
	&= W_{2n}^{(2i + 1)}\left( y_{0}^{2n - 1},\left\lbrack F_{0}^{- 1}F_{s}u_{0},u_{1}^{2i} \right\rbrack \middle| u_{2i + 1} \right)
\end{align*}

Similarly,
\(W_{2n}^{(2i)}\left( y_{0}^{2n - 1},u_{0}^{2i - 1} \middle| u_{2i},s \right) = W_{2n}^{(2i)}\left( y_{0}^{2n - 1},\left\lbrack F_{0}^{- 1}F_{s}u_{0},u_{1}^{2i - 1} \right\rbrack \middle| u_{2i} \right)\).

Namely, \(p_{m + 1}\) is true.

With 1) and 2) we know \(p_{m}\) is true for any
\(m \in \mathbb{N}^{*}\). Proof of Lemma 1 is finished.

\emph{Lemma 4:} On any bijective FSMCs
\(\left( \mathcal{S},\mathbb{B},F,W \right)\), at any code length
\(n \in \left\{ 2^{m} \middle| m \in \mathbb{N}^{*} \right\}\), each
sub-channel \(W_{n}^{(i)},i \in \{ 0,1,\ldots,n - 1\}\) and
\({\widetilde{W}}_{\left\langle i \right\rangle}^{\left( i^{*} \right)}\)
viz. the \(i^{*}\)-th sub-channel of conventional polar code on
\(W^{\left\langle i \right\rangle}\) satisfy A-condition, i.e.,
\(W_{n}^{(i)}\overset{A}{\sim}{\widetilde{W}}_{\left\langle i \right\rangle}^{\left( i^{*} \right)}\).

\emph{Proof:}

Let \(p_{m}\) be the statement ``On any bijective FSMCs
\(\left( \mathcal{S},\mathbb{B},F,W \right)\), at code length \(n = 2^{m}\), each
sub-channel \(W_{n}^{(i)},i \in \{ 0,1,\ldots,n - 1\}\) and
\({\widetilde{W}}_{\left\langle i \right\rangle}^{\left( i^{*} \right)}\)
viz. the \(i^{*}\)-th sub-channel of conventional polar code on
\(W^{\left\langle i \right\rangle}\) satisfy A-condition, i.e.,
\(W_{n}^{(i)}\overset{A}{\sim}{\widetilde{W}}_{\left\langle i \right\rangle}^{\left( i^{*} \right)}\)''.
We give an induction on \(m\).

\begin{enumerate}
	\def\labelenumi{\arabic{enumi})}
	\item
	For \(m = 1\), \(n = 2\), the synthesized channel is \(W_{2}\):
\end{enumerate}
\begin{align*}
	W_{2}^{(1)}\left( y_{0}^{1},u_{0} \middle| u_{1} \right) &= \frac{1}{2}W\left( y_{0} \middle| u_{0} + u_{1} \right)W\left( y_{1} \middle| u_{0} \right) \\
	&= W\left( y_{0} \middle| u_{0} + u_{1} \right)  \frac{1}{2}W\left( y_{1} \middle| u_{0} \right)
\end{align*}

Denote
\(j = 0,d\left( u_{0}^{1} \right) = u_{0} + u_{1},w(y_{\overline{j}},u_{0}) = \frac{1}{2}W\left( y_{\overline{j}} \middle| u_{0} \right)\),
then
\(W_{2}^{(1)}\left( y_{0}^{1},u_{0} \middle| u_{1} \right) = W\left( y_{j} \middle| d\left( u_{0}^{1} \right) \right)w(y_{\overline{j}},u_{0})\), trivially \(d(u_0^1)\) is bijective to \(u_1\),
and
\(\sum_{y_{\overline{j}}\mathcal{\in Y,}u_{0}\mathbb{\in B}}^{}{w(y_{\overline{j}},u_{0})} = \sum_{y_{\overline{j}}\mathcal{\in Y,}u_{0}\mathbb{\in B}}^{}{\frac{1}{2}W\left( y_{\overline{j}} \middle| u_{0} \right)} = 1\),
so according to Definition 5,
\(W_{2}^{(1)}\overset{A}{\sim}W = {\widetilde{W}}_{1}^{(0)}\);
similarly,
\(W_{2}^{(0)}\overset{A}{\sim}W = {\widetilde{W}}_{1}^{(0)}\).

\begin{enumerate}
	\def\labelenumi{\arabic{enumi})}
	\setcounter{enumi}{1}
	\item
	For any \(m \geq 1\), assume that \(p_{m}\) is true, then for
	\(p_{m + 1}\):
\end{enumerate}

For the first sub-channel pair
\(\left\lbrack W_{2n}^{(0)},W_{2n}^{(1)} \right\rbrack\)
\begin{align*}
	&W_{2n}^{(1)}\left( y_{0}^{2n - 1},u_{0} \middle| u_{1} \right) \\
	&= \frac{1}{2}\sum_{t \in \mathcal{S}}^{}{W_{n}^{(0)}\left( y_{0}^{n - 1},t \middle| u_{0} + u_{1} \right)W_{n}^{(0)}\left( y_{n}^{2n - 1} \middle| u_{1},t \right)} \\
	&= \frac{1}{2}W_{n}^{(0)}\left( y_{0}^{n - 1} \middle| u_{0} + u_{1} \right)W_{n}^{(0)}\left( y_{n}^{2n - 1} \middle| u_{1},u_{0} + u_{1} \right) \\
	&\overset{Le. 3}{=}\frac{1}{2}W_{n}^{(0)}\left( y_{0}^{n - 1} \middle| u_{0} + u_{1} \right)W_{n}^{(0)}\left( y_{n}^{2n - 1} \middle| F_{0}^{- 1}F_{u_{0} + u_{1}}u_{1} \right) \\
	&\overset{Co. 2}{=}\frac{1}{2}W_{n}^{(0)}\left( y_{0}^{n - 1} \middle| u_{0} + u_{1} \right)W_{n}^{(0)}\left( y_{n}^{2n - 1} \middle| u_{0} \right)
\end{align*}

According to the hypothesis that \(p_{m}\) is true, indicating
\(W_{n}^{(0)}\overset{A}{\sim}{\widetilde{W}}_{\left\langle 0 \right\rangle}^{\left( 0^{*} \right)} = {\widetilde{W}}_{1}^{(0)} = W\),
we know there exists \(j,d\) that
\[W_{n}^{(0)}\left( y_{0}^{n - 1} \middle| u_{0} + u_{1} \right) = W\left( y_{j} \middle| d\left( u_{0} + u_{1} \right) \right)w\left( y_{\overline{j}} \right)\]
\[W_{n}^{(0)}\left( y_{n}^{2n - 1} \middle| u_{0} \right) = W\left( y_{n + j} \middle| d\left( u_{0} \right) \right)w\left( y_{\overline{j} + n} \right)\]
so
\begin{align*}
	&W_{2n}^{(1)}\left( y_{0}^{2n - 1},u_{0} \middle| u_{1} \right) \\
	&= \frac{1}{2}W\left( y_{j} \middle| d\left( u_{0} + u_{1} \right) \right)w\left( y_{\overline{j}} \right)W\left( y_{n + j} \middle| d\left( u_{0} \right) \right)w\left( y_{\overline{j} + n} \right) \\
	&= W\left( y_{j} \middle| d\left( u_{0} + u_{1} \right) \right)  \frac{1}{2}W\left( y_{n + j} \middle| d\left( u_{0} \right) \right)w\left( y_{\overline{j}} \right)w\left( y_{\overline{j} + n} \right)
\end{align*}

Denote \(d_1(u_0^1)=d(u_0+u_1)\), and
\(w^{\prime}\left( y_{\overline{j}},u_{0} \right) = \frac{1}{2}W\left( y_{n + j} \middle| d\left( u_{0} \right) \right)w\left( y_{\overline{j}} \right)w\left( y_{\overline{j} + n} \right)\),
we have
\[W_{2n}^{(1)}\left( y_{0}^{2n - 1},u_{0} \middle| u_{1} \right) = W\left( y_{j} \middle| d_1(u_0^1) \right)w^{\prime}\left( y_{\overline{j}},u_{0} \right)\]

Trivially, \(d_1(u_0^1)\) is bijective to \(u_1\), and
\begin{align*}
	&\sum_{y_{\overline{j}} \in \mathcal{Y}^{2n - 1},u_{0}\mathbb{\in B}}^{}{w^{\prime}\left( y_{\overline{j}},u_{0} \right)} \\
	&= \sum_{y_{\overline{j}} \in \mathcal{Y}^{2n - 1},u_{0}\mathbb{\in B}}^{}{\frac{1}{2}W\left( y_{n + j} \middle| d\left( u_{0} \right) \right)w\left( y_{\overline{j}} \right)w\left( y_{\overline{j} + n} \right)} \\
	&= \sum_{y_{n + j}\mathcal{\in Y,}u_{0}\mathbb{\in B}}^{}\frac{1}{2}W\left( y_{n + j} \middle| d\left( u_{0} \right) \right) \\
	&\cdot \sum_{y_{\overline{j}} \in \mathcal{Y}^{n - 1}}^{}w\left( y_{\overline{j}} \right)\sum_{y_{\overline{j} + n} \in \mathcal{Y}^{n - 1}}^{}w\left( y_{\overline{j} + n} \right) \\
	&= 1
\end{align*}
so according to Definition 5,
\(W_{2n}^{(1)}\overset{A}{\sim}W = {\widetilde{W}}_{1}^{(0)} = {\widetilde{W}}_{\left\langle 0 \right\rangle}^{\left( 0^{*} \right)}\);
similarly,
\(W_{2n}^{(0)}\overset{A}{\sim}W = {\widetilde{W}}_{1}^{(0)} = {\widetilde{W}}_{\left\langle 0 \right\rangle}^{\left( 0^{*} \right)}\).

For non-first sub-channels
\(\left\lbrack W_{2n}^{(2i)},W_{2n}^{(2i + 1)} \right\rbrack,i \in \left\{ 1,\ldots,n - 1 \right\}\),
there is
\begin{align*}
	&W_{2n}^{(2i + 1)}\left( y_{0}^{2n - 1},u_{0}^{2i} \middle| u_{2i + 1} \right) \\
	&= \frac{1}{2}\sum_{t \in \mathcal{S}}^{}W_{n}^{(i)}\left( y_{0}^{n - 1},v_{0,0}^{i - 1},t \middle| u_{2i} + u_{2i + 1} \right)\\
	&\cdot W_{n}^{(i)}\left( y_{n}^{2n - 1},v_{1,0}^{i - 1} \middle| u_{2i + 1},t \right) \\
	&= \frac{1}{2}W_{n}^{(i)}\left( y_{0}^{n - 1},v_{0,0}^{i - 1} \middle| u_{2i} + u_{2i + 1} \right)\\
	&\cdot W_{n}^{(i)}\left( y_{n}^{2n - 1},v_{1,0}^{i - 1} \middle| u_{2i + 1},u_{0} + u_{1} \right)\\
	&\overset{Le. 3}{=}\frac{1}{2}W_{n}^{(i)}\left( y_{0}^{n - 1},v_{0,0}^{i - 1} \middle| u_{2i} + u_{2i + 1} \right)\\
	&\cdot W_{n}^{(i)}\left( y_{n}^{2n - 1},\left\lbrack F_{0}^{- 1}F_{u_{0} + u_{1}}v_{1,\ 0},v_{1,1}^{i - 1} \right\rbrack \middle| u_{2i + 1} \right)\\
	&\overset{Co. 2}{=}\frac{1}{2}W_{n}^{(i)}\left( y_{0}^{n - 1},v_{0,0}^{i - 1} \middle| u_{2i} + u_{2i + 1} \right)\\
	&\cdot W_{n}^{(i)}\left( y_{n}^{2n - 1},\left\lbrack u_{0},v_{1,1}^{i - 1} \right\rbrack \middle| u_{2i + 1} \right)
\end{align*}
where
\(v_{0,0}^{i - 1} = \left\lbrack u_{2j} + u_{2j + 1} \right\rbrack_{j \in \left\{ 0,1,\ldots,i - 1 \right\}},v_{1,0}^{i - 1} = \left\lbrack u_{2j + 1} \right\rbrack_{j \in \left\{ 0,1,\ldots,i - 1 \right\}}\).

According to the hypothesis that \(p_{m}\) is true, indicating
\(W_{n}^{(i)}\overset{A}{\sim}{\widetilde{W}}_{\left\langle i \right\rangle}^{\left( i^{*} \right)}\),
we know there exists \(\mathcal{J,}\rho,d\) that
\begin{align*}
	&W_{n}^{(i)}\left( y_{0}^{n - 1},v_{0,0}^{i - 1} \middle| u_{2i} + u_{2i + 1} \right) \\
	&= {\widetilde{W}}_{\left\langle i \right\rangle}^{\left( i^{*} \right)}\left( y_{\mathcal{J}},\rho\left( v_{0,0}^{i - 1} \right) \middle| d\left( \left\lbrack v_{0,0}^{i - 1},u_{2i} + u_{2i + 1} \right\rbrack \right) \right)w\left( y_{\overline{\mathcal{J}}},v_{0,0}^{i - 1} \right)
\end{align*}
\begin{align*}
	&W_{n}^{(i)}\left( y_{n}^{2n - 1},\left\lbrack u_{0},v_{1,1}^{i - 1} \right\rbrack \middle| u_{2i + 1} \right) \\
	&= {\widetilde{W}}_{\left\langle i \right\rangle}^{\left( i^{*} \right)}\left( y_{\mathcal{J +}n},\rho\left( \left\lbrack u_{0},v_{1,1}^{i - 1} \right\rbrack \right) \middle| d\left( \left\lbrack u_{0},v_{1,1}^{i - 1},u_{2i + 1} \right\rbrack \right) \right)\\
	&\cdot w\left( y_{\overline{\mathcal{J}} + n},\left\lbrack u_{0},v_{1,1}^{i - 1} \right\rbrack \right)
\end{align*}
so
\begin{align*}
	&W_{2n}^{(2i + 1)}\left( y_{0}^{2n - 1},u_{0}^{2i} \middle| u_{2i + 1} \right) \\
	&= \frac{1}{2}{\widetilde{W}}_{\left\langle i \right\rangle}^{\left( i^{*} \right)}\left( y_{\mathcal{J}},\rho\left( v_{0,0}^{i - 1} \right) \middle| d\left( \left\lbrack v_{0,0}^{i - 1},u_{2i} + u_{2i + 1} \right\rbrack \right) \right)w\left( y_{\overline{\mathcal{J}}},v_{0,0}^{i - 1} \right)\\
	&\cdot {\widetilde{W}}_{\left\langle i \right\rangle}^{\left( i^{*} \right)}\left( y_{\mathcal{J +}n},\rho\left( \left\lbrack u_{0},v_{1,1}^{i - 1} \right\rbrack \right) \middle| d\left( \left\lbrack u_{0},v_{1,1}^{i - 1},u_{2i + 1} \right\rbrack \right) \right)\\
	&\cdot w\left( y_{\overline{\mathcal{J}} + n},\left\lbrack u_{0},v_{1,1}^{i - 1} \right\rbrack \right)\\
	&\overset{Co. 1}{=}\frac{1}{2}{\widetilde{W}}_{\left\langle i \right\rangle}^{\left( i^{*} \right)}\left( y_{\mathcal{J}},\rho\left( v_{0,0}^{i - 1} \right) \middle| d_{0}\left( v_{0,0}^{i - 1} \right) + u_{2i} + u_{2i + 1} \right)\\
	&\cdot {\widetilde{W}}_{\left\langle i \right\rangle}^{\left( i^{*} \right)}\left( y_{\mathcal{J +}n},\rho\left( \left\lbrack u_{0},v_{1,1}^{i - 1} \right\rbrack \right) \middle| d_{0}\left( \left\lbrack u_{0},v_{1,1}^{i - 1} \right\rbrack \right) + u_{2i + 1} \right)\\
	&\cdot w\left( y_{\overline{\mathcal{J}}},v_{0,0}^{i - 1} \right)w\left( y_{\overline{\mathcal{J}} + n},\left\lbrack u_{0},v_{1,1}^{i - 1} \right\rbrack \right)
\end{align*}

Denote
\[\mathcal{J}_{1} = \left\lbrack \mathcal{J,J +}n \right\rbrack\]
\begin{align*}
	&\rho_{1}\left( u_{0}^{2i} \right) =\left\lbrack \left\lbrack \rho\left( v_{0,0}^{i - 1} \right),\rho\left( \left\lbrack u_{0},v_{1,1}^{i - 1} \right\rbrack \right) \right\rbrack R_{2i^{*}}^{- 1}(I_{i^{*}} \otimes F), \right. \\
	&\left. d_{0}\left( v_{0,0}^{i - 1} \right) + d_{0}\left( \left\lbrack u_{0},v_{1,1}^{i - 1} \right\rbrack \right) + u_{2i} \right\rbrack
\end{align*}
\[d_{1}\left( u_{0}^{2i + 1} \right) = d\left( \left\lbrack u_{0},v_{1,1}^{i - 1},u_{2i + 1} \right\rbrack \right)\]
\[w_{1}\left( y_{\mathcal{J}_{1}},u_{0}^{2i} \right) = \ w\left( y_{\overline{\mathcal{J}}},v_{0,0}^{i - 1} \right)w\left( y_{\overline{\mathcal{J}} + n},\left\lbrack u_{0},v_{1,1}^{i - 1} \right\rbrack \right)\]
and consider it is trivial that
\(2\left\langle i \right\rangle = \left\langle 2i + 1 \right\rangle,\ \left( 2i^{*} + 1 \right) = (2i + 1)^{*}\),
we have
\begin{align*}
	&W_{2n}^{(2i + 1)}\left( y_{0}^{2n - 1},u_{0}^{2i} \middle| u_{2i + 1} \right) \\
	&= {\widetilde{W}}_{\left\langle 2i + 1 \right\rangle}^{\left( (2i + 1)^{*} \right)}\left( y_{\mathcal{J}_{1}},\rho_{1}\left( u_{0}^{2i} \right) \middle| d_{1}\left( u_{0}^{2i + 1} \right) \right)w_{1}\left( y_{{\overline{\mathcal{J}}}_{1}},u_{0}^{2i} \right)
\end{align*}

Trivially, \(d_{1}(u_{0}^{2i + 1})\) is bijective to \(u_{2i+1}\); and as with \(w_{1}\), we are to verify that for any
\(a_{0}^{2i^{*}} \in \mathbb{B}^{2i^{*} + 1}\),
\(\sum_{y_{{\overline{\mathcal{J}}}_{1}} \in \mathcal{Y}^{2l},u_{0}^{2i} \in \mathbb{B}^{2i + 1}}^{}{\delta\left( \rho_{1}\left( u_{0}^{2i} \right) - a_{0}^{j - 1} \right)w_{1}\left( y_{\mathcal{J}_{1}},u_{0}^{2i} \right)} = 1\),
where \(l = n - \left\langle i \right\rangle\):

To begin with, we perform a transformation on
\(\delta\left( \rho_{1}\left( u_{0}^{2i} \right) - a_{0}^{j - 1} \right)\):
From the definition of \(\rho_{1}\) we know
\begin{align*}
	&\rho_{1}\left( u_{0}^{2i} \right) = a_{0}^{2i^{*}} \\
	&\leftrightarrow \left\lbrack \left\lbrack \rho\left( v_{0,0}^{i - 1} \right),\rho\left( \left\lbrack u_{0},v_{1,1}^{i - 1} \right\rbrack \right) \right\rbrack R_{2i^{*}}^{- 1}(I_{i^{*}} \otimes F), \right.\\
	&\left. d_{0}\left( v_{0,0}^{i - 1} \right) + d_{0}\left( \left\lbrack u_{0},v_{1,1}^{i - 1} \right\rbrack \right) + u_{2i} \right\rbrack = a_{0}^{2i^{*}} \\
	&\leftrightarrow \rho\left( v_{0,0}^{i - 1} \right) = a_{0}^{2i^{*} - 1}\left( I_{i^{*}} \otimes F \right)R_{2i^{*}}\left\lbrack I_{i^{*}},O_{i^{*} \times i^{*}} \right\rbrack^{T} \\
	&\land \rho\left( \left\lbrack u_{0},v_{1,1}^{i - 1} \right\rbrack \right) = a_{0}^{2i^{*} - 1}\left( I_{i^{*}} \otimes F \right)R_{2i^{*}}\left\lbrack O_{i^{*} \times i^{*}},\ I_{i^{*}} \right\rbrack^{T} \\
	&\land u_{2i} = d_{0}\left( v_{0,0}^{i - 1} \right) + d_{0}\left( \left\lbrack u_{0},v_{1,1}^{i - 1} \right\rbrack \right) + a_{0}^{2i^{*}}
\end{align*}
so
\begin{align*}
	&\delta\left( \rho_{1}\left( u_{0}^{2i} \right) - a_{0}^{2i^{*}} \right) \\
	&= \delta\left( \rho\left( v_{0,0}^{i - 1} \right) - a_{0}^{2i^{*} - 1}\left( I_{i^{*}} \otimes F \right)R_{2i^{*}}\left\lbrack I_{i^{*}},O_{i^{*} \times i^{*}} \right\rbrack^{T} \right) \\
	&\cdot \delta\left( \rho\left( \left\lbrack u_{0},v_{1,1}^{i - 1} \right\rbrack \right) - a_{0}^{2i^{*} - 1}\left( I_{i^{*}} \otimes F \right)R_{2i^{*}}\left\lbrack O_{i^{*} \times i^{*}},\ I_{i^{*}} \right\rbrack^{T} \right) \\
	&\cdot \delta\left( u_{2i} - \left( d_{0}\left( v_{0,0}^{i - 1} \right) + d_{0}\left( \left\lbrack u_{0},v_{1,1}^{i - 1} \right\rbrack \right) + a_{0}^{2i^{*}} \right) \right)
\end{align*}

Denote
\(g_{1} = a_{0}^{2i^{*} - 1}\left( I_{i^{*}} \otimes F \right)R_{2i^{*}}\left\lbrack I_{i^{*}},O_{i^{*} \times i^{*}} \right\rbrack^{T},g_{2} = a_{0}^{2i^{*} - 1}\left( I_{i^{*}} \otimes F \right)R_{2i^{*}}\left\lbrack O_{i^{*} \times i^{*}},\ I_{i^{*}} \right\rbrack^{T}\),
we have
\begin{align*}
	&\sum_{y_{{\overline{\mathcal{J}}}_{1}} \in \mathcal{Y}^{2l},u_{0}^{2i} \in \mathbb{B}^{2i + 1}}^{}{\delta\left( \rho_{1}\left( u_{0}^{2i} \right) - a_{0}^{j - 1} \right)w_{1}\left( y_{\mathcal{J}_{1}},u_{0}^{2i} \right)} \\
	&= \sum_{y_{{\overline{\mathcal{J}}}_{1}} \in \mathcal{Y}^{2l},u_{0}^{2i} \in \mathbb{B}^{2i + 1}}^{}\delta\left( \rho\left( v_{0,0}^{i - 1} \right) - g_{1} \right)\delta\left( \rho\left( \left\lbrack u_{0},v_{1,1}^{i - 1} \right\rbrack \right) - g_{2} \right) \\
	&\cdot \delta\left( u_{2i} - \left( d_{0}\left( v_{0,0}^{i - 1} \right) + d_{0}\left( \left\lbrack u_{0},v_{1,1}^{i - 1} \right\rbrack \right) + a_{0}^{2i^{*}} \right) \right)w_{1}\left( y_{\mathcal{J}_{1}},u_{0}^{2i} \right) \\
	&= \sum_{y_{{\overline{\mathcal{J}}}_{1}} \in \mathcal{Y}^{2l},u_{0}^{2i} \in \mathbb{B}^{2i + 1}}^{}\delta\left( \rho\left( v_{0,0}^{i - 1} \right) - g_{1} \right)\delta\left( \rho\left( \left\lbrack u_{0},v_{1,1}^{i - 1} \right\rbrack \right) - g_{2} \right) \\
	&\cdot \delta\left( u_{2i} - \left( d_{0}\left( v_{0,0}^{i - 1} \right) + d_{0}\left( \left\lbrack u_{0},v_{1,1}^{i - 1} \right\rbrack \right) + a_{0}^{2i^{*}} \right) \right)w\left( y_{\overline{\mathcal{J}}},v_{0,0}^{i - 1} \right) \\
	&\cdot w\left( y_{\overline{\mathcal{J}} + n},\left\lbrack u_{0},v_{1,1}^{i - 1} \right\rbrack \right)
\end{align*}

Now we perform a change of variables: denote
\(b_{0}^{i - 1} = v_{0,0}^{i - 1},c_{0}^{i - 1} = \left\lbrack u_{0},v_{1,1}^{i - 1} \right\rbrack\),
it is trivial that
\(\left\lbrack b_{0}^{i - 1},c_{0}^{i - 1},u_{2i} \right\rbrack\) is
one-to-one mapped to \(u_{0}^{2i}\), so change \(u_{0}^{2i}\) to
\(\left\lbrack b_{0}^{i - 1},c_{0}^{i - 1},u_{2i} \right\rbrack\), we
have
\begin{align*}
	&\sum_{y_{{\overline{\mathcal{J}}}_{1}} \in \mathcal{Y}^{2l},u_{0}^{2i} \in \mathbb{B}^{2i + 1}}^{}{\delta\left( \rho_{1}\left( u_{0}^{2i} \right) - a_{0}^{j - 1} \right)w_{1}\left( y_{\mathcal{J}_{1}},u_{0}^{2i} \right)} \\
	&= \sum_{\left\lbrack y_{\overline{\mathcal{J}}},y_{\overline{\mathcal{J}} + n} \right\rbrack \in \mathcal{Y}^{2l},\left\lbrack b_{0}^{i - 1},c_{0}^{i - 1},u_{2i} \right\rbrack \in \mathbb{B}^{2i + 1}}^{}\delta\left( \rho\left( b_{0}^{i - 1} \right) - g_{1} \right) \\
	&\cdot \delta\left( \rho\left( c_{0}^{i - 1} \right) - g_{2} \right)\delta\left( u_{2i} - \left( d_{0}\left( b_{0}^{i - 1} \right) + d_{0}\left( c_{0}^{i - 1} \right) + a_{0}^{2i^{*}} \right) \right) \\
	&\cdot w\left( y_{\overline{\mathcal{J}}},b_{0}^{i - 1} \right)w\left( y_{\overline{\mathcal{J}} + n},c_{0}^{i - 1} \right) \\
	&= \sum_{y_{\overline{\mathcal{J}}} \in \mathcal{Y}^{l},b_{0}^{i - 1} \in \mathbb{B}^{i}}^{}\delta\left( \rho\left( b_{0}^{i - 1} \right) - g_{1} \right)w\left( y_{\overline{\mathcal{J}}},b_{0}^{i - 1} \right) \\
	&\cdot \sum_{y_{\overline{\mathcal{J}} + n} \in \mathcal{Y}^{l},c_{0}^{i - 1} \in \mathbb{B}^{i}}^{}\delta\left( \rho\left( c_{0}^{i - 1} \right) - g_{2} \right)w\left( y_{\overline{\mathcal{J}} + n},c_{0}^{i - 1} \right) \\
	&\cdot \sum_{u_{2i}\mathbb{\in B}}^{}{\delta\left( u_{2i} - \left( d_{0}\left( b_{0}^{i - 1} \right) + d_{0}\left( c_{0}^{i - 1} \right) + a_{0}^{2i^{*}} \right) \right)} \\
	&= 1
\end{align*}
briefly
\begin{align*}
	\sum_{y_{{\overline{\mathcal{J}}}_{1}} \in \mathcal{Y}^{2l},u_{0}^{2i} \in \mathbb{B}^{2i + 1}}^{}{\delta\left( \rho_{1}\left( u_{0}^{2i} \right) - a_{0}^{j - 1} \right)w_{1}\left( y_{\mathcal{J}_{1}},u_{0}^{2i} \right)} = 1
\end{align*}

So according to Definition 5,
\(W_{2n}^{(2i + 1)}\overset{A}{\sim}{\widetilde{W}}_{\left\langle 2i + 1 \right\rangle}^{\left( (2i + 1)^{*} \right)}\);
similarly,
\(W_{2n}^{(2i)}\overset{A}{\sim}{\widetilde{W}}_{\left\langle 2i \right\rangle}^{\left( (2i)^{*} \right)}\).

Namely, \(p_{m + 1}\) is true.

With 1) and 2) we know \(p_{m}\) is true for any
\(m \in \mathbb{N}^{*}\). Proof of Lemma 4 is finished.

\emph{Proof to Proposition 2:}

According to Lemma 4, on any bijective FSMCs
\(\left( \mathcal{S},\mathbb{B},F,W \right)\), at any code length
\(n \in \left\{ 2^{m} \middle| m \in \mathbb{N}^{*} \right\}\), each
sub-channel \(W_{n}^{(i)},i \in \{ 0,1,\ldots,n - 1\}\) and
\({\widetilde{W}}_{\left\langle i \right\rangle}^{\left( i^{*} \right)}\)
viz. the \(i^{*}\)-th sub-channel of conventional polar code on
\(W^{\left\langle i \right\rangle}\) satisfy A-condition, i.e.,
\(W_{n}^{(i)}\overset{A}{\sim}{\widetilde{W}}_{\left\langle i \right\rangle}^{\left( i^{*} \right)}\).
With Lemma 1 indicating that channels satisfying A-condition have
identical capacity, we know
\(I\left( W_{n}^{(i)} \right) = I\left( {\widetilde{W}}_{\left\langle i \right\rangle}^{\left( i^{*} \right)} \right)\),
namely
\(I_{\left( \mathcal{P}_{c},\mathcal{F}_{b} \right),\ n}^{(i)} = I_{\left( \mathcal{P}_{c},W^{\left\langle i \right\rangle} \right),\ \left\langle i \right\rangle}^{\left( i^{*} \right)}\),
which illustrates Proposition 2 is true. Proof to Proposition 2 is
finished.

\subsection{Proof of Theorem 1}

To prove Theorem 1, we firstly put up with a condition about two
channels denoted as \emph{B-condition} in Definition 6, and prove it
acts as a sufficient condition for identical capacity in Lemma 5. Then
we infer a conclusion in Lemma 6 to prove that B-condition is satisfied
between sub-channels of the proposed polar code on bijective FSMC and those of conventional polar code on memoryless channels, in
Lemma 7. Finally with major premise Lemma 5 and minor premise Lemma 7,
we finish the proof to Theorem 1.

And for convenience we define:
\[W_{n}^{(i)} := W_{\left( \mathcal{P}_{s},\mathcal{F}_{b} \right),\ n}^{(i)},{\widetilde{W}}_{n}^{(i)} := W_{\left( \mathcal{P}_{c},W^{n} \right),\ n}^{(i)}\]
\[I_{n}^{(i)} := I\left( W_{n}^{(i)} \right),{\widetilde{I}}_{n}^{(i)} := I\left( {\widetilde{W}}_{n}^{(i)} \right)\]
\[F_{s}^{- 1}t := the\ u\ that\ F_{s}u = t\]
\[F_{s}^{n}u_{0}^{n - 1} := \left\{ \begin{aligned}
	& F_{s}u_{0} & ,n = 1 \\
	& F_{F_{s}^{n - 1}u_{0}^{n - 2}}u_{n - 1} & ,n \geq 2
\end{aligned} \right.\ \]
\[{\widehat{G}}_{n} := {\overline{G}}_{n}(0\ldots n - 2,\ 1\ldots n - 1)\]
where \(F_{s}u\) means \(F(s,u)\), just as Section III denotes; and
\({\overline{G}}_{n}\) refers to the generator matrix of last-pair swapping
polar code at code length \(n\), just as Section IV-B denotes.

\emph{Definition 6:} B-condition. For any channel with likelihood
probability distribution
\(w_{1}\left( y_{0}^{n - 1},u_{0}^{i - 1} \middle| u_{i} \right),w_{2}\left( y_{0}^{n - 1},u_{0}^{i - 1} \middle| u_{i} \right),y_{0}^{n - 1} \in \mathcal{Y}^{n},u_{0}^{i} \in \mathbb{B}^{i + 1},n \in \mathbb{N}^{*},i\mathbb{\in N}\),
\(\mathcal{Y}\) is any set, iff there exist a sequence \(\mathcal{J}\)
that
\(\left\{ \mathcal{J} \right\} = \left\{ 0,\ldots,n - 1 \right\},\left| \mathcal{J} \right| = n\),
a bijection \(\rho:\mathbb{B}^{i} \mapsto \mathbb{B}^{i}\) and a mapping
\(d\left( u_{0}^{i} \right):\mathbb{B}^{i}\mathbb{\mapsto B}\) bijective
to \(u_{i}\), such that
\[w_{1}\left( y_{0}^{n - 1},u_{0}^{i - 1} \middle| u_{i} \right) = w_{2}\left( y_{\mathcal{J}},\rho\left( u_{0}^{i - 1} \right) \middle| d\left( u_{0}^{i} \right) \right)\]
call \(w_{1}\) and \(w_{2}\) satisfy B-condition, denoted as
\(w_{1}\overset{B}{\sim}w_{2}\).

B-condition indicates that variables in likelihood probability of
\(w_{1}\) and \(w_{2}\) are one-to-one mapped to each other, so through
the summation of capacity calculation they get identical sum, namely
they have identical capacity, as Lemma 5 proves.

\emph{Lemma 5:} Channels satisfying B-condition have identical capacity.
Namely, for any \(w_{1},w_{2}\), if \(w_{1}\overset{B}{\sim}w_{2}\),
\(I\left( w_{1} \right) = I\left( w_{2} \right)\).

\emph{Proof:}

\(w_{1}\overset{B}{\sim}w_{2}\), so according to Definition 6, there
exists a sequence \(\mathcal{J}\) that
\(\left\{ \mathcal{J} \right\} = \left\{ 0,\ldots,n - 1 \right\},\left| \mathcal{J} \right| = n\),
a bijection \(\rho:\mathbb{B}^{i} \mapsto \mathbb{B}^{i}\) and a mapping
\(d\left( u_{0}^{i} \right):\mathbb{B}^{i}\mathbb{\mapsto B}\) bijective
to \(u_{i}\), such that

\[w_{1}\left( y_{0}^{n - 1},u_{0}^{i - 1} \middle| u_{i} \right) = w_{2}\left( y_{\mathcal{J}},\rho\left( u_{0}^{i - 1} \right) \middle| d\left( u_{0}^{i} \right) \right)\]

Therefore
\begin{align*}
	&w_{1}\left( y_{0}^{n - 1},u_{0}^{i - 1} \middle| 0 \right) + w_{1}\left( y_{0}^{n - 1},u_{0}^{i - 1} \middle| 1 \right) \\
	&= w_{2}\left( y_{\mathcal{J}},\rho\left( u_{0}^{i - 1} \right) \middle| 0 \right) + w_{2}\left( y_{\mathcal{J}},\rho\left( u_{0}^{i - 1} \right) \middle| 1 \right)
\end{align*}

So mutual information
\begin{align*}
	&\eta_{1}\left( y_{0}^{n - 1},u_{0}^{i - 1};u_{i} \right) \\
	&= \log_{2}\frac{2w_{1}\left( y_{0}^{n - 1},u_{0}^{i - 1} \middle| u_{i} \right)}{w_{1}\left( y_{0}^{n - 1},u_{0}^{i - 1} \middle| 0 \right) + w_{1}\left( y_{0}^{n - 1},u_{0}^{i - 1} \middle| 1 \right)} \\
	&= \log_{2}\frac{2w_{2}\left( y_{\mathcal{J}},\rho\left( u_{0}^{i - 1} \right) \middle| d\left( u_{0}^{i} \right) \right)}{w_{2}\left( y_{\mathcal{J}},\rho\left( u_{0}^{i - 1} \right) \middle| 0 \right) + w_{2}\left( y_{\mathcal{J}},\rho\left( u_{0}^{i - 1} \right) \middle| 1 \right)} \\
	&= \eta_{2}\left( y_{\mathcal{J}},\rho\left( u_{0}^{i - 1} \right);d\left( u_{0}^{i} \right) \right)
\end{align*}

Define \(p\) as probability distribution of \(w_{1}\), and \(q\) as that
of \(w_{2}\), and define
\(Z_{0}^{n - 1} = Y_{\mathcal{J}},\ V_{0}^{i} = \left\lbrack \rho\left( U_{0}^{i - 1} \right),d\left( U_{0}^{i} \right) \right\rbrack\),
then
\begin{align*}
	I\left( w_{1} \right) &= \mathbb{E}_{p}\left\lbrack \eta_{1}\left( Y_{0}^{n - 1},U_{0}^{i - 1};U_{i} \right) \right\rbrack \\
	&= \mathbb{E}_{p}\left\lbrack \eta_{2}\left( Y_{\mathcal{J}},\rho\left( U_{0}^{i - 1} \right);d\left( U_{0}^{i} \right) \right) \right\rbrack \\
	&= \mathbb{E}_{p}\left\lbrack \eta_{2}\left( Z_{0}^{n - 1},V_{0}^{j - 1};V_{j} \right) \right\rbrack
\end{align*}
\begin{align*}
	&p_{Z_{0}^{n - 1}V_{0}^{i}}\left( z_{0}^{n - 1},v_{0}^{i} \right) \\
	&= p_{Y_{0}^{n - 1}U_{0}^{i}}\left( z_{\mathcal{J}^{- 1}},\rho^{- 1}\left( v_{0}^{i - 1} \right),d^{- 1}\left( v_{i};\rho^{- 1}\left( v_{0}^{i - 1} \right) \right) \right) \\
	&= \frac{1}{2}w_{1}\left( z_{\mathcal{J}^{- 1}},\rho^{- 1}\left( v_{0}^{i - 1} \right) \middle| d^{- 1}\left( v_{i};\rho^{- 1}\left( v_{0}^{i - 1} \right) \right) \right) \\
	&= \frac{1}{2}w_{2}\left( z_{0}^{n - 1},v_{0}^{i - 1} \middle| v_{i} \right) \\
	&= q_{Z_{0}^{n - 1}V_{0}^{i}}\left( z_{0}^{n - 1},v_{0}^{i} \right)
\end{align*}
so
\begin{align*}
	I\left( w_{1} \right) &= \mathbb{E}_{p}\left\lbrack \eta_{2}\left( Z_{0}^{n - 1},V_{0}^{j - 1};V_{j} \right) \right\rbrack \\
	&= \mathbb{E}_{q}\left\lbrack \eta_{2}\left( Z_{0}^{n - 1},V_{0}^{j - 1};V_{j} \right) \right\rbrack = I\left( w_{2} \right)
\end{align*}

Lemma 6 given below gives the relationship of likelihood probability of
each sub-channel with each initial state, i.e.,
\(W_{n}^{(j)}\left( y_{0}^{n - 1},u_{0}^{j - 1} \middle| u_{j},s \right)\)
and that with initial state 0, i.e.,
\(W_{n}^{(j)}\left( y_{0}^{n - 1},{\widetilde{u}}_{0}^{j - 1} \middle| {\widetilde{u}}_{j} \right)\),
namely
\(W_{n}^{(j)}\left( y_{0}^{n - 1},{\widetilde{u}}_{0}^{j - 1} \middle| {\widetilde{u}}_{j} \right) = W_{n}^{(j)}\left( y_{0}^{n - 1},{\widetilde{u}}_{0}^{j - 1} \middle| {\widetilde{u}}_{j},0 \right)\),
which indicating that through a transformation on \(u_{0}^{j}\) to
produce \({\widetilde{u}}_{0}^{j}\), the former equals the latter.

\emph{Lemma 6:} On any bijective FSMCs
\(\left( \mathcal{S},\mathbb{B},F,W \right)\), for every code length
\(n = 2^{m},m \in \mathbb{N}^{*}\), transition probability of each
sub-channel \(W_{n}^{(i)},i \in \left\{ 0,\ldots,n - 1 \right\}\)
satisfies
\begin{align*}
	&W_{n}^{(i)}\left( y_{0}^{n - 1},u_{0}^{i - 1} \middle| u_{i},s \right) \\
	&= \left\{
	\begin{aligned}
		& W_{n}^{(i)}\left( y_{0}^{n - 1},u_{0}^{i - 1} \middle| u_{i} \right) & ,i \in \left\{ 0,\ldots,n - 2 \right\} \\
		& W_{n}^{(i)}\left( y_{0}^{n - 1},u_{0}^{i - 1} \middle| F_{0}^{- 1}F_{s}u_{i} \right) & ,i = n - 1
	\end{aligned} \right.
\end{align*}

\emph{Proof:} The last sub-channel
\begin{align*}
	&W_{n}^{(n - 1)}\left( y_{0}^{n - 1},u_{0}^{n - 2} \middle| u_{n - 1},s \right) \\
	&= \frac{1}{2^{n - 1}}p_{Y_{0}^{n - 1}|U_{0}^{n - 1}S_{- 1}}\left( y_{0}^{n - 1} \middle| u_{0}^{n - 1},s \right) \\
	&= \frac{1}{2^{n - 1}}W^{n}\left( y_{0}^{n - 1} \middle| F_{s}^{n}\left( u_{0}^{n - 1}G_{n} \right) \right)
\end{align*}

Considering \({\overline{G}}_{n} = \begin{bmatrix}
	\boldsymbol{0} & {\widehat{G}}_{n} \\
	1 & \boldsymbol{0}^{T}
\end{bmatrix}\), there is
\(u_{0}^{n - 1}{\overline{G}}_{n} = \left\lbrack u_{0}^{n - 2},u_{n - 1} \right\rbrack\begin{bmatrix}
	\boldsymbol{0} & {\widehat{G}}_{n} \\
	1 & \boldsymbol{0}^{T}
\end{bmatrix} = \left\lbrack u_{n - 1},u_{0}^{n - 2}{\widehat{G}}_{n} \right\rbrack\),
so
\begin{align*}
	&F_{s}^{n}\left( u_{0}^{n - 1}G_{n} \right) = F_{s}^{n}\left( \left\lbrack u_{n - 1},u_{0}^{n - 2}{\widehat{G}}_{n} \right\rbrack \right) \\
	&= F_{F_{s}u_{n - 1}}^{n - 1}\left( u_{0}^{n - 2}{\widehat{G}}_{n} \right) = F_{F_{0}F_{0}^{- 1}F_{s}u_{n - 1}}^{n - 1}\left( u_{0}^{n - 2}{\widehat{G}}_{n} \right) \\
	&= F_{0}^{n}\left( \left\lbrack u_{0}^{n - 2},F_{0}^{- 1}F_{s}u_{n - 1} \right\rbrack{\overline{G}}_{n} \right)
\end{align*}
then
\begin{align*}
	&W^{n}\left( y_{0}^{n - 1} \middle| F_{s}^{n}\left( u_{0}^{n - 1}{\overline{G}}_{n} \right) \right) \\
	&= W^{n}\left( y_{0}^{n - 1} \middle| F_{s}^{n}\left( \left\lbrack u_{0}^{n - 2},F_{0}^{- 1}F_{s}u_{n - 1} \right\rbrack{\overline{G}}_{n} \right) \right)
\end{align*}

Namely
\begin{align*}
	&W_{n}^{(n - 1)}\left( y_{0}^{n - 1},u_{0}^{n - 2} \middle| u_{n - 1},s \right) \\
	&= W_{n}^{(n - 1)}\left( y_{0}^{n - 1},u_{0}^{n - 2} \middle| F_{0}^{- 1}F_{s}u_{n - 1} \right)
\end{align*}

And each non-last sub-channel
\begin{align*}
	&W_{n}^{(i)}\left( y_{0}^{n - 1},u_{0}^{i - 1} \middle| u_{i},s \right) \\
	&= \sum_{u_{i + 1}^{n - 1}}^{}{W_{n}^{(n - 1)}\left( y_{0}^{n - 1},u_{0}^{n - 2} \middle| u_{n - 1},s \right)} \\
	&= \sum_{u_{i + 1}^{n - 1}}^{}{W_{n}^{(n - 1)}\left( y_{0}^{n - 1},u_{0}^{n - 2} \middle| F_{0}^{- 1}F_{s}u_{n - 1} \right)} \\
	&= \sum_{u_{i + 1}^{n - 2}}^{}{\sum_{u_{n - 1}}^{}{W_{n}^{(n - 1)}\left( y_{0}^{n - 1},u_{0}^{n - 2} \middle| F_{0}^{- 1}F_{s}u_{n - 1} \right)}} \\
	&= \sum_{u_{i + 1}^{n - 2}}^{}{\sum_{s^{\prime}\mathcal{\in S}}^{}{W_{n}^{(n - 1)}\left( y_{0}^{n - 1},u_{0}^{n - 2} \middle| F_{0}^{- 1}s^{\prime} \right)}} \\
	&= \sum_{u_{i + 1}^{n - 2}}^{}{\sum_{u_{n - 1}\in \mathbb{B}}^{}{W_{n}^{(n - 1)}\left( y_{0}^{n - 1},u_{0}^{n - 2} \middle| u_{n - 1} \right)}} \\
	&= \sum_{u_{i + 1}^{n - 1}}^{}{W_{n}^{(n - 1)}\left( y_{0}^{n - 1},u_{0}^{n - 2} \middle| u_{n - 1} \right)} \\
	&= W_{n}^{(i)}\left( y_{0}^{n - 1},u_{0}^{i - 1} \middle| u_{i} \right)
\end{align*}

So proof to Lemma 6 is finished.

\emph{Lemma 7:} On any bijective FSMCs
\(\left( \mathcal{S},\mathbb{B},F,W \right)\), at any code length
\(n \in \left\{ 2^{m} \middle| m \in \mathbb{N}^{*} \right\}\), each
sub-channel \(W_{n}^{(i)},i \in \{ 0,1,\ldots,n - 1\}\) and
\({\widetilde{W}}_{n}^{(i)}\) viz. corresponding sub-channel of
conventional polar code on \(W^{n}\) satisfy B-condition, i.e.,
\(W_{n}^{(i)}\overset{B}{\sim}{\widetilde{W}}_{n}^{(i)}\).

\emph{Proof:}

Let \(p_{m}\) be the statement ``On any bijective FSMCs
\(\left( \mathcal{S},\mathbb{B},F,W \right)\), at code length \(n = 2^{m}\), each
sub-channel \(W_{n}^{(i)},i \in \{ 0,1,\ldots,n - 1\}\) and
\({\widetilde{W}}_{n}^{(i)}\) viz. corresponding sub-channel of
conventional polar code on \(W^{n}\) satisfy B-condition, i.e.,
\(W_{n}^{(i)}\overset{B}{\sim}{\widetilde{W}}_{n}^{(i)}\)''. We give an
induction on \(m\).

\begin{enumerate}
	\def\labelenumi{\arabic{enumi})}
	\item
	For \(m = 1\), there is
\end{enumerate}
\begin{align*}
	&W_{2}^{(1)}\left( y_{0}^{1},u_{0} \middle| u_{1} \right) = \frac{1}{2}W\left( y_{0} \middle| F_{0}u_{1} \right)W\left( y_{1} \middle| F_{0}^{2}\left\lbrack u_{1},u_{0} \right\rbrack \right) \\
	&= \frac{1}{2}W\left( y_{0} \middle| F_{0}u_{1} \right)W\left( y_{1} \middle| F_{F_{0}u_{1}}u_{0} \right)\\
	&\overset{Co. 2}{=}\frac{1}{2}W\left( y_{0} \middle| u_{1} + F_{00} \right)W\left( y_{1} \middle| u_{0} + u_{1} \right) \\
	&= \frac{1}{2}W\left( y_{1} \middle| u_{0} + u_{1} \right)W\left( y_{0} \middle| u_{1} + F_{00} \right) \\
	&= {\widetilde{W}}_{2}^{(1)}\left( \left\lbrack y_{1},y_{0} \right\rbrack,u_{0} + F_{00} \middle| u_{1} + F_{00} \right)
\end{align*}

Denote
\(\mathcal{J =}\lbrack 1,0\rbrack,\rho\left( u_{0} \right) = u_{0} + F_{00},d\left( u_{0}^{1} \right) = u_{1} + F_{00}\), then \(W_{2}^{(1)}\left( y_{0}^{1},u_{0} \middle| u_{1} \right) = {\widetilde{W}}_{2}^{(1)}\left( y_{\mathcal{J}},\rho(u_0) \middle| d(u_0^1) \right)\), trivially \(\rho\) is a bijection, and \(d(u_0^1)\) is bijective to \(u_1\), so
according to Definition 6 we know
\(W_{2}^{(1)}\overset{B}{\sim}{\widetilde{W}}_{2}^{(1)}\); similarly,
\(W_{2}^{(0)}\overset{B}{\sim}{\widetilde{W}}_{2}^{(0)}\).

\begin{enumerate}
	\def\labelenumi{\arabic{enumi})}
	\setcounter{enumi}{1}
	\item
	For any \(m \geq 1\), suppose that \(p_{m}\) is true, then
\end{enumerate}

For each non-last sub-channel pair, i.e.,
\(\left\lbrack W_{2n}^{(2i)},W_{2n}^{(2i + 1)} \right\rbrack,\ i \in \left\{ 0,1,\ldots,n - 2 \right\}\)
\begin{align*}
	&W_{2n}^{(2i + 1)}\left( y_{0}^{2n - 1},u_{0}^{2i} \middle| u_{2i + 1} \right) \\
	&= \frac{1}{2}\sum_{t \in \mathcal{S}}^{}W_{n}^{(i)}\left( y_{0}^{n - 1},v_{0,0}^{i - 1},t \middle| u_{2i} + u_{2i + 1} \right) \\
	&\cdot W_{n}^{(i)}\left( y_{n}^{2n - 1},v_{1,0}^{i - 1} \middle| u_{2i + 1},t \right)\\
	&\overset{Le. 6}{=}\frac{1}{2}\sum_{t \in \mathcal{S}}^{}W_{n}^{(i)}\left( y_{0}^{n - 1},v_{0,0}^{i - 1},t \middle| u_{2i} + u_{2i + 1} \right) \\
	&\cdot W_{n}^{(i)}\left( y_{n}^{2n - 1},v_{1,0}^{i - 1} \middle| u_{2i + 1} \right) \\
	&= \frac{1}{2}W_{n}^{(i)}\left( y_{0}^{n - 1},v_{0,0}^{i - 1} \middle| u_{2i} + u_{2i + 1} \right)\\
	&\cdot W_{n}^{(i)}\left( y_{n}^{2n - 1},v_{1,0}^{i - 1} \middle| u_{2i + 1} \right)
\end{align*}
where
\(v_{0,0}^{i - 1} = \left\lbrack u_{2j} + u_{2j + 1} \right\rbrack_{j \in \left\{ 0,1,\ldots,i - 1 \right\}},v_{1,0}^{i - 1} = \left\lbrack u_{2j + 1} \right\rbrack_{j \in \left\{ 0,1,\ldots,i - 1 \right\}}\).

And according to the hypothesis that \(p_{m}\) is true, indicating
\(W_{n}^{(i)}\overset{B}{\sim}{\widetilde{W}}_{n}^{(i)}\), we know there
exists \(\mathcal{J,}\rho,d\) that
\(W_{n}^{(i)}\left( y_{0}^{n - 1},v_{0,0}^{i - 1} \middle| u_{2i} + u_{2i + 1} \right) = {\widetilde{W}}_{n}^{(i)}\left( y_{\mathcal{J}},\rho\left( v_{0,0}^{i - 1} \right) \middle| d\left( \left\lbrack v_{0,0}^{i - 1},u_{2i} + u_{2i + 1} \right\rbrack \right) \right)\),
\(W_{n}^{(i)}\left( y_{n}^{2n - 1},v_{1,0}^{i - 1} \middle| u_{2i + 1} \right) = {\widetilde{W}}_{n}^{(i)}\left( y_{\mathcal{J +}n},\rho\left( v_{1,0}^{i - 1} \right) \middle| d\left( \left\lbrack v_{1,0}^{i - 1},u_{2i + 1} \right\rbrack \right) \right)\),
so
\begin{align*}
	&W_{2n}^{(2i + 1)}\left( y_{0}^{2n - 1},u_{0}^{2i} \middle| u_{2i + 1} \right) \\
	&= \frac{1}{2}{\widetilde{W}}_{n}^{(i)}\left( y_{\mathcal{J}},\rho\left( v_{0,0}^{i - 1} \right) \middle| d\left( \left\lbrack v_{0,0}^{i - 1},u_{2i} + u_{2i + 1} \right\rbrack \right) \right)\\
	&\cdot {\widetilde{W}}_{n}^{(i)}\left( y_{\mathcal{J +}n},\rho\left( v_{1,0}^{i - 1} \right) \middle| d\left( \left\lbrack v_{1,0}^{i - 1},u_{2i + 1} \right\rbrack \right) \right)\\
	&\overset{Co. 1}{=}\frac{1}{2}{\widetilde{W}}_{n}^{(i)}\left( y_{\mathcal{J}},\rho\left( v_{0,0}^{i - 1} \right) \middle| u_{2i} + u_{2i + 1} + d_{0}\left( v_{0,0}^{i - 1} \right) \right)\\
	&\cdot {\widetilde{W}}_{n}^{(i)}\left( y_{\mathcal{J +}n},\rho\left( v_{1,0}^{i - 1} \right) \middle| u_{2i + 1} + d_{0}\left( v_{1,0}^{i - 1} \right) \right) \\
	&= \frac{1}{2}{\widetilde{W}}_{n}^{(i)}\left( y_{\mathcal{J}},\rho\left( v_{0,0}^{i - 1} \right) \middle| u_{2i} + d_{0}\left( v_{0,0}^{i - 1} \right) + d_{0}\left( v_{1,0}^{i - 1} \right) \right. \\
	&\left. + u_{2i + 1} + d_{0}\left( v_{1,0}^{i - 1} \right) \right)\\
	&\cdot {\widetilde{W}}_{n}^{(i)}\left( y_{\mathcal{J +}n},\rho\left( v_{1,0}^{i - 1} \right) \middle| u_{2i + 1} + d_{0}\left( v_{1,0}^{i - 1} \right) \right)
\end{align*}

Denote
\[\mathcal{J}_{1} = \left\lbrack \mathcal{J,J +}n \right\rbrack\]
\begin{align*}
	&\rho_{1}\left( u_{0}^{2i} \right) = \left\lbrack \left\lbrack \rho\left( v_{0,0}^{i - 1} \right),\rho\left( v_{1,0}^{i - 1} \right) \right\rbrack R_{2i}^{- 1}\left( I_{i} \otimes F \right), \right. \\
	&\left. u_{2i} + d_{0}\left( v_{0,0}^{i - 1} \right) + d_{0}\left( v_{1,0}^{i - 1} \right) \right\rbrack
\end{align*}
\[d_{1}\left( u_{0}^{2i + 1} \right) = d\left( \left\lbrack v_{1,0}^{i - 1},u_{2i + 1} \right\rbrack \right)\]
we have
\begin{align*}
	&W_{2n}^{(2i + 1)}\left( y_{0}^{2n - 1},u_{0}^{2i} \middle| u_{2i + 1} \right) \\
	&= {\widetilde{W}}_{2n}^{(2i + 1)}\left( y_{\mathcal{J}_{1}},\rho_{1}\left( u_{0}^{2i} \right) \middle| d_{1}\left( u_{0}^{2i + 1} \right) \right)
\end{align*}

Trivially \(\rho_1\) is a bijection, and \(d_1(u_0^{2i+1})\) is bijective to \(u_{2i+1}\), so according to Definition 6,
\(W_{2n}^{(2i + 1)}\overset{B}{\sim}{\widetilde{W}}_{2n}^{(2i + 1)}\);
similarly,
\(W_{2n}^{(2i)}\overset{B}{\sim}{\widetilde{W}}_{2n}^{(2i)}\).

For the last sub-channel pair
\(\left\lbrack W_{2n}^{(2n - 2)},W_{2n}^{(2n - 1)} \right\rbrack\)
\begin{align*}
	&W_{2n}^{(2n - 1)}\left( y_{0}^{2n - 1},u_{0}^{2n - 2} \middle| u_{2n - 1} \right) \\
	&= \frac{1}{2}\sum_{t \in \mathcal{S}}^{}W_{n}^{(n - 1)}\left( y_{0}^{n - 1},v_{0,0}^{n - 2},t \middle| u_{2n - 1} \right)\\
	&\cdot W_{n}^{(n - 1)}\left( y_{n}^{2n - 1},v_{1,0}^{n - 2} \middle| u_{2n - 2},t \right) \\
	&= \frac{1}{2}W_{n}^{(n - 1)}\left( y_{0}^{n - 1},v_{0,0}^{n - 2} \middle| u_{2n - 1} \right)\\
	&\cdot W_{n}^{(n - 1)}\left( y_{n}^{2n - 1},v_{1,0}^{n - 2} \middle| u_{2n - 2},F_{0}^{n}\left( \left\lbrack v_{0,0}^{n - 2},u_{2n - 1} \right\rbrack{\overline{G}}_{n} \right) \right) \\
	&= \frac{1}{2}W_{n}^{(n - 1)}\left( y_{0}^{n - 1},v_{0,0}^{n - 2} \middle| u_{2n - 1} \right)\\
	&\cdot W_{n}^{(n - 1)}\left( y_{n}^{2n - 1},v_{1,0}^{n - 2} \middle| u_{2n - 2},F_{0}^{n}\left\lbrack u_{2n - 1},v_{0,0}^{n - 2}{\widehat{G}}_{n} \right\rbrack \right)\\
	&\overset{Le. 6}{=}\frac{1}{2}W_{n}^{(n - 1)}\left( y_{0}^{n - 1},v_{0,0}^{n - 2} \middle| u_{2n - 1} \right)\\
	&\cdot W_{n}^{(n - 1)}\left( y_{n}^{2n - 1},v_{1,0}^{n - 2} \middle| F_{0}^{- 1}F_{F_{0}^{n}\left\lbrack u_{2n - 1},v_{0,0}^{n - 2}{\widehat{G}}_{n} \right\rbrack}u_{2n - 2} \right) \\
	&= \frac{1}{2}W_{n}^{(n - 1)}\left( y_{0}^{n - 1},v_{0,0}^{n - 2} \middle| u_{2n - 1} \right)\\
	&\cdot W_{n}^{(n - 1)}\left( y_{n}^{2n - 1},v_{1,0}^{n - 2} \middle| F_{0}^{- 1}F_{0}^{n + 1}\left\lbrack u_{2n - 1},v_{0,0}^{n - 2}{\widehat{G}}_{n},u_{2n - 2} \right\rbrack \right)\\
	&\overset{Co. 2}{=}\frac{1}{2}W_{n}^{(n - 1)}\left( y_{0}^{n - 1},v_{0,0}^{n - 2} \middle| u_{2n - 1} \right)\\
	&\cdot W_{n}^{(n - 1)}\left( y_{n}^{2n - 1},v_{1,0}^{n - 2} \middle| u_{2n - 1} + v_{0,0}^{n - 2}{\widehat{G}}_{n} + u_{2n - 2} \right)
\end{align*}

And according to the hypothesis that \(p_{m}\) is true, indicating
\(W_{n}^{(n - 1)}\overset{B}{\sim}{\widetilde{W}}_{n}^{(n - 1)}\), we
know there exists \(\mathcal{J,}\rho,d\) that
\begin{align*}
	&W_{n}^{(n - 1)}\left( y_{0}^{n - 1},v_{0,0}^{n - 2} \middle| u_{2n - 1} \right) \\
	&= {\widetilde{W}}_{n}^{(n - 1)}\left( y_{\mathcal{J}},\rho\left( v_{0,0}^{n - 2} \right) \middle| d\left( \left\lbrack v_{0,0}^{n - 2},u_{2n - 1} \right\rbrack \right) \right)
\end{align*}
\begin{align*}
	&W_{n}^{(n - 1)}\left( y_{n}^{2n - 1},v_{1,0}^{n - 2} \middle| u_{2n - 2} + v_{0,0}^{n - 2}{\widehat{G}}_{n} + u_{2n - 1} \right) \\
	&= {\widetilde{W}}_{n}^{(n - 1)}\left( y_{\mathcal{J +}n},\rho\left( v_{1,0}^{n - 2} \right) \middle|\right.\\
	&\left. d\left( \left\lbrack v_{1,0}^{n - 2},u_{2n - 2} + v_{0,0}^{n - 2}{\widehat{G}}_{n} + u_{2n - 1} \right\rbrack \right) \right)
\end{align*}
so
\begin{align*}
	&W_{2n}^{(2n - 1)}\left( y_{0}^{2n - 1},u_{0}^{2n - 2} \middle| u_{2n - 1} \right) \\
	&= \frac{1}{2}{\widetilde{W}}_{n}^{(n - 1)}\left( y_{\mathcal{J}},\rho\left( v_{0,0}^{n - 2} \right) \middle| d\left( \left\lbrack v_{0,0}^{n - 2},u_{2n - 1} \right\rbrack \right) \right)\\
	&\cdot {\widetilde{W}}_{n}^{(n - 1)}\left( y_{\mathcal{J +}n},\rho\left( v_{1,0}^{n - 2} \right) \middle| \right.\\
	&\left. d\left( \left\lbrack v_{1,0}^{n - 2},u_{2n - 2} + v_{0,0}^{n - 2}{\widehat{G}}_{n} + u_{2n - 1} \right\rbrack \right) \right)\\
	&\overset{Co. 1}{=}\frac{1}{2}{\widetilde{W}}_{n}^{(n - 1)}\left( y_{\mathcal{J}},\rho\left( v_{0,0}^{n - 2} \right) \middle| u_{2n - 1} + d_{0}\left( v_{0,0}^{n - 2} \right) \right)\\
	&\cdot {\widetilde{W}}_{n}^{(n - 1)}\left( y_{\mathcal{J +}n},\rho\left( v_{1,0}^{n - 2} \right) \middle| u_{2n - 2} + v_{0,0}^{n - 2}{\widehat{G}}_{n} \right. \\
	&\left. + u_{2n - 1} + d_{0}\left( v_{1,0}^{n - 2} \right) \right) \\
	&= \frac{1}{2}{\widetilde{W}}_{n}^{(n - 1)}\left( y_{\mathcal{J +}n},\rho\left( v_{1,0}^{n - 2} \right) \middle| u_{2n - 2} + v_{0,0}^{n - 2}{\widehat{G}}_{n} + u_{2n - 1} \right. \\
	&\left. + d_{0}\left( v_{1,0}^{n - 2} \right) \right) \cdot {\widetilde{W}}_{n}^{(n - 1)}\left( y_{\mathcal{J}},\rho\left( v_{0,0}^{n - 2} \right) \middle| u_{2n - 1} + d_{0}\left( v_{0,0}^{n - 2} \right) \right) \\
	&= \frac{1}{2}{\widetilde{W}}_{n}^{(n - 1)}\left( y_{\mathcal{J +}n},\rho\left( v_{1,0}^{n - 2} \right) \middle| u_{2n - 2} + v_{0,0}^{n - 2}{\widehat{G}}_{n} + d_{0}\left( v_{0,0}^{n - 2} \right)  \right. \\
	&\left. + d_{0}\left( v_{1,0}^{n - 2} \right) + u_{2n - 1} + d_{0}\left( v_{0,0}^{n - 2} \right) \right)\\
	&\cdot {\widetilde{W}}_{n}^{(n - 1)}\left( y_{\mathcal{J}},\rho\left( v_{0,0}^{n - 2} \right) \middle| u_{2n - 1} + d_{0}\left( v_{0,0}^{n - 2} \right) \right)
\end{align*}

Denote
\[\mathcal{J}_{1} = \left\lbrack \mathcal{J +}n,\mathcal{J} \right\rbrack\]
\begin{align*}
	&\rho_{1}\left( u_{0}^{2n - 2} \right) = \left\lbrack \left\lbrack \rho\left( v_{1,0}^{n - 2} \right),\rho\left( v_{0,0}^{n - 2} \right) \right\rbrack R_{2n - 2}^{- 1}\left( I_{n - 1} \otimes F \right),\right.\\
	&\left. u_{2n - 2} + v_{0,0}^{n - 2}{\widehat{G}}_{n} + d_{0}\left( v_{0,0}^{n - 2} \right) + d_{0}\left( v_{1,0}^{n - 2} \right) \right\rbrack
\end{align*}
\[d_{1}\left( u_{0}^{2n - 1} \right) = d\left( \left\lbrack v_{0,0}^{n - 2},u_{2n - 1} \right\rbrack \right)\]
we have
\begin{align*}
	&W_{2n}^{(2n - 1)}\left( y_{0}^{2n - 1},u_{0}^{2n - 2} \middle| u_{2n - 1} \right) \\
	&= {\widetilde{W}}_{2n}^{(2n - 1)}\left( y_{\mathcal{J}_{1}},\rho_{1}\left( u_{0}^{2n - 2} \right) \middle| d_{1}\left( u_{0}^{2n - 1} \right) \right)
\end{align*}

Trivially \(\rho_1\) is a bijection, and \(d_1(u_0^{2n-1})\) is bijective to \(u_{2n-1}\), so according to Definition 6,
\(W_{2n}^{(2n - 1)}\overset{B}{\sim}{\widetilde{W}}_{2n}^{(2n - 1)}\);
similarly,\(\ W_{2n}^{(2n - 2)}\overset{B}{\sim}{\widetilde{W}}_{2n}^{(2n - 2)}\).

In summary,
\(W_{2n}^{(j)}\overset{B}{\sim}{\widetilde{W}}_{2n}^{(j)},\forall j \in \left\{ 0,\ldots,2n - 1 \right\}\),
namely \(p_{m + 1}\) is true.

With 1) and 2) we know \(p_{m}\) is true for all
\(m \in \mathbb{N}^{*}\), namely Lemma 7 is true.

\emph{Proof to Theorem 1:}

According to Lemma 7, on any bijective FSMCs
\(\left( \mathcal{S},\mathbb{B},F,W \right)\), at any code length
\(n \in \left\{ 2^{m} \middle| m \in \mathbb{N}^{*} \right\}\), each
sub-channel \(W_{n}^{(i)},i \in \{ 0,1,\ldots,n - 1\}\) and
\({\widetilde{W}}_{n}^{(i)}\) viz. corresponding sub-channel of
conventional polar code on \(W^{n}\) satisfy B-condition, i.e.,
\(W_{n}^{(i)}\overset{B}{\sim}{\widetilde{W}}_{n}^{(i)}\). With Lemma 5
indicating that channels satisfying B-condition have identical capacity,
we know
\(I\left( W_{n}^{(i)} \right) = I\left( {\widetilde{W}}_{n}^{(i)} \right)\),
namely
\(I_{\left( \mathcal{P}_{s},\mathcal{F}_{b} \right),\ n}^{(i)} = I_{\left( \mathcal{P}_{c},W^{\left\langle i \right\rangle} \right),\ n}^{(i)}\),
which illustrates Theorem 1 is true. Proof to Theorem 1 is finished.

\subsection{Proof of Theorem 2}

Define:
\[{\overline{I}}_{n}^{(i)} := I_{\left( \mathcal{P}_{c},\mathcal{F}_{b} \right),\ n}^{(i)},\ {\widetilde{I}}_{n}^{(i)} := I_{\left( \mathcal{P}_{c},W^{n} \right),\ n}^{(i)},I_{n}^{(i)} := I_{\left( \mathcal{P}_{s},\mathcal{F}_{b} \right),\ n}^{(i)}\]
\[{\overline{\mathcal{I}}}_{n} := \mathcal{I}_{\left( \mathcal{P}_{c},\mathcal{F}_{b} \right),\ n},{\widetilde{\mathcal{I}}}_{n} := \mathcal{I}_{\left( \mathcal{P}_{c},W^{n} \right),\ n},\mathcal{I}_{n} := \mathcal{I}_{\left( \mathcal{P}_{s},\mathcal{F}_{b} \right),\ n}\]
\[{\overline{\sigma}}_{n}^{2} := \sigma^{2}\left( {\overline{\mathcal{I}}}_{n} \right),{\widetilde{\sigma}}_{n}^{2} := \sigma^{2}\left( {\widetilde{\mathcal{I}}}_{n} \right),\sigma_{n}^{2} := \sigma^{2}\left( \mathcal{I}_{n} \right)\]

To prove Theorem 2, we put up with a mathematical lemma, i.e.,
Lemma 8:

\emph{Lemma 8:} For a sample sequence \(X\) and any partition of it
\(\left\lbrack \mathbb{B}_{k} \right\rbrack_{k \in \mathcal{K}}\mathcal{,K :=}\left\{ 0,1,\ldots,K - 1 \right\}\),
the variance of sequence \(X\) is the sum of weighted variance of each
subsequence's average and weighted average of each subsequence's
variance, namely
\[\sigma^{2}(X) = \sigma^{2}\left( \mu_{\mathcal{K}};r_{\mathcal{K}} \right) + \mu\left( \sigma_{\mathcal{K}}^{2};r_{\mathcal{K}} \right)\]
where
\[\mu_{\mathcal{K}} := \left\lbrack \mu_{k} \right\rbrack_{k \in \mathcal{K}} = \left\lbrack \mu\left( \mathbb{B}_{k} \right) \right\rbrack_{k \in \mathcal{K}}\]
\[\sigma_{\mathcal{K}}^{2} := \left\lbrack \sigma_{k}^{2} \right\rbrack_{k \in \mathcal{K}} = \left\lbrack \sigma^{2}\left( \mathbb{B}_{k} \right) \right\rbrack_{k \in \mathcal{K}}\]
\[r_{\mathcal{K}} := \left\lbrack r_{k} \right\rbrack_{k \in \mathcal{K}} = \left\lbrack \frac{|\mathbb{B}_{k}|}{|X|} \right\rbrack_{k \in \mathcal{K}}\]
and \(\mu,\sigma^{2}\) represent sample mean and variance, respectively.

\emph{Proof:} Average of sequence
\[\mu(X) = \frac{1}{N}\sum_{k = 0}^{K - 1}{N_{k}\mu\left( \mathbb{B}_{k} \right)} = \sum_{k = 0}^{K - 1}{\frac{N_{k}}{N}\mu\left( \mathbb{B}_{k} \right)} = \sum_{k = 0}^{K - 1}{r_{k}\mu_{k}}\]
and variance
\[\sigma^{2}(X) = \mu\left( \left( X - \mu(X) \right)^{2} \right) = \sum_{k = 0}^{K - 1}r_{k}\mu\left( \left( \mathbb{B}_{k} - \mu(X) \right)^{2} \right)\]
where
\begin{align*}
	&\left( \mathbb{B}_{k} - \mu(X) \right)^{2} = \left( \mathbb{B}_{k} - \mu_{k} + \mu_{k} - \mu(X) \right)^{2} \\
	&= \left( \mathbb{B}_{k} - \mu_{k} \right)^{2} + \left( \mu_{k} - \mu(X) \right)^{2} + 2(\mathbb{B}_{k} - \mu_{k})(\mu_{k} - \mu(X))
\end{align*}
then
\begin{align*}
	&\mu\left( \left( \mathbb{B}_{k} - \mu(X) \right)^{2} \right) \\
	&= \mu\left( \left( \mathbb{B}_{k} - \mu_{k} \right)^{2} \right) + \mu\left( \left( \mu_{k} - \mu(X) \right)^{2} \right) + 0 \\
	&= \sigma_{k}^{2} + \left( \mu_{k} - \mu(X) \right)^{2}
\end{align*}

So
\begin{align*}
	\sigma^{2}(X) &= \sum_{k = 0}^{K - 1}r_{k}\sigma_{k}^{2} + \sum_{k = 0}^{K - 1}r_{k}\left( \mu_{k} - \mu(X) \right)^{2} \\
	&= \mu\left( \sigma_{\mathcal{K}}^{2};r_{\mathcal{K}} \right) + \sigma^{2}\left( \mu_{\mathcal{K}};r_{\mathcal{K}} \right) \\
	&= \sigma^{2}\left( \mu_{\mathcal{K}};r_{\mathcal{K}} \right) + \mu\left( \sigma_{\mathcal{K}}^{2};r_{\mathcal{K}} \right)
\end{align*}

Proof of Lemma 8 is finished.

\emph{Proof to Theorem 2:}

For any \({\widetilde{\mathcal{I}}}_{n}\), where
\(n = 2^{m},m \in \mathbb{N}^{*}\), we make a partition
\(\mathcal{L}_{k} = \left\lbrack {\widetilde{I}}_{n}^{(2k)},{\widetilde{I}}_{n}^{(2k + 1)} \right\rbrack,k \in \mathcal{K}_{0} := \left\{ 0,\ldots,\frac{n}{2} - 1 \right\}\),
from Section III-A in \cite{ref1} we know
\(\mu_{k} = \mu\left( \mathcal{L}_{k} \right) = \frac{1}{2}\left( {\widetilde{I}}_{n}^{(2k)} + {\widetilde{I}}_{n}^{(2k + 1)} \right) = {\widetilde{I}}_{n/2}^{(k)}\).
So with Lemma 8 we have
\({\widetilde{\sigma}}_{n}^{2} := \sigma^{2}\left( {\widetilde{\mathcal{I}}}_{n} \right) = \sigma^{2}\left( \mu_{\mathcal{K}};r_{\mathcal{K}} \right) + \mu\left( \sigma_{\mathcal{K}}^{2};r_{\mathcal{K}} \right) = {\widetilde{\sigma}}_{n/2} + \mu\left( \sigma_{\mathcal{K}}^{2};r_{\mathcal{K}} \right) \geq {\widetilde{\sigma}}_{n/2}\),
namely \({\widetilde{\mathcal{I}}}_{n}\) is increasing about code length
\(n\).

And from Proposition 2 we know
\({\overline{I}}_{n}^{(i)} = {\widetilde{I}}_{\left\langle i \right\rangle}^{\left( i^{*} \right)}\),
so make a partition
\(\mathcal{J}_{k} = \left\lbrack {\overline{I}}_{n}^{(i)} \right\rbrack_{i \in \left\{ 2^{k},\ldots,2^{k + 1} - 1 \right\}},k \in \mathcal{K :=}\left\{ 0,\ldots,m - 1 \right\}\),
there are
\(\mathcal{J}_{k}\lbrack j\rbrack = {\overline{I}}_{n}^{\left( j + 2^{k} \right)} = {\widetilde{I}}_{\left\langle j + 2^{k} \right\rangle}^{\left( \left( j + 2^{k} \right)^{*} \right)} = {\widetilde{I}}_{2^{k}}^{(j)},j \in \left\{ 0,\ldots,2^{k} - 1 \right\}\),
namely \(\mathcal{J}_{k} = {\widetilde{\mathcal{I}}}_{2^{k}}\). So
\(\mu_{k} = \mu\left( {\widetilde{\mathcal{I}}}_{2^{k}} \right) = I,\sigma_{k}^{2} = \sigma^{2}\left( {\widetilde{\mathcal{I}}}_{2^{k}} \right) = {\widetilde{\sigma}}_{2^{k}}\),
from above conclusion that we have
\({\widetilde{\sigma}}_{2^{k}} \leq {\widetilde{\sigma}}_{2^{m - 1}}\),
namely \(\sigma_{k}^{2} \leq {\widetilde{\sigma}}_{2^{m - 1}}\). With
Lemma 8, there is
\begin{align*}
	{\overline{\sigma}}_{n}^{2} &:= \sigma^{2}\left( {\overline{\mathcal{I}}}_{n} \right) = \sigma^{2}\left( \mu_{\mathcal{K}};r_{\mathcal{K}} \right) + \mu\left( \sigma_{\mathcal{K}}^{2};r_{\mathcal{K}} \right) \\
	&= \mu\left( \sigma_{\mathcal{K}}^{2};r_{\mathcal{K}} \right) \leq {\widetilde{\sigma}}_{2^{m - 1}} \leq {\widetilde{\sigma}}_{2^{m}} = {\widetilde{\sigma}}_{n}^{2}
\end{align*}

As with the strict inequality, from Section III-A in \cite{ref1} we know if
\(I(W) \in (0,1)\), in partition \(\mathcal{L}_{k}\) there are
\({\widetilde{I}}_{n}^{(2k)} < {\widetilde{I}}_{n}^{(2k + 1)},{\widetilde{\sigma}}_{k}^{2} = \sigma^{2}\left( \mathcal{L}_{k} \right) = \sigma^{2}\left( \left\lbrack {\widetilde{I}}_{n}^{(2k)},{\widetilde{I}}_{n}^{(2k + 1)} \right\rbrack \right) > 0,\forall k \in \mathcal{K}_{0}\),
so
\({\widetilde{\sigma}}_{n}^{2} = {\widetilde{\sigma}}_{n/2} + \mu\left( \sigma_{\mathcal{K}}^{2};r_{\mathcal{K}} \right) > {\widetilde{\sigma}}_{n/2}\),
and in partition \(\mathcal{J}_{k}\) there are
\({\overline{\sigma}}_{n}^{2} \leq {\widetilde{\sigma}}_{2^{m - 1}} < {\widetilde{\sigma}}_{2^{m}} = {\widetilde{\sigma}}_{n}^{2}\);
similarly if \(I(W) \in \left\{ 0,1 \right\}\),
\({\overline{\sigma}}_{n}^{2} = {\widetilde{\sigma}}_{n}^{2}\).
Therefore, \(\sigma_{n}^{2} < {\widetilde{\sigma}}_{n}^{2}\) iff
\(I(W) \in (0,1)\). Then with Theorem 1 indicating that
\(\sigma_{n}^{2} = {\widetilde{\sigma}}_{n}^{2}\), proof to Theorem 2 is
finished.

\end{document}